\documentclass[11pt,a4paper]{article}
\pdfoutput=1
\usepackage{a4wide}
\usepackage{amsfonts}
\usepackage{amssymb}
\usepackage{amsmath}
\usepackage{graphicx}
\usepackage{booktabs}
\usepackage{array}
\usepackage{color}
\usepackage{ulem}
\usepackage[small,bf]{caption}
\usepackage{cite}
\usepackage[usenames,dvipsnames]{xcolor}
\usepackage{subfigure}
\usepackage{verbatim}
\usepackage{dsfont}

\def\1{\mathbf{1}}
\def\3{\mathbf{3}}
\def\2{\mathbf{2}}

\allowdisplaybreaks[1]

\numberwithin{equation}{section}

\newcounter{mysubequation}[equation]

\definecolor{pink}{rgb}{1.,.2,.8}

\begin{document}

\begin{titlepage}

\vspace*{-15mm}
\begin{flushright}
MPP-2015-124 \\
TTP15-022
\end{flushright}
\vspace*{0.7cm}

\begin{center}
{ \bf\LARGE Renormalisation Group Corrections to\\\vspace{0.1cm} Neutrino Mass Sum Rules}
\\[8mm]
Julia Gehrlein$^{\, a,}$\footnote{E-mail: \texttt{julia.gehrlein@student.kit.edu}},
Alexander Merle$^{\, b,}$\footnote{E-mail: \texttt{amerle@mpp.mpg.de}},
Martin Spinrath$^{\, a,}$\footnote{E-mail: \texttt{martin.spinrath@kit.edu}}
\\[1mm]
\end{center}
\vspace*{0.50cm}
\centerline{$^a$ \it Institut f\"ur Theoretische Teilchenphysik, Karlsruhe Institute of Technology,}
\centerline{\it Engesserstra\ss{}e 7, D-76131 Karlsruhe, Germany}
\centerline{$^b$ \it Max-Planck-Institut f\"ur Physik (Werner-Heisenberg-Institut),}
\centerline{\it F\"ohringer Ring 6, D-80805 M\"unchen, Germany}
\vspace*{0.2cm}

\vspace*{1.20cm}

\begin{abstract}
\noindent
Neutrino mass sum rules are an important class of predictions in flavour models relating the Majorana phases to the neutrino masses. This leads, for instance, to enormous restrictions on the effective mass as probed in experiments on neutrinoless double beta decay. While up to now these sum rules have in practically all cases been taken to hold exactly, we will go here beyond that. After a discussion of the types of corrections that could possibly appear and elucidating on the theory behind neutrino mass sum rules, we estimate and explicitly compute the impact of radiative corrections, as these appear in general and thus hold for whole groups of models. We discuss all neutrino mass sum rules currently present in the literature, which together have realisations in more than 50~explicit neutrino flavour models. We find that, while the effect of the renormalisation group running can be visible, the qualitative features do not change. This finding strongly backs up the solidity of the predictions derived in the literature, and it thus marks a very important step in deriving testable and reliable predictions from neutrino flavour models.
\end{abstract}

\end{titlepage}
\setcounter{footnote}{0}

\section{Introduction}

The last two decades have lifted neutrino physics to new heights, as experiments have considerably driven the field. Nowadays it can be seen as an established fact that the neutrino mass and flavour bases are different~\cite{Gonzalez-Garcia:2014bfa}. In a nutshell, a certain flavour (say, an electron neutrino $\nu_e$) does not have a well-defined mass but is instead a superposition of three active neutrino mass eigenstates $\nu_{1,2,3}$. Mathematically, this change of the basis is described by a mixing matrix called the Pontecorvo-Maki-Nakagawa-Sakata (PMNS) matrix,
\begin{align}
U_\text{PMNS}&=R_{23}U_{13}R_{12}P_{0}\nonumber\\&=
\begin{pmatrix}
c_{12}c_{13}&s_{12}c_{13}&s_{13}\mathrm{e}^{-\mathrm{i}\delta}\\
-s_{12}c_{23}-c_{12}s_{23}s_{13}\mathrm{e}^{\mathrm{i}\delta}&c_{12}c_{23}-s_{12}s_{23}s_{13}\mathrm{e}^{\mathrm{i}\delta}&s_{23}c_{13}\\
s_{12}s_{23}-c_{12}c_{23}s_{13}\mathrm{e}^{\mathrm{i}\delta}&-c_{12}s_{23}-s_{12}c_{23}s_{13}\mathrm{e}^{\mathrm{i}\delta}&c_{23}c_{13}\\
\end{pmatrix}
P_{0},
\label{eq:U2}
\end{align}
where $\delta$ is the Dirac CP-phase and $P_{0}$=diag$(\mathrm{e}^{-\mathrm{i}\phi_1/2},\mathrm{e}^{-\mathrm{i}\phi_2/2},1)$ is a diagonal matrix containing the two Majorana phases $\phi_{1,2}$. While experimentally we have a relatively clear picture about mixing in the lepton sector, in the sense that we have by now measured all three mixing angles $\theta_{12,13,23}$ to some precision, we have no idea why their values are so large compared to those of the quark sector~\cite{pdg}. The most popular idea to explain these obvious patterns is to relate them to the properties of discrete symmetry groups~\cite{flavour-reviews}, although alternative ideas such as, e.g., a random mass pattern~\cite{Haba:2000be} or a transmission from other sectors~\cite{Adulpravitchai:2009re} do exist.

One basic problem of neutrino flavour models based on discrete symmetries is their indistinguishability at low energies: would the models predict mixing angles far from their experimental values, we would consider the models to be excluded. If, however, they predict values within the current $3\sigma$ ranges, we have a hard time distinguishing them unless the experimental precision is considerably improved, which is not to be expected very soon. If on the other hand the models predict correlations between certain low-energy observables, these could be used as additional handles. Well-known examples for testable correlations are {\it neutrino mixing sum rules}~\cite{Mixing-SR}, such as $s_{23} - \frac{1}{\sqrt{2}} = -\frac{s_{13}}{2} \cos \delta$, which, e.g., allow to predict the Dirac CP phase from some of the mixing angles. The type of correlation we would like to study here, instead, are so-called {\it neutrino mass sum rules}, which relate the three (complex) neutrino mass eigenvalues to each other. Mass sum rules have been studied for quite some time~\cite{Altarelli:2008bg,Chen:2009um,Barry:2010zk,Bazzocchi:2009da,Altarelli:2009kr,Hirsch:2008rp}, but it was only in recent years that systematic analyses have been presented~\cite{Barry:2010yk,SR11,King:2013psa}. These analyses clearly show that, among all observables, the effective neutrino mass $|m_{ee}|$ as measured in neutrinoless double beta decay ($0\nu\beta\beta$)~\cite{Rodejohann:2011mu} can be modified most significantly\footnote{The reason for other observables like the effective electron-neutrino mass square $m_\beta^2$ or the sum $\Sigma$ of all light neutrino masses not to be affected very strongly is that they contain no Majorana phases and thus are {\it only} constrained by ``half'' of the information contained in a neutrino mass sum rule, which is intrinsically a complex equation. Furthermore, both these observables are comparatively insensitive to changes in the smallest neutrino mass, if that is small by itself; they may however be affected in cases where a sum rule forbids a certain mass ordering.} -- in a way that several groups of models could be excluded in the near future, despite the uncertainties involved, in particular if in addition to a new limit or even a measurement of $0\nu\beta\beta$ information on the neutrino mass ordering (i.e., whether normal, $m_1 < m_2 < m_3$, or inverted, $m_3 < m_1 < m_2$) was available~\cite{King:2013psa}. Turning the logic round, future experiments could even ``gauge'' their definitions of stages with increasing exposure using the predictions from sum rules~\cite{SR-GERDA}.

What all previous studies have in common is that they treated neutrino mass sum rules as if they were exact to each order, i.e., as if they would be perfectly known.\footnote{Ref.~\cite{Barry:2010yk} did study the effect of possible deviations of a few per cent, however, in that case the deviation was assumed to be proportional to one of the light neutrino masses, which is not only unmotivated but also introduces an undesired measure-dependence.} This leads to relatively strong predictions such as some sum rules excluding a particular mass ordering even in the case where neutrinos are very close in mass, such that the two orderings should be hardly distinguishable~\cite{King:2013psa}. However, in general neutrino mass sum rules are very unlikely to hold exactly, since several types of corrections could appear. In particular two types of corrections are evident:
\begin{itemize}

\item First of all, mass matrices in concrete models are typically only computed at leading order. Thus, when taking into account next-to-leading order (NLO) corrections, a mass sum rule may be destroyed (although counterexamples are known~\cite{Cooper:2012bd}). Further corrections may arise, too, e.g.\ from normalising non-canonical kinetic terms correctly. However, all these types of corrections are difficult to analyse in a unified manner, since they are very model-dependent. Furthermore, it is not a priori clear how to treat the case of an ``approximate'' sum rule in an accurate way.

\item Second, as any quantity in a quantum field theory, the values of neutrino masses vary depending on the energy scale, a fact known as {\it renormalisation group evolution (RGE)} or, in a less formal manner, simply dubbed {\it running}. The running may also affect the validity of a sum rule, and in particular it may modify the allowed regions and/or open up or close down the consistency of the sum rule for a particular mass ordering. The good point with these types of corrections is that, while also they are in principle model dependent, one can at least study their effect on classes of models which can be effectively described by the Weinberg operator~\cite{Weinberg:1979sa}, i.e., where all right-handed neutrinos are so heavy as to be integrated out at a relatively high energy scale, such that their exact mass spectrum does not affect the low-energy mass matrices significantly. Note that we assume neutrinos here to be Majorana particles.

\end{itemize}
In this work we will focus on the effect of the second type of corrections, the radiative corrections, since for them practical consequences can be derived for many cases without having to resort to a model-by-model investigation. While this approach of course implies that we intrinsically disregard the first type of corrections, we would like to stress once more that in some cases the NLO corrections do {\it not} modify the sum rules~\cite{Cooper:2012bd}. While up to now no general criteria for this behaviour to appear are known, it is at least clear that it can potentially happen in which case the RGE corrections are the only general corrections that exist. Thus we will in what follows investigate the effect that RGE corrections have on the ranges predicted for $|m_{ee}|$ from neutrino mass sum rules.

This paper is organised as follows: in Sec.~\ref{sec:SumRules} we shortly review how to derive predictions from neutrino mass sum rules, before explaining in more detail both the general effect of radiative corrections and our numerical approach in Sec.~\ref{sec:RGE-Effects}. The results for all known sum rules, along with an accompanying discussion, can be found in Sec.~\ref{sec:results}. We then conclude in Sec.~\ref{sec:conc}. Some subtleties related to computing roots of complex numbers, which are decisive in order to derive the correct predictions from those sum rules involving square roots, are discussed in Appendix~\ref{sec:squareroot}.

\section{\label{sec:SumRules} Reviewing neutrino mass sum rules}

In this section we want to briefly review how neutrino mass sum rules can be parametrised and how they can be interpreted as a prediction for the Majorana phases as a function of the (physical) neutrino masses. A very detailed description can be found in Ref.~\cite{King:2013psa}, while a more pedagogical introduction to the broader topic of neutrino flavour models, featuring several example models that partially also lead to sum rules, can be gained by studying the known reviews~\cite{flavour-reviews}.

The essential feature of a flavour model incorporating a (light) neutrino mass sum rule
is that the eigenvalues of the light neutrino mass matrix depend on two complex parameters only.
Typically this can arise from any neutrino mass generation mechanism in which
the structure of one mass matrix is generated by two flavon\footnote{A flavon is a scalar field that is a total singlet under the standard model gauge symmetry, but it transforms non-trivially under the family symmetry and thus breaks it spontaneously when obtaining a vacuum expectation value (VEV).} couplings while all
other matrices only have a single scale which can be factored out.\footnote{We want to remark that there are also other cases imagineable
and indeed known. For instance, the model \cite{Gehrlein:2014wda} has only
one flavon relevant for the light neutrino masses which receives a VEV which
depends on two complex mass parameters.}
For illustrational purposes we will discuss now exactly such a two flavon case,
where we suppose that in a certain model, the light neutrino mass matrix $m_\nu$ is generated by a type~I seesaw mechanism~\cite{seesaw}, such that $m_\nu = - m_D M_R^{-1} m_D^T$ with $m_D$ ($M_R$) being the Dirac (right-handed neutrino) mass matrix.
While in the most general case, the Dirac mass matrix $m_D$ is practically arbitrary (apart from being at most of electroweak size) and the Majorana mass matrix $M_R$ is only symmetric, in a concrete flavour model their structure may be further constrained. For example, the different generations of right-handed neutrinos and charged leptons could be chosen to transform under particular representations of a family symmetry, such that the mass matrices are given by
\begin{equation}
m_D = \left(\begin{array}{ccc}
1 & 0 & 0
\\
0 & 0 & 1
\\
0 & 1 & 0
\end{array}\right) y v \; \; {\rm and}\; \;
M_R
= \left( \begin{array}{ccc}
2\alpha_s + \alpha_0 & -\alpha_s & -\alpha_s \\
-\alpha_s & 2\alpha_s & -\alpha_s + \alpha_0 \\
-\alpha_s & -\alpha_s + \alpha_0 & 2\alpha_s
\end{array}\right) \Lambda \; ,
\label{eq:matrices}
\end{equation}
where $y$ is a Yukawa coupling, $v$ is the electroweak VEV, and $\Lambda$ is a the mass scale of the right-handed neutrinos (see Ref.~\cite{King:2013psa} for details). Note that $M_R$ does have a particular structure depending on the two dimensionless couplings $\alpha_{s,0}$. These couplings arise as ratios of two different flavon VEVs and the breaking scale of the family symmetry, and the representation of the flavons determines in which entries of $M_R$ the two parameters show up. The Dirac mass matrix $m_D$, in turn, does not involve any flavon coupling and its size is entirely determined by the product of the Yukawa coupling and the electroweak VEV. While it does have some non-trivial structure owing to the family symmetry,  the mass scale $y v$ can be factored out of this matrix. This is the decisive point: looking at the resulting light neutrino mass matrix, 
\begin{equation}
 m_\nu = -\frac{y^2 v^2}{\alpha_0 (\alpha_0 + 3 \alpha_s) \Lambda}
 \begin{pmatrix}
 \alpha_0 + \alpha_s & \alpha_s & \alpha_s\\
 \alpha_s & \alpha_s (1-3b) & \alpha_s + b\alpha_0 \\
 \alpha_s & \alpha_s + b\alpha_0 & \alpha_s (1-3b) \\ 
 \end{pmatrix},
 \label{eq:nu-matrix}
\end{equation}
with $b \equiv \alpha_0 / (\alpha_0 - 3 \alpha_s)$, depends only on {\it two} paramaters (namely $\alpha_s$ and $\alpha_0$) in what its structure is concerned, while any mass scales can be factored out.\footnote{Such a structure can only be achieved if exactly one type of matrix in the seesaw formula is generated by precisely two flavon couplings, while all other mass matrices need at most one flavour (so that the VEV can be factored out).} Computing the complex eigenvalues $(\tilde{m}_{1}, \tilde{m}_{2}, \tilde{m}_{3}) = \left(\frac{1}{3\alpha_{s}+\alpha_{0}}, \; \frac{1}{\alpha_{0}}, \; \frac{1}{3\alpha_{s}-\alpha_{0}}\right) \frac{y^{2} v^{2}}{\Lambda}$ of this neutrino mass matrix, one can see immediately that a sum rule of the form
\begin{equation}
 \frac{1}{\tilde{m}_{1}} - \frac{1}{\tilde{m}_{3}} = \frac{2}{\tilde{m}_{2}}
 \label{eq:example_eigenvalues}
\end{equation}
is implied (see Sec.~\ref{sec:SR7} for a discussion of exactly this sum rule). Note that, in fact, the only true requirement on the light neutrino mass matrix to fulfill a neutrino mass sum rule is the dependence on exactly two parameters, up to an overall scale. While of course this structure is somewhat related to the mixing pattern predicted by a certain model, there is no direct constraint that the mixing would have on the sum rule. In particular, part of the mixing could be induced by the charged lepton sector which is not involved in Eq.~\eqref{eq:nu-matrix}. Note futher that, at least in the case at hand, there is in fact also a mass sum rule for the eigenvalues $(M_1,M_2,M_3) = (\alpha_0 + 3 \alpha_s, \alpha_0, -\alpha_0 + 3 \alpha_s) \Lambda$ of the right-handed Majorana mass matrix $M_R$: $M_1 - M_3 = 2 M_2$, again because it depends on exactly two parameters up to an absolute scale. While such a relation might have interesting implications at high energies, it is experimentally hardly accessible at low-energies, as long as the right-handed neutrinos are very heavy. As a final note, we should mention that typically the parameters playing the role of $\alpha_s$ and $\alpha_0$ are {\it complex} numbers, which is exactly what yields to a neutrino mass sum rule imposing a relation between the two light neutrino Majorana CP phases, while however their exact values are not constrained since the prefactor in Eq.~\eqref{eq:nu-matrix} will be complex in general.

The most general mass sum rule can be written as:\footnote{Note that we are using conventions different from the ones used in~\cite{King:2013psa}. We do this in order to match the conventions used in {\tt REAP}/{\tt MPT}~\cite{Antusch:2005gp}, which is our tool of choice to compute the RGE corrections.}
\begin{align}
A_1 \tilde{m}_1^d \text{e}^{\text{i}\chi_1}+A_2 \tilde{m}_2^d \text{e}^{\text{i}\chi_2}+A_3 \tilde{m}_3^d \text{e}^{\text{i}\chi_3}=0~,
\label{eq:mass_sumrule}
\end{align}
where $\tilde{m}_{i}$ labels the {\it complex} mass eigenvalues (i.e., including the Majorana phases), $d\neq 0$, and $A_i > 0$. The phase $\chi_i \in[0,2 \pi)$ originates from the sum rule itself, i.e., it contains both the phase of $A_i$ and a possible minus sign. Following similar steps as in~\cite{King:2013psa} we can rewrite Eq.~\eqref{eq:mass_sumrule} in a more convenient form.  First we express the complex masses in terms of the Majorana phases $\alpha_i \in [0,2\pi)$ and the physical mass eigenvalues $m_i \ge 0$ as $\tilde{m}_{i}=m_i\text{e}^{\text{i}\alpha_i}$. Dividing Eq.~\eqref{eq:mass_sumrule} by $A_3$ and using the abbreviations $c_i \equiv A_i/A_3 \geq 0$ and $\tilde{\alpha}_{i} \equiv \chi_i+d \alpha_i$, we end up with:
\begin{align}
c_1 m_1^d \text{e}^{\text{i}\tilde{\alpha}_{1}}+
c_2 m_2^d \text{e}^{\text{i} \tilde{\alpha}_{2}}+
m_3^d \text{e}^{\text{i} \tilde{\alpha}_{3}}=0~.
\label{eq:mass_sumrule2}
\end{align}
Now we can multiply Eq.~\eqref{eq:mass_sumrule2} by $\text{e}^{-\text{i}\tilde{\alpha}_{3}}=\text{e}^{-\text{i}(\chi_3+d \alpha_3)}$ and define 
\begin{align}
\Delta \chi_{i3} \equiv \chi_i-\chi_3~.
\end{align}
With the use of the Majorana phases $-\phi_{i}=\tilde{\alpha}_{i}-\tilde{\alpha}_{3}$, which appear in Eq.~\eqref{eq:U2}, we end up with our final sum rule:
 \begin{align}
 s(m_1,m_2,&m_3,\phi_1,\phi_2;c_1,c_2,d,\Delta \chi_{13},\Delta \chi_{23}) \equiv \nonumber\\
 &c_1 \left(m_1 \text{e}^{-\text{i}\phi_{1}}\right)^d 
\text{e}^{\text{i}\Delta \chi_{13}}+
c_2 \left(m_2 \text{e}^{-\text{i}\phi_{2}}\right)^d 
\text{e}^{\text{i}\Delta \chi_{23}}
+m_3^d \stackrel{!}{=} 0~.
\label{eq:parametrisation_SR}
\end{align}
Similarly as in~\cite{King:2013psa,SR11,Barry:2010yk}, we interpret the mass sum rules geometrically, since this equation describes the sum of three vectors which form a triangle in the complex plane, see Fig.~\ref{fig:dreieck_sum}. Via the law of cosines we can express the angles $\alpha \equiv -d \phi_{2}+\Delta \chi_{23}$ and $\beta \equiv -d \phi_1+\Delta \chi_{13}$ in terms of the masses:
\begin{figure}
\centering
\includegraphics[width=10cm]{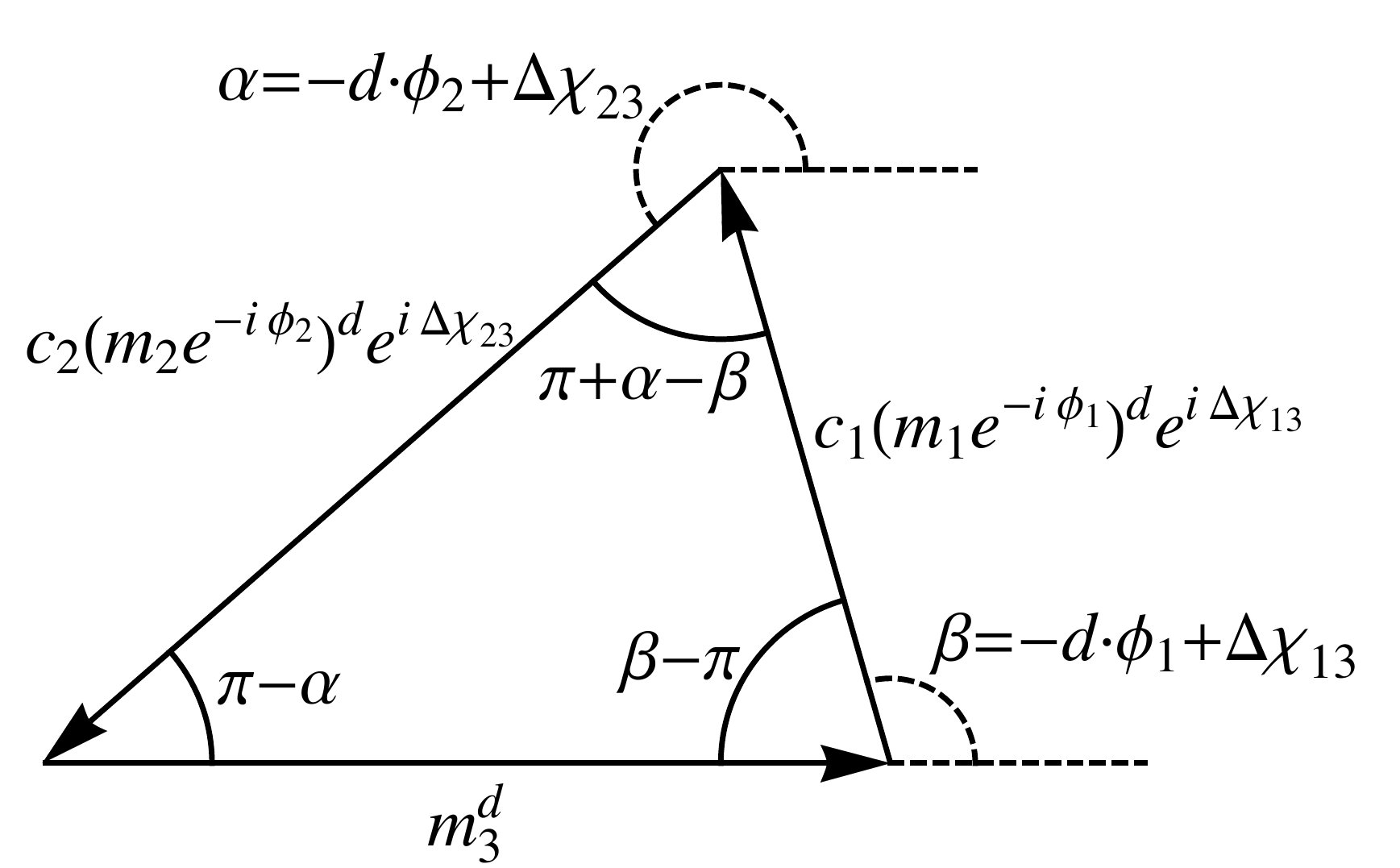}
\caption{Geometrical interpretation of the mass sum rule.}
\label{fig:dreieck_sum}
\end{figure}
\begin{align}
\cos\alpha&=\frac{c_1^2 m_1^{2d}-c_ 2^2 m_2^{2d}-m_3^{2d}}{2 c_2 m_2^d m_3^d}~, \label{eq:cosa}\\
\cos\beta&=\frac{c_2^2 m_2^{2d}-c_ 1^2 m_1^{2d}-m_3^{2d}}{2 c_1 m_1^d m_3^d}~. \label{eq:cosb}
\end{align}
These equations decide about the validity of a mass sum rule since the right-hand side has to be in the interval $[-1,1]$ to obtain real values for $\alpha$ and $\beta$ or, respectively, $\phi_1$ and $\phi_2$. Since the cosine is an even function we obtain two solutions for $\phi_i$ by taking its inverse. This is connected to the fact that the orientation of the triangle in the complex plane is irrelevant. Nevertheless one encounters a subtlety when the sum rule involves square roots of the masses. Since the square root  of a complex number is not uniquely defined, the orientation of the triangle in the complex plane in fact turns out to be important (see Appendix~\ref{sec:squareroot}).

The basic reasons for the rise of sum rules is that, in most neutrino mass models, the structure of the physical light neutrino mass matrix arises from products of several mass matrices. The most generic example is again a type~I seesaw mechanism~\cite{seesaw}, where the light neutrino mass matrix $m_\nu = - m_D M_R^{-1} m_D^T$ is generated from a multiplication of the Dirac mass matrix $m_D$ and the heavy neutrino mass matrix $M_R$. If, in this product, the structure of $m_D$ is generated by two flavon couplings, whereas the scale of $M_R$ can be factored out, this would lead to a sum rules featuring a square root, such as $\sqrt{\tilde m_1}\pm\sqrt{\tilde m_3}=2\sqrt{\tilde m_2}$. On the contrary, if the structure of $M_R$ is generated by two flavon couplings, whereas $m_D$ only features a single scale, the resulting sum rule would feature an inverse power of one, like $2/\tilde m_2=1/\tilde m_1+1/\tilde m_3$ or $1/\tilde m_1 + 1/\tilde m_2 = 1/\tilde m_3$. This scheme extends to basically all kinds of neutrino mass models, see Ref.~\cite{King:2013psa} and in the last Ref.~\cite{flavour-reviews} for more detailed explanations. In Tab.~\ref{tab:overview_SR} we have collected all the sum rules we found in the literature and which we will discuss in the following with their parameters $c_1, c_2, d, \Delta \chi_{13}$, and $\Delta \chi_{23}$ that characterise them according to Eq.~\eqref{eq:parametrisation_SR}.

\begin{table}
\centering
\begin{tabular}{c c c c c c c}
\toprule
Sum rule & References & $c_1$&$c_2$&$d$&$\Delta \chi_{13}$&$\Delta \chi_{23}$ \\
\midrule
	1& \cite{Barry:2010zk,Bazzocchi:2009da,Ding:2010pc,Ma:2005sha,Ma:2006wm,Honda:2008rs,Brahmachari:2008fn,Kang:2015xfa,SR1} &$1$&$1$&$1$&$\pi$&$\pi$\\
	2& \cite{SR2} &$1$&$2$&$1$&$\pi$&$\pi$\\
	3& \cite{Barry:2010zk,Ma:2005sha,Ma:2006wm,Honda:2008rs,Brahmachari:2008fn,Altarelli:2005yx,Chen:2009um,Chen:2009gy,Kang:2015xfa,SR3} &$1$&$2$&$1$&$\pi$&$0$\\
	4& \cite{SR4} &$1/2$&$1/2$&$1$&$\pi$&$\pi$\\
	5& \cite{SR5} &$\tfrac{2}{\sqrt{3}+1}$&$\tfrac{\sqrt{3}-1}{\sqrt{3}+1}$&$1$&$0$&$\pi$\\
	6& \cite{Barry:2010zk,Bazzocchi:2009da,Ding:2010pc,Cooper:2012bd,SR6,Gehrlein:2014wda } &$1$&$1$&$-1$&$\pi$&$\pi$\\
	7& \cite{Barry:2010zk,Altarelli:2005yx,Chen:2009um,Chen:2009gy,Altarelli:2009kr,SR7,Altarelli:2008bg} &$1$&$2$&$-1$&$\pi$&$0$\\
	8& \cite{SR8} &$1$&$2$&$-1$&$0$&$\pi$\\
	9& \cite{SR9} &$1$&$2$&$-1$&$\pi$&$\pi/2,3\pi/2$\\
	10& \cite{SR10,Hirsch:2008rp} &$1$&$2$&$1/2$&$\pi,0,\pi/2$ &$0,\pi,\pi/2$ \\
	11& \cite{SR11} &$1/3$&$1$&$1/2$&$\pi$&$0$\\
	12& \cite{SR12} &$1/2$&$1/2$&$-1/2$&$\pi$&$\pi$\\
	\bottomrule
\end{tabular}
\caption{Summary table of the sum rule we will analyse in the following. The parameters $c_1, c_2, d, \Delta \chi_{13}$, and $\Delta \chi_{23}$ that characterise them are defined in Eq.~\eqref{eq:parametrisation_SR}. In sum rule 9 and 10 two possible signs appear which lead to two possible values of $\Delta \chi_{i3}$.
Note that sum rule 10 with $\Delta \chi_{13} = \Delta \chi_{23} = \pi/2$ is the sum rule in \cite{Hirsch:2008rp} which was wrongly interpreted before.
}
\label{tab:overview_SR}
\end{table}

We do not want to explicitly discuss any models which lead to the sum rules here. Note that, however, there is no model which predicts sum rule~11, but it was shown in~\cite{SR11} that such a sum rule can be predicted in models with a type~I seesaw mechanism.

\section{\label{sec:RGE-Effects} The implications of renormalisation group running and how to compute them}

In this section, we will estimate how renormalisation group running may affect a neutrino mass sum rule. In particular, we will answer the question whether it is possible for a radiatively corrected sum rule to allow for mass orderings which are forbidden at tree-level (i.e., if the exact sum rule holds exactly). We then discuss why for some sum rules we can expect sizeable RGE effects even for a small mass scale which one would not expect from the RGE running of the parameters itself. But before we come to these two points, we start with some general remarks.

\subsection{\label{sec:basics}The general effect of radiative corrections}

The running of neutrino masses and mixing parameters is already known for quite some time, see, e.g.,\cite{Antusch:2003kp}. Naively one might expect that the running of the mixing parameters is small and that visible effects only happen if we have a large Yukawa coupling. In the SM there is no large Yukawa coupling in the lepton sector but in the MSSM, for large $\tan \beta$, the Yukawa couplings can even be of $\mathcal{O}(1)$. This expectation will be confirmed later where we find only small corrections in the SM, while for the MSSM with $\tan \beta \gtrsim 20$ the corrections start to become interesting.

On top of this effect there can be an additional enhancement of the RGE effects induced by a parametric enhancement. Some corrections, e.g., for the mixing angles and phases, are proportional to $m^2 / \Delta m^2$, where $m$ labels the lightest neutrino mass and $\Delta m^2$ is one of the neutrino mass squared differences. In the quasi-degenerate mass regime this easily yields an enhancement of  $\mathcal{O}(100)$.

As discussed in~\cite{Antusch:2003kp}, the masses themselves run mostly due to the Higgs wave
function renormalisation which includes the top Yukawa coupling but which is flavourblind and
therefore will not matter for us. But the $\tan \beta$ enhancement and the parametric enhancement
for a large neutrino mass scale will directly induce large RGE effects for the Majorana phases, such
that their low-energy values can be very different from the constrained high-energy values.\footnote{There is some subtle issue involving the mass ordering. Just by the running of the
masses itself the ordering will hardly flip. Nevertheless, by running $\theta_{12}$, for
instance, could turn negative at high energies which one might compensate by exchanging
the first and second mass state and hence $\Delta m_{21}^2$ becomes negative at the high
scale which would also lead to kinks and jumps in the running of the angles and phases. To avoid confusion with this effect we have chosen conventions where the mass hierarchy is
always preserved.
}

This can be easily seen in our numerical results later on but before we get there we want to discuss two other questions which can be understood better by estimates instead of extensive numerical parameter scans.

\subsection{\label{sec:forbidden}Trying to reconstitute forbidden mass orderings}

Some sum rules are only viable for one mass ordering at tree-level as it was already noted before~\cite{Barry:2010yk,SR11,King:2013psa}. For instance, in what we labelled sum rule~2, where $(c_1, c_2, d, \Delta \chi_{13}, \Delta \chi_{23}) = (1, 2, 1, \pi, \pi)$, only normal mass ordering is allowed. A natural question is whether RGE corrections can reconstitute the inverted ordering. To answer this question, we can have a closer look at Eq.~\eqref{eq:cosa}. For the RGE-corrected value of $\cos \alpha$, we expand the masses and find:
\begin{align}
 \cos\alpha^{\text{tree}} + \delta (\cos\alpha) &= \frac{(m_1 + \delta m_1)^{2} - 4 (m_2 + \delta m_2)^{2}-(m_3 + \delta m_3)^{2}}{4 (m_2 + \delta m_2)(m_3 + \delta m_3)} \nonumber\\
 &\approx \frac{m_1^{2} - 4 m_2^{2}-m_3^{2}}{4 m_2 m_3}
 + \frac{m_1  }{2 m_2 m_3} \delta m_1
 - \frac{ \left( m_1^2 + 4 m_2^2 - m_3^2 \right) }{4 m_2^2 m_3} \delta m_2 \nonumber\\
& - \frac{ \left({m_1}^2-4
   {m_2}^2+{m_3}^2\right)}{4 m_2 m_3^2} \delta m_3 \;,
   \label{eq:a-corr}
\end{align}
where the $m_i$ are the low energy neutrino masses and the $\delta m_i$
their respective RGE corrections.
We want to obtain inverted ordering, where $m_3 < m_1 < m_2$. This implies that
\begin{align}
m_1^{2} - 4 m_2^{2} - m_3^{2} &< - (3 m_2^2 + m_3^2) \quad \land \quad \frac{1}{m_2 m_3} < \frac{1}{m_3^2} \\
&\Rightarrow \cos\alpha^{\text{tree}} = \frac{ \left( m_1^2 - 4 m_2^2 - m_3^2 \right) }{4 m_2 m_3} < - \frac{1}{4} \left( 3 \frac{m_2^2}{m_3^2} + 1 \right)  < - 1 \;.
\end{align}
Thus, inverted ordering is ruled out on tree-level or low energies. Now, if $\delta (\cos\alpha)$ is sufficiently positive at high energies, we might nevertheless realise this regime. The correction $\delta m_i$ to the $i$-th neutrino mass eigenvalue can be estimated as~\cite{Antusch:2003kp}:
\begin{align} 
\delta m_1 &= \frac{1}{16 \pi^2} \left( \alpha_{\text{RGE}} + 2 C y_\tau^2 s_{12}^2 s_{23}^2 \right) m_1 \log \frac{\mu}{M_Z} + \mathcal{O}(\theta_{13}) \;,\\
\delta m_2 &= \frac{1}{16 \pi^2} \left( \alpha_{\text{RGE}} + 2 C y_\tau^2 c_{12}^2 s_{23}^2 \right) m_2  \log \frac{\mu}{M_Z}+ \mathcal{O}(\theta_{13}) \;,\\
\delta m_3 &= \frac{1}{16 \pi^2} \left( \alpha_{\text{RGE}} + 2 C y_\tau^2 c_{23}^2 \right) m_3 \log \frac{\mu}{M_Z} + \mathcal{O}(\theta_{13}) \;,
\end{align}
where we have assumed the parameters to be constant at leading order, so that we integrate the $\beta$ function between the $Z$-scale $M_Z$ and $\mu > M_Z$. Note that $\alpha_{\text{RGE}} \approx 3$ is a function of gauge and Yukawa couplings, while $C =  -3/2$ in the Standard Model (SM) and $C = 1$ in the minimal supersymmetric Standard Model (MSSM). 

We can plug this back into Eq.~\eqref{eq:a-corr} and use $\theta_{23} \approx \pi/4$ and $\sin \theta_{12} \approx 1/\sqrt{3}$ to find:
\begin{align}
\delta (\cos\alpha) \approx - \frac{C y_\tau^2}{192 \pi^2}  \frac{3 m_1^2 - 4 m_2^2 + m_3^2}{m_2 m_3}  \log \frac{\mu}{M_Z}  \;.
\end{align}
The first thing to note is that the dependence on $\alpha_{\text{RGE}}$ drops out (which is true for all sum rules). Because of $3 m_1^2 - 4 m_2^2 + m_3^2 < 0$, the corrections are negative for the MSSM and hence they make $\cos \alpha$ even smaller. For the SM, on the other hand, they would have the right sign -- but then we would need $\frac{C y_\tau^2}{192 \pi^2} \log \frac{\mu}{M_Z}$ to be of $\mathcal{O}(1)$, which implies $\mu$ to be way beyond the Planck scale. Hence, we conclude that the inverted ordering {\it cannot} be reconstituted by RGE corrections for the second sum rule.

Similar estimates can be done for the other sum rules with missing mass orderings (i.e., sum rules 2, 3, 4, 5, 10, 12). In fact, for sum rule~12 ($c_1=1/2$, $c_2=1/2$, $d=-1/2$, $\Delta \chi_{13}=\pi$, $\Delta \chi_{23}=\pi$), the MSSM corrections would have the right sign so that we could hope to reconstitute the missing ordering in that case, but according to our estimate we would need for a neutrino mass scale below 1~eV a $\tan \beta$ value of more than 500, which practically excludes this possibility as well.

\subsection{\label{sec:smallmasses}Impact of the RGE corrections for a small mass scale}

By looking at the formulas for the RGE effects on the masses and hence on $\cos \alpha$, one might think that they have barely an impact for a small mass scale since the RGEs are proportional to the mass scale itself.  But we will show that indeed also a small mass scale can lead to significant corrections for $\cos \alpha$, which happens due to a parametric enhancement of the form $\sqrt{ \Delta m^2/m^2}$ that is large for small masses.

As an example we have a look at sum rule~1 with an inverted ordering ($m = m_3$). The expression for  $\cos\alpha$ for sum rule 1 reads 
\begin{align}
 \cos\alpha^{\text{tree}} + \delta (\cos\alpha) &\approx\frac{m_1^2-m_2^2-m_3^2}{2 m_2 m_3}+\frac{C y_\tau^2}{96  \pi^2} \log\frac{\mu}{M_Z}\left(\frac{-3 m_1^2+m_2^2-m_3^2}{m_2 m_3}\right)~,
\end{align}
where we have already used the previous estimates and $\theta_{23} \approx \pi/4$ and $\sin \theta_{12} \approx 1/\sqrt{3}$. Expressing the masses in terms of $m_3$ via the mass squared differences and neglecting small terms of order $\sqrt{ m_3^2/\Delta m^2}$, we end up with a negative tree-level value:
\begin{align}
\cos \alpha ^{\text{tree}}\approx-\frac{\Delta m_{21}^2}{2 m_3 \sqrt{|\Delta m_{32}^2|}}~.
\end{align}
For the correction term we find:
\begin{align}
 \delta (\cos\alpha) \approx - \frac{C y_\tau^2}{96  \pi^2} \log\frac{\mu}{M_Z}\left(\frac{2 |\Delta m_{32}^2|}{m_3 \sqrt{|\Delta m_{32}^2|}  }\right)~.
 \end{align}
From the tree-level term we get a lower bound on $m_3$, given by $m_3=7.6 \cdot 10^{-4}$~eV. The correction has a negative sign in the MSSM -- just as the leading order term -- which means that the corrections increase the lower bound on the masses for this sum rule. For the mass where $\cos \alpha^\text{tree} = -1$, the correction further decreases the value of $\cos\alpha$ and hence it will be forbidden. In addition, we see that the corrections are enhanced for a small mass scale due to the small values in the denominator. Therefore the total $\cos \alpha$ gets very sensitive to small changes in $m_3$ (due to RG corrections) and hence the allowed range for the Majorana phases at the high scale becomes larger.

This correction differs from the leading order term by a factor of $A= 4\frac{C y_\tau^2}{96  \pi^2} \log\frac{\mu}{M_Z} \frac{|\Delta m_{32}^2|}{\Delta m_{21}^2}$. To get a better feeling for the size of the effect of the corrections we plug in $\tan \beta=50$, $m = 8 \cdot 10^{-4}$~eV, and $\mu = M_S = 10^{13}$~GeV to get:
\begin{align}
  A \approx 0.83~.
\end{align}
This implies an increase of the lower bound of $m_3$ by $83\%$. For sum rule~4 we find in a similar way a $83\%$ correction to $m_3$, as well. Other sum rules in both orderings do not cover such small mass scales, and hence we do not find such an enhancement of the RGE corrections for them. 

Later on, in our numerical scans, we will obtain a lower bound for $m_3$ for sum rule~1 of about $m\approx 9.1 \cdot 10^{-4}$~eV, which is a {\it much smaller effect than our estimate suggests}. But indeed for the parameter point we mentioned our one-step  approximation for the $\beta$ functions is not very good because also the angles, especially $\theta_{12}$, run significantly.  Nevertheless, with our estimates we can easily understand the apparent broadening of the allowed region for sum rules~1 and~4 which we will see later on.

\subsection{\label{sec:numerical}The numerical approach}

Since the constraints on the mixing parameters are satisfied at different energy scales, their experimental ranges, cf.\ Tab.~\ref{tab:exp_parameters}, restrict the mixing angles and the mass squared differences at a low energy scale $M_Z$.\footnote{To be precise, the actual experiments performed detect neutrinos of even lower energies, between MeV and GeV depending on the source. However, it is a well-known fact that the change of these parameters between the scale $M_Z$ and low energies is negligible, provided that the particle content is that of the SM.}

In contrast, the mass sum rules in fact constrain the Majorana phases as functions of the lightest neutrino mass which we will label in the following simply as $m$ at a high energy scale, where we assume the sum rule to be predicted by the respective flavour model. As a generic choice we set this scale equal to the seesaw scale~\cite{seesaw}, $M_S \approx 10^{13}$~GeV, and we then employ a running procedure between $M_S$ and $M_Z$. Choosing a different scale instead would not change our results dramatically, as long as it is not different from $M_S$ by several orders of magnitude, so that the running would extend over a considerably larger or smaller energy range.

In our scans we will present results for the SM extended minimally by the Weinberg operator to accommodate neutrino masses, as well as for the MSSM plus the Weinberg operator. The most relevant supersymmetry (SUSY) parameter for the running is $\tan \beta$, which we have chosen to be 30 or 50 in our scans, while the exact mass spectrum of the SUSY particles plays hardly any role. We have fixed the SUSY scale, where we switch from SM to MSSM RGEs, to 1~TeV but -- again -- the dependence on this scale is only logarithmic and hence very weak. Furthermore we have neglected the SUSY threshold corrections for the masses and mixing parameters~\cite{SUSYthresholds}. Both for small $\tan \beta$ and in the SM the running is small, and hence the results in these two cases would be very similar. In fact the SM results will look very similar to the results without RGE effects at all, cf.\ Ref.~\cite{King:2013psa}, as to be expected from the small size of the relevant corrections.

In order to perform our numerical computations, we have made use of the \texttt{REAP}/\texttt{MPT} package~\cite{Antusch:2005gp}. To do this we run the parameters up to the high scale $M_S$ and calculate there the modulus of the sum rule [i.e., of the left-hand side of Eq.~\eqref{eq:parametrisation_SR}], which is minimised with respect to the low energy Majorana phases such that, if the sum rule holds, the minimum of the modulus should yield a numerical zero. The mixing angles and mass squared differences are varied within their experimental $3 \sigma$-ranges,\footnote{In principle we could have employed a similar procedure using their best-fit values at low energies. However, since the continuous comparison between low- and high-energy scales is numerically expensive, this would blow up the computational time without significant gain.} while $\delta$ and $m$ are free parameters at the low scale. We vary the Dirac CP phase $\delta$ between 0 and $2 \pi$, since it has not been measured yet,\footnote{Note that the global fits indicate some finite range at $1\sigma$ level, however, this is not a significant tendency at the moment.} and we have also scanned over values for the lightest mass between $1\cdot 10^{-4}$ and $0.15$~eV. The upper bound on $m$ is chosen in accordance with the cosmological bound  on the sum of the neutrino masses~\cite{Planck:2015xua}:
\begin{equation}
\sum m_{\nu} < 0.17\text{ eV,}
\label{eq:kosmoschranke}
\end{equation} 
although we should note that what is displayed is the average limit taken between the two mass orderings.

\begin{table}
\centering
\begin{tabular}{lcc} 
\toprule
Parameter & best-fit ($\pm 1\sigma$) & $ 3\sigma$ range\\ 
\midrule 
$\theta_{12}$ in $^{\circ}$ & $ 33.48^{+0.78}_{-0.75}$& $31.29\rightarrow 35.91$\\[0.5 pc]
$\theta_{13}$ in $^{\circ}$ & $ 8.50^{+0.20}_{-0.21}\oplus 8.51^{+0.20}_{-0.21} $& $7.85\rightarrow 9.10 \oplus 7.87\rightarrow 9.11$\\[0.5 pc]
$\theta_{23}$ in $^{\circ}$ & $ 42.3^{+3.0}_{-1.6}\oplus 49.5^{+1.5}_{-2.2}$ & $38.2\rightarrow 53.3 \oplus 38.6\rightarrow 53.3$\\[0.5 pc]
$\delta$  in $^{\circ}$&$251^{+67}_{-59}$&$0\rightarrow 360$\\
\midrule
$\Delta m_{21}^{2}$ in $10^{-5}$~eV$^2$ & $7.50^{+0.19}_{-0.17}$ & $7.02\rightarrow 8.09$\\[0,5 pc]
$\Delta m_{31}^{2}$ in $10^{-3}$~eV$^2$~(NO) &$2.457^{+0.047}_{-0.047}$&$2.317\rightarrow 2.607$\\[0,5 pc]
$\Delta m_{32}^{2}$ in $10^{-3}$~eV$^2$~(IO) &$-2.449^{+0.048}_{-0.047}$&$-2.590\rightarrow -2.307$\\
\bottomrule
\end{tabular}
\caption{The best-fit values and the 3$\sigma$ ranges for the parameters taken from~\cite{Gonzalez-Garcia:2014bfa}. The two minima for both $\theta_{13}$ and $\theta_{23}$ correspond to normal and inverted mass ordering, respectively.}
\label{tab:exp_parameters}
\end{table}

One might wonder if the running of the Majorana phases is sufficiently large to alter the results obtained in previous studies~\cite{King:2013psa}. We will show that the running of the Majorana phases is negligibly small in the SM, whereas we do see a substantial effect in the MSSM. Furthermore we will show that the running is not identical for $\phi_1$ and $\phi_2$.
 
The running of the Majorana phases is given by~\cite{Antusch:2005gp}:
\begin{align}
\dot{\phi}_1&=\frac{C y_{\tau}}{4 \pi^2}\left(m_3 \cos (2\theta_{23}) \frac{m_1 s_{12}^2 \sin \phi_1+m_2 c_{12}^2 \sin \phi_2}{\Delta m^2_{32}}+\frac{m_1 m_2 c_{12}^2 s_{23}^2 
 \sin (\phi_1-\phi_2)}{\Delta m_{21}^2}\right)+ \mathcal{O}(\theta_{13})~,\\
 \dot{\phi}_2&=\frac{C y_{\tau}}{4 \pi^2}\left(m_3 \cos (2\theta_{23}) \frac{m_1 s_{12}^2 \sin \phi_1+m_2 c_{12}^2 \sin \phi_2}{\Delta m^2_{32}}+\frac{m_1 m_2 s_{12}^2 s_{23}^2 
 \sin (\phi_1-\phi_2)}{\Delta m_{21}^2}\right)+ \mathcal{O}(\theta_{13})~,
 \label{eq:running_majoranas}
\end{align}
where we have used the abbreviations $s_{ij}\equiv \sin \theta_{ij}$ and $c_{ij}\equiv \cos \theta_{ij}$, and we have neglected factors proportional to the small quantity $\tfrac{\Delta m_{21}^2}{\Delta m_{32}^2}$. The two formulas are identical except for the factor $c_{12}^2$ for $\dot{\phi}_1$ instead of $s_{12}^2$ for $\dot{\phi}_2$ in the last term. This small difference is nevertheless crucial for the difference in the running of the phases. Since $c_{12}^2$ is about two times larger than  $s_{12}^2$ for the best-fit value of $\theta_{12}$, and since this term is additionally enhanced compared to the first term in~\eqref{eq:running_majoranas} due to the small mass difference in the denominator, the running of $\phi_1$ is considerably stronger than the running of $\phi_2$. With increasing mass scale, the RGE effects drive $\theta_{12}$ to smaller values. This further enlarges the difference in the running of the Majorana phases, since  $s_{12}^2$ is decreasing whereas $c_{12}^2$ is increasing with smaller values of $\theta_{12}$.
 
\begin{figure}[t]
\begin{tabular}{lr}
\includegraphics[width=6.6cm]{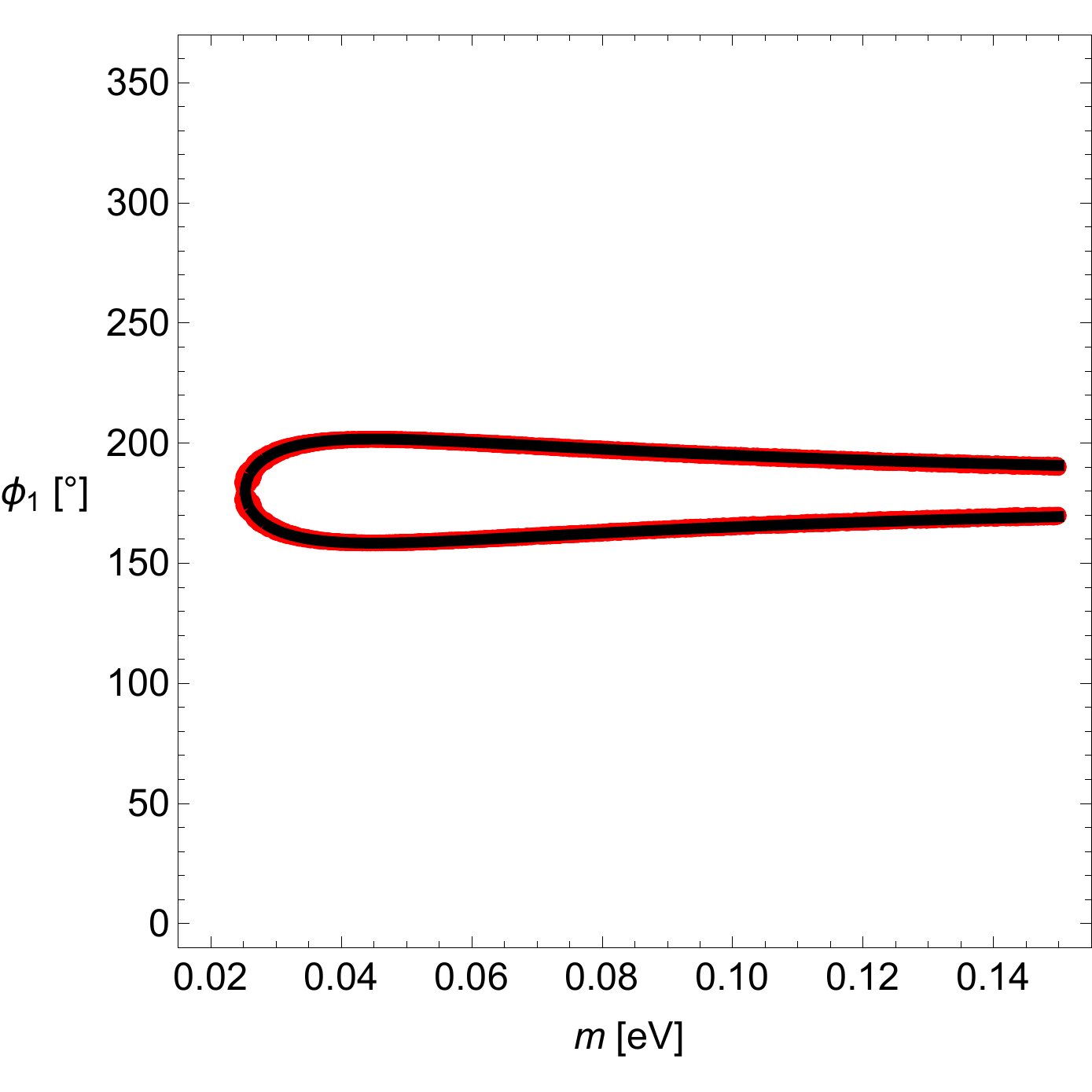} &
\includegraphics[width=6.6cm]{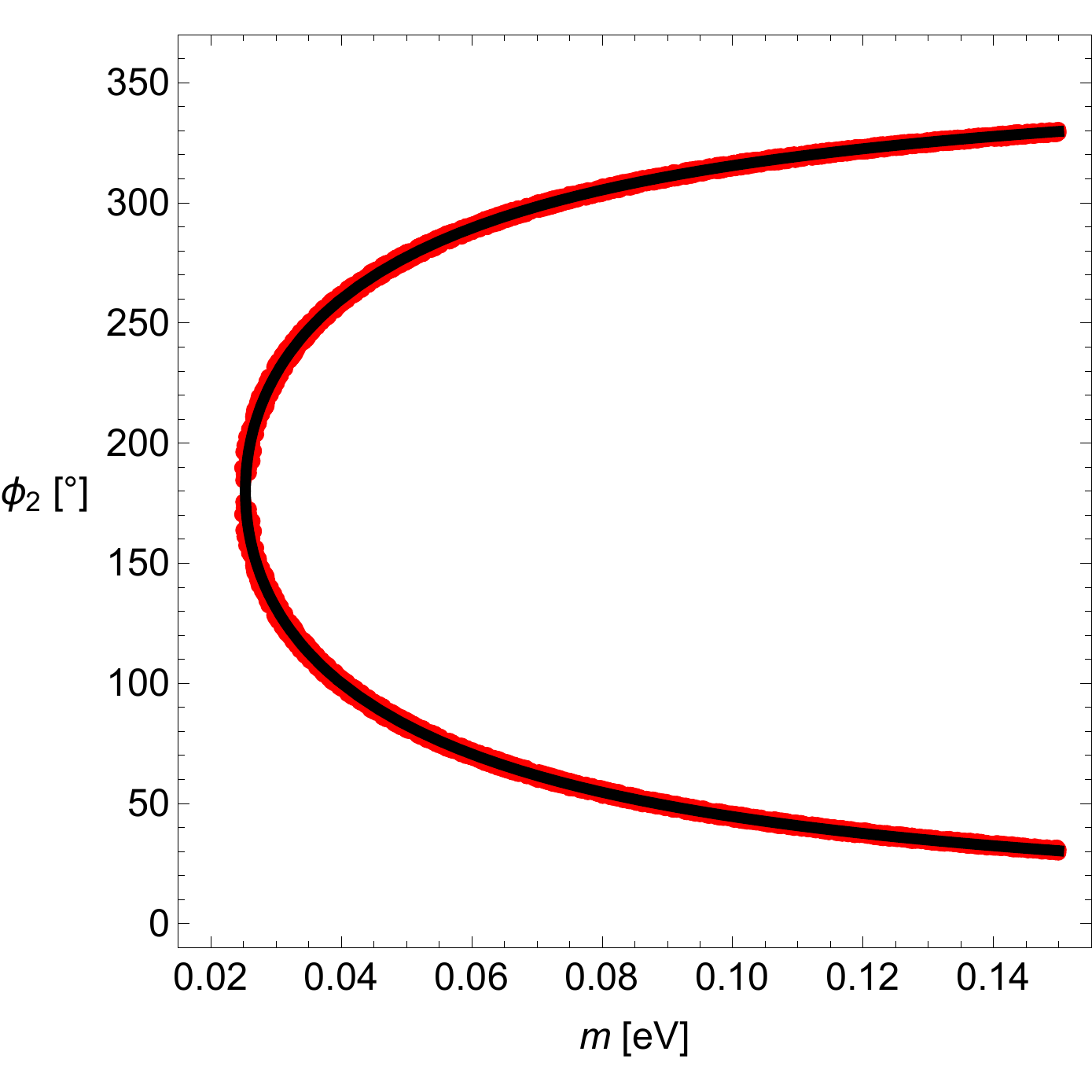}\\
\includegraphics[width=6.6cm]{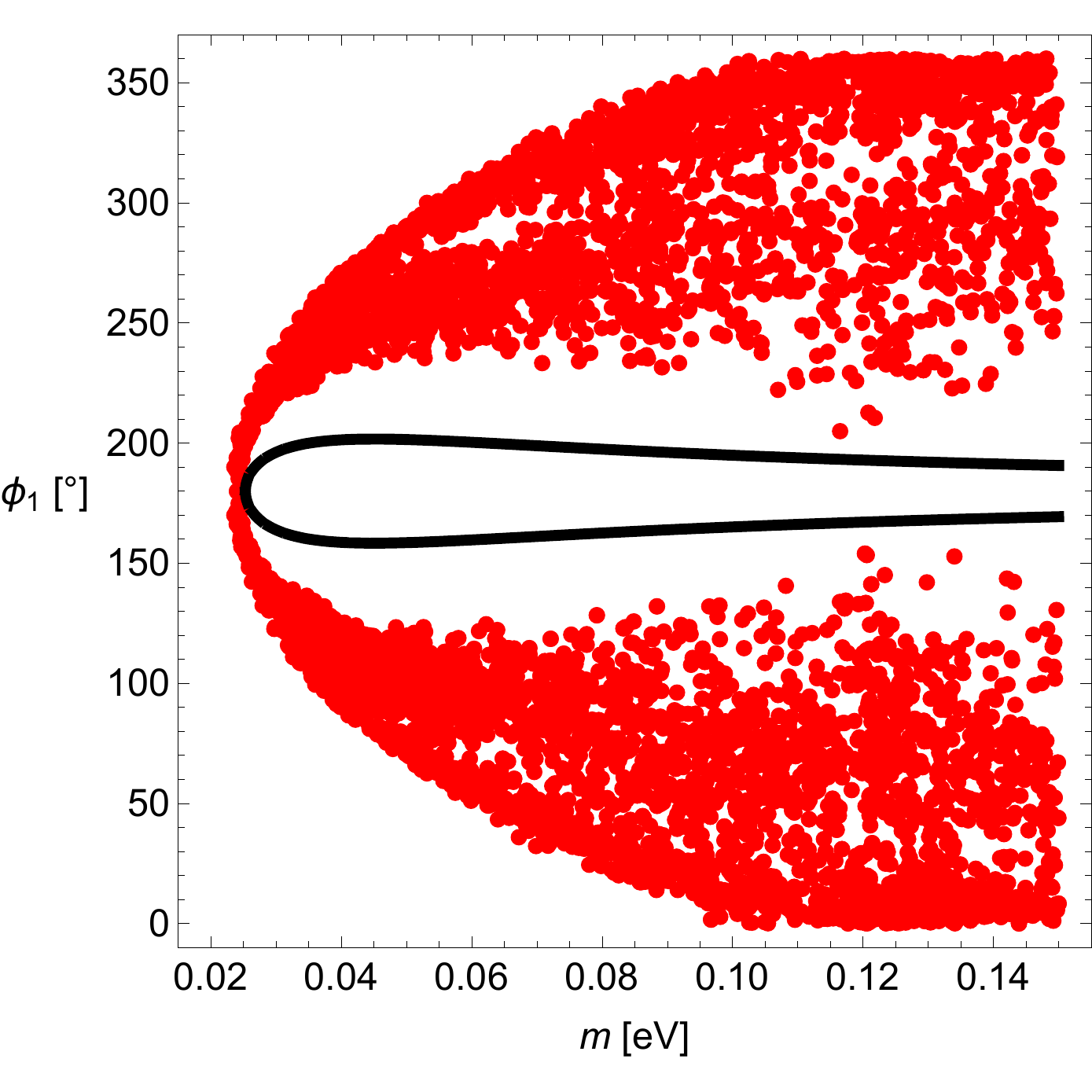} &
\includegraphics[width=6.6cm]{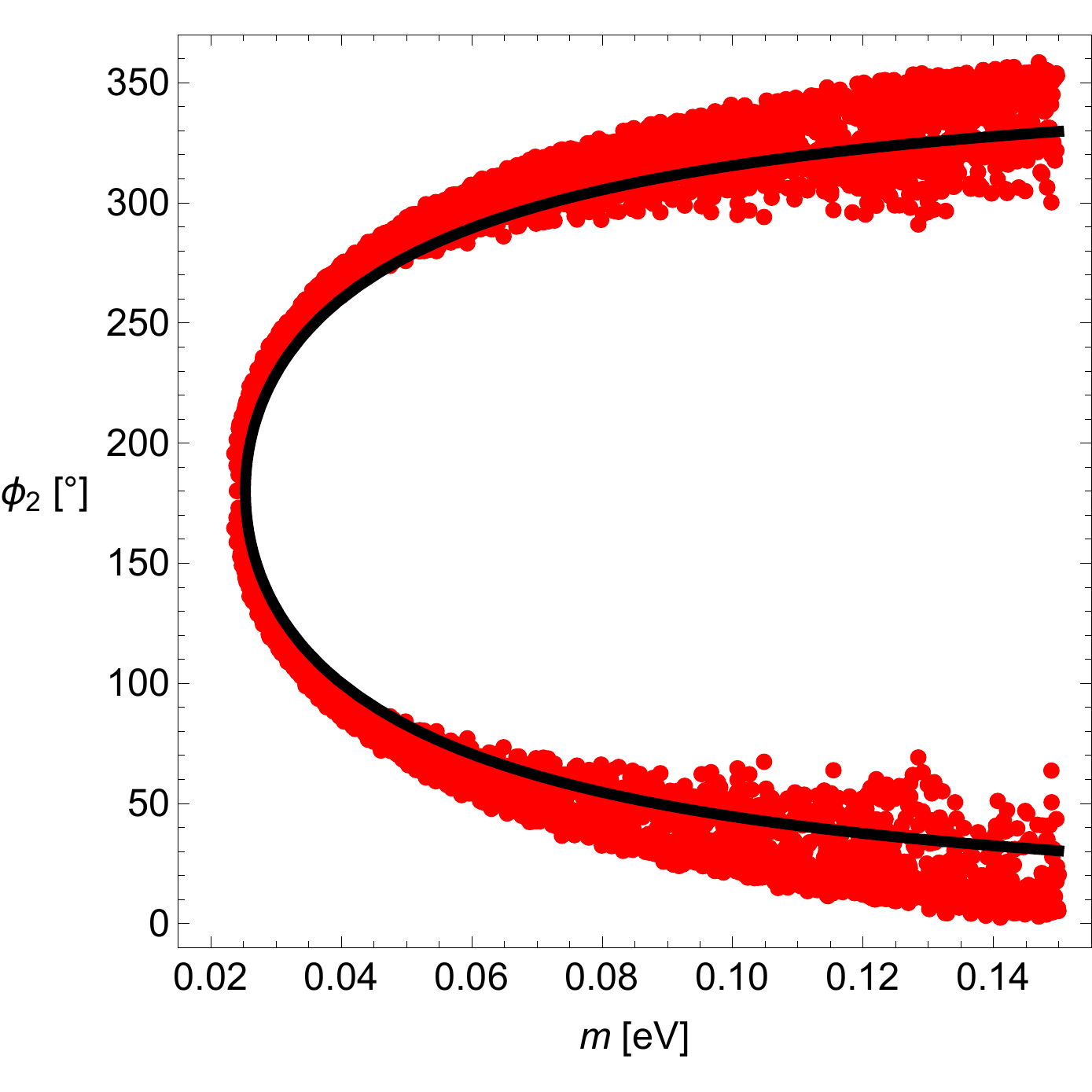}
\end{tabular}
\caption{Predicted values of the Majorana phases in the SM (upper plots) and in the MSSM with $\tan \beta=50$ (lower plots) as a function of $m$ (which is $m_3$ in the case of sum rule~5). The black lines represent the predicted low scale values of the Majorana phases without taking RGE corrections into account, while the red points are the results of our numerical approach.}
\label{fig:running_phases}
\end{figure}
 
Due to the enhancement in the $\beta$-functions of the MSSM governed by $\tan \beta$, we see this difference best in the running of the phases in the MSSM. As a typical example we compare the predicted low scale values of the Majorana phases as a function of $m$ coming from sum rule~5 within the SM and the MSSM for $\tan \beta=50$ in Fig.~\ref{fig:running_phases}. The black lines represent the predicted values for the phases without taking the RGE corrections into account, while the red points are the results from our numerical approach as described above. For $\phi_1$ the red points strongly deviate from the black lines whereas the points for $\phi_2$ gather around the black lines which supports our argument.

Since the Majorana phases themselves are not directly measurable in the near future, we will present in the following section our results in terms of predictions for the allowed range of the effective neutrino mass $|m_{ee}|$ as potentially measured in $0\nu\beta\beta$. This observable is explicitly given by:
\begin{equation}
|m_{ee}|=\left|m_{1} U_{e1}^{2}+m_{2} U_{e2}^{2}+m_{3} U_{e3}^{2}\right|=\left| m_{1}c_{12}^{2}c_{13}^{2}\text{e}^{-\text{i}\phi_{1}}+m_{2}s_{12}^{2}c_{13}^{2}\text{e}^{-\text{i}\phi_{2}}+m_{3}s_{13}^{2}\text{e}^{-2 \text{i} \delta}\right|.
\label{eq:mee}
\end{equation}
In all cases, as explained above, we will compute the predictions for the SM and for the MSSM (with $\tan\beta = 30$ and $50$), the latter of which can lead to considerably different predictions.

\section{\label{sec:results}Numerical results for concrete sum rules}

In this section we employ the procedure as described in the previous section to obtain allowed ranges for the smallest neutrino mass eigenvalue $m$ and $|m_{ee}|$ for all sum rules we found in the literature. Note that the numerical values obtained in this section may be limited by the finite statistics of our numerics. Furthermore, the oscillation parameters have been updated since the time when Ref.~\cite{King:2013psa} has been written, which accounts for the small differences we obtain compared to that reference. In our plots regions with inverted mass ordering are drawn in yellow while regions with normal mass ordering are drawn in blue.

\subsection{\label{sec:SR1} Sum Rule 1: $\mathbf{\tilde m_1+ \tilde m_2= \tilde m_3}$}

The parameters for this sum rule are $(d, c_1, c_2, \Delta \chi_{13}, \Delta \chi_{23})=(1, 1, 1, \pi, \pi)$, and the corresponding plots look like:
\begin{center}
\begin{tabular}{lll}
\hspace{-1cm}
\includegraphics[width=5.4cm]{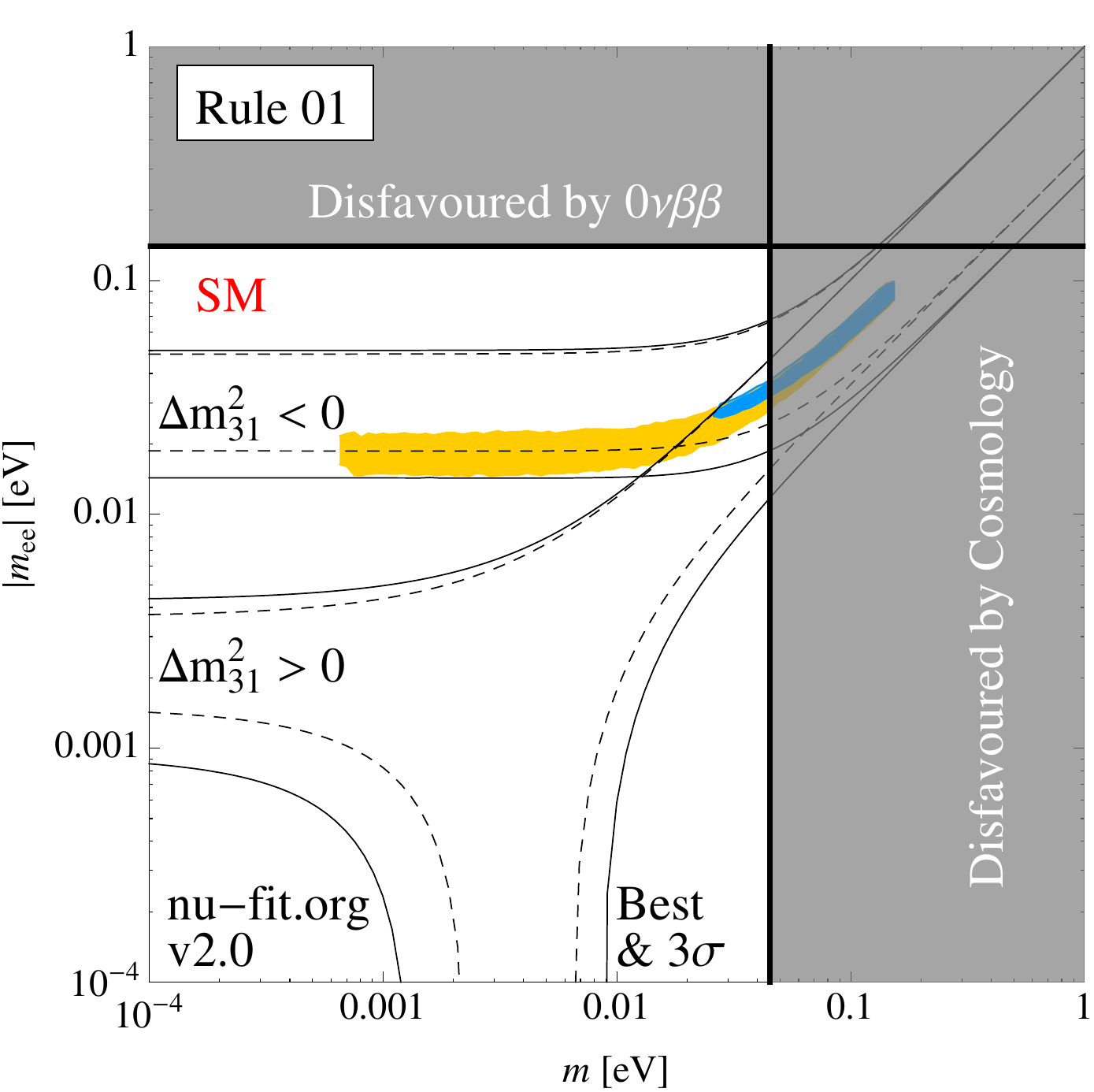} &
\includegraphics[width=5.4cm]{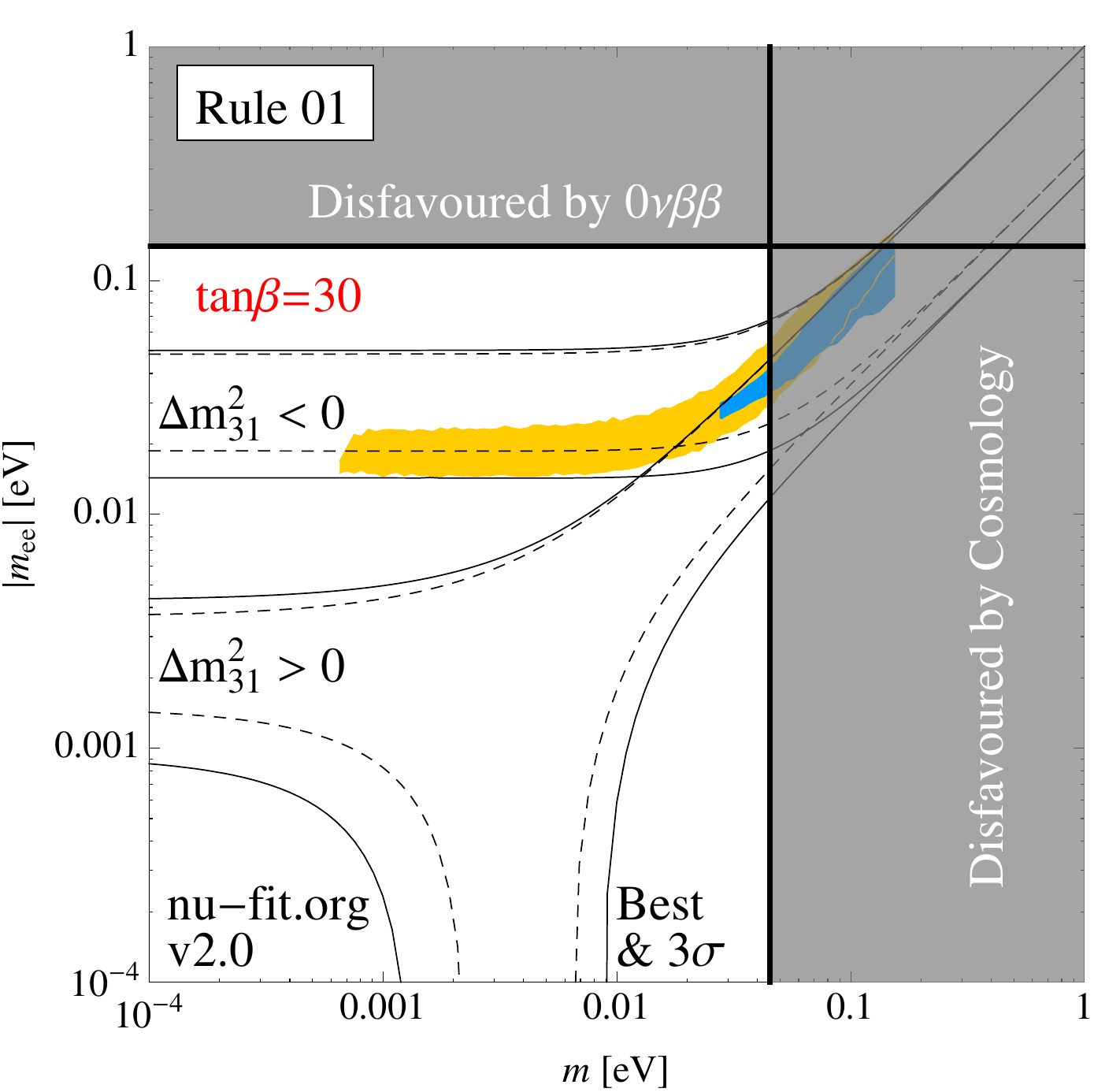} &
\includegraphics[width=5.4cm]{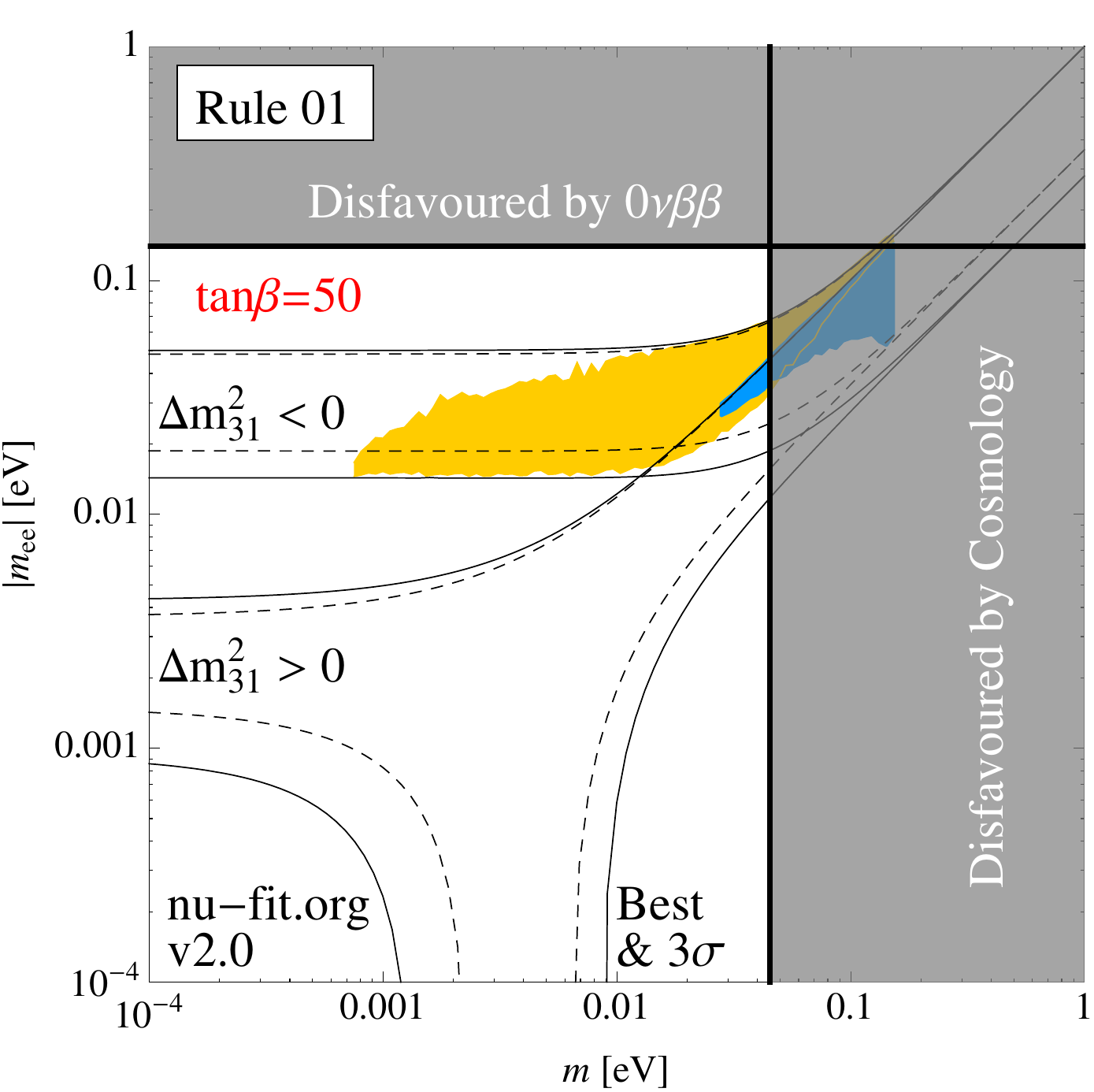}
\end{tabular}
\end{center}
This sum rule yields $(m_{\rm min}, |m_{ee}|_{\rm min}) = (0.026,0.026)$~eV ($(0.00065,0.015)$~eV) for normal (inverted) mass ordering, if running with the SM particle content is applied, which is consistent with the values obtained in Ref.~\cite{King:2013psa} (see discussion in Sec.~7.7 therein). For $\tan \beta = 30$ ($50$), the values change to $(0.028,0.026)$~eV ($(0.028,0.026)$~eV) for NO and to $(0.00065,0.014)$~eV ($(0.00075,0.015)$~eV) for IO, respectively.

\subsection{\label{sec:SR2} Sum Rule 2: $\mathbf{\tilde m_1= \tilde m_3-2 \tilde m_2}$}

The parameters for this sum rule are $(d, c_1, c_2, \Delta \chi_{13}, \Delta \chi_{23})=(1, 1, 2, \pi, \pi)$, and the corresponding plots look like:
\begin{center}
\begin{tabular}{lll}
\hspace{-1cm}
\includegraphics[width=5.4cm]{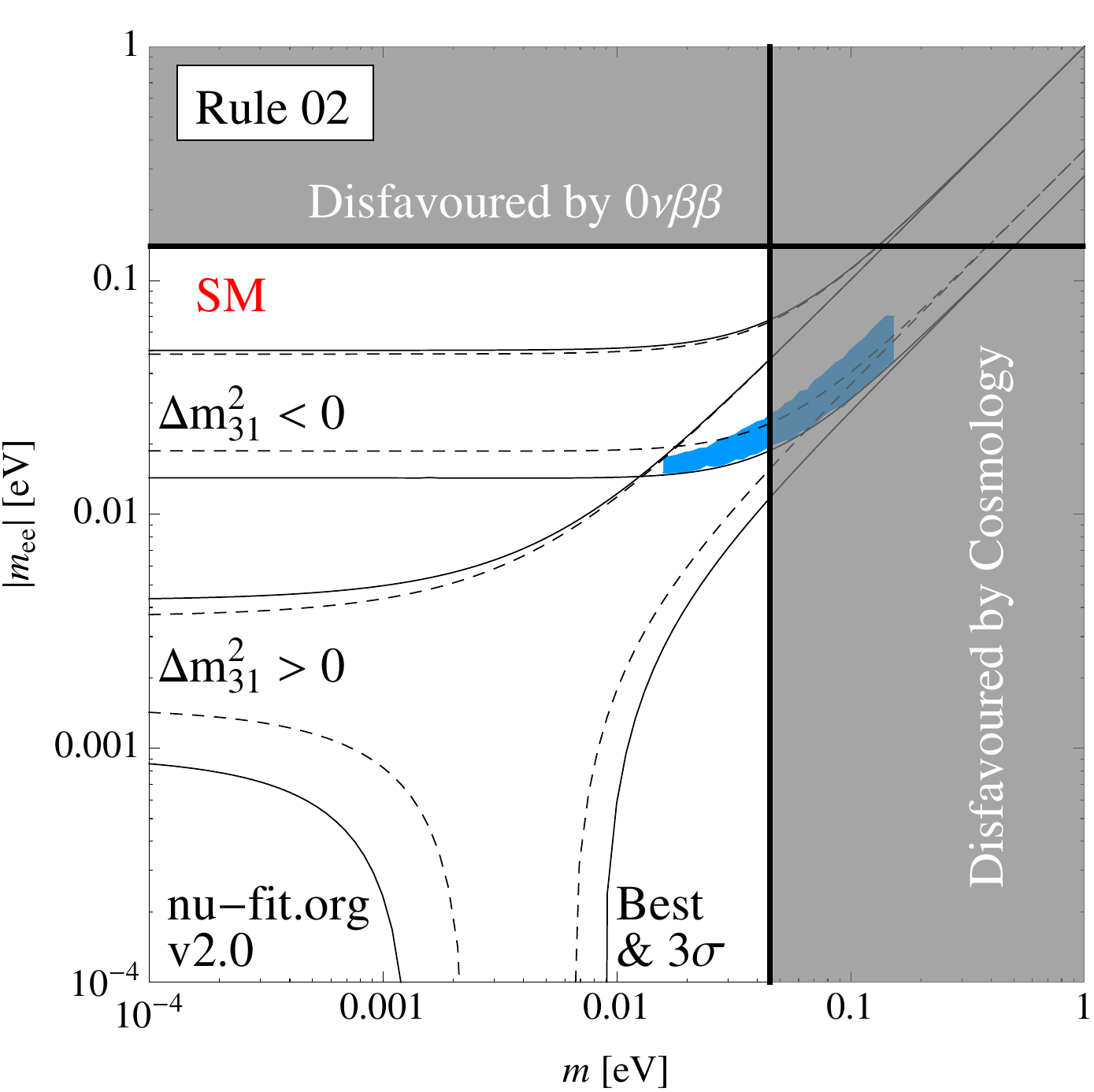} &
\includegraphics[width=5.4cm]{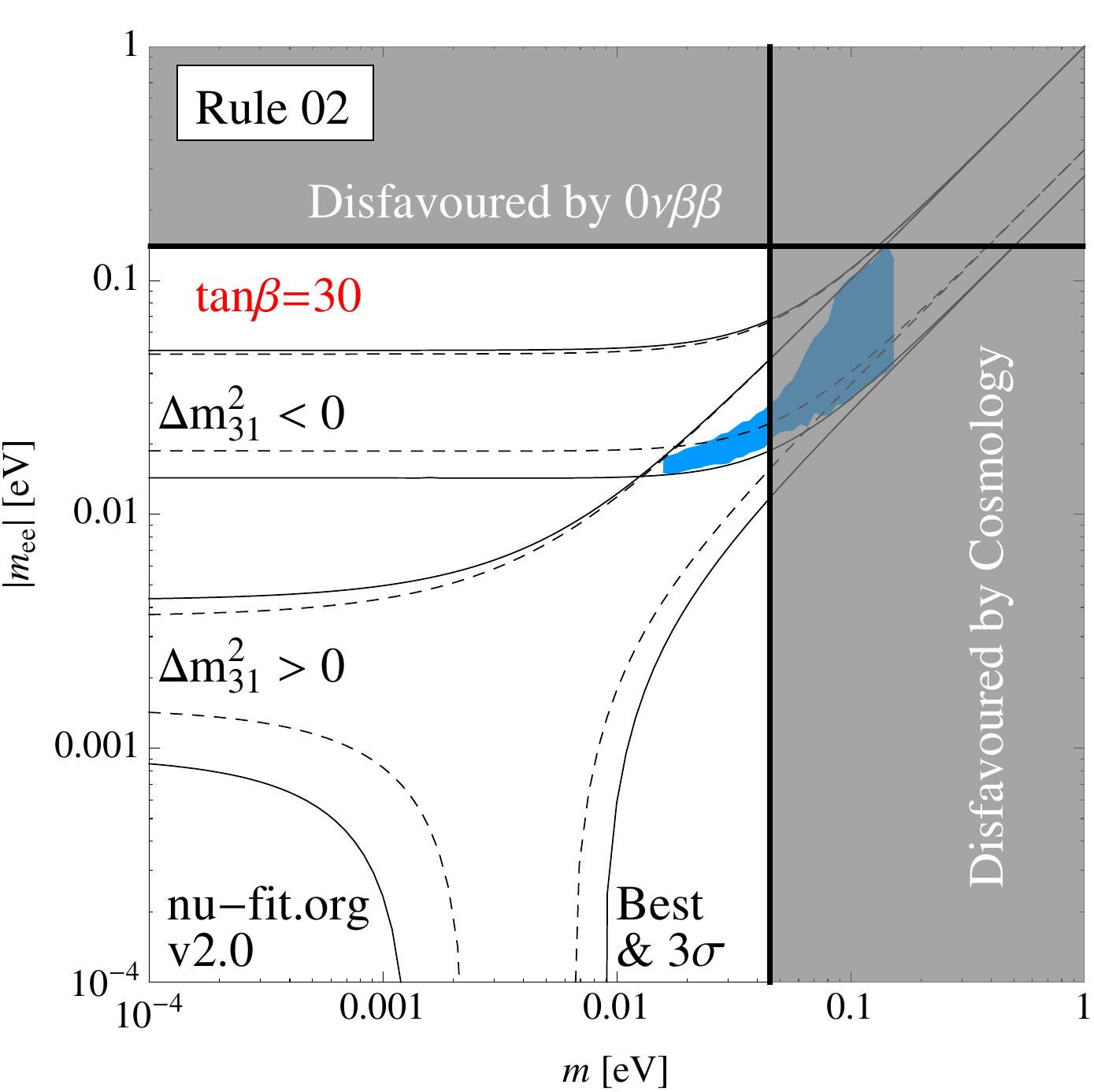} &
\includegraphics[width=5.4cm]{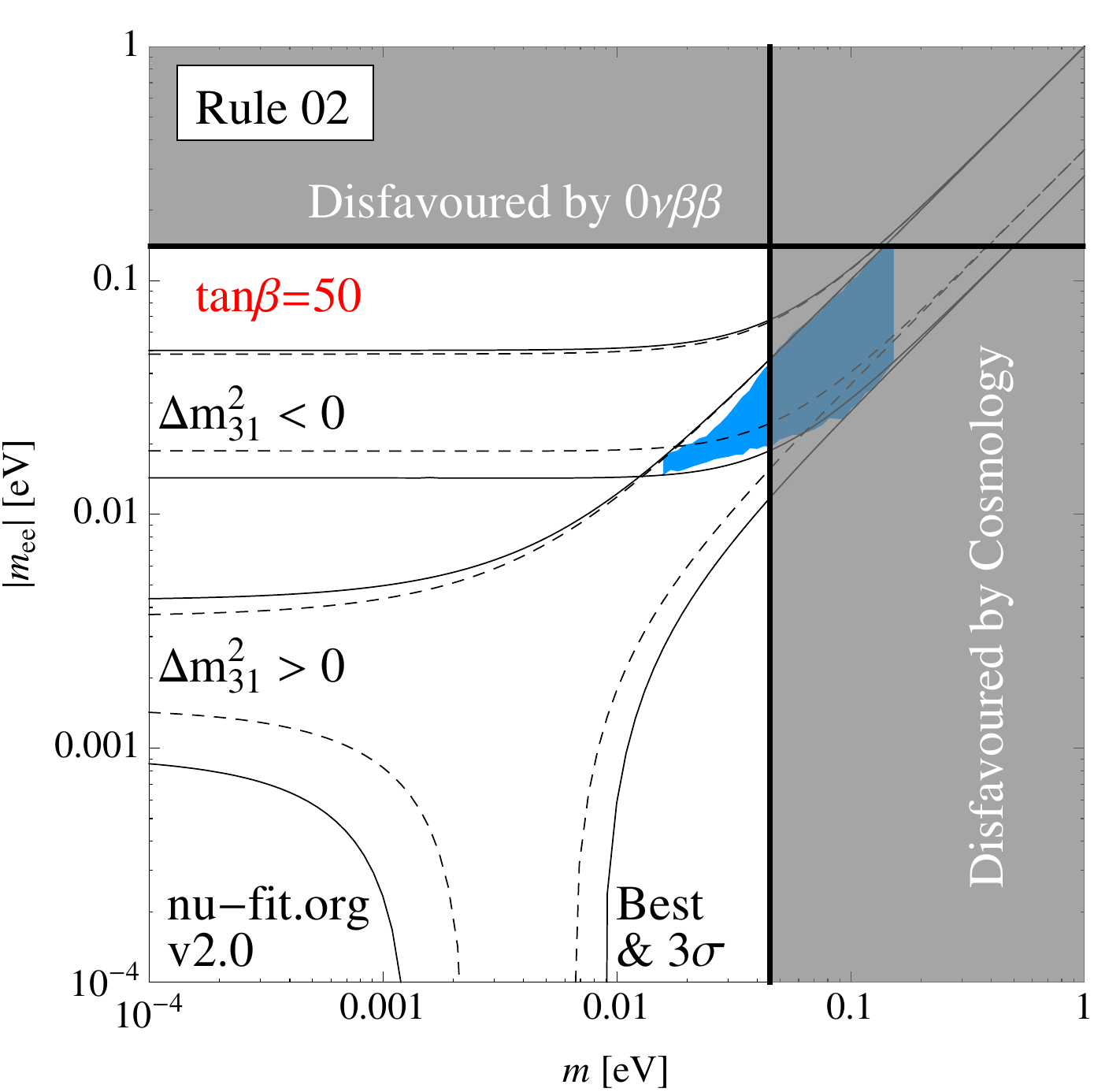}
\end{tabular}
\end{center}
This sum rule predicts normal ordering only, and with the SM particle content it yields $(m_{\rm min}, |m_{ee}|_{\rm min}) = (0.016,0.015)$~eV, if the running is applied. These numbers are consistent with the values obtained in Ref.~\cite{King:2013psa} (see discussion in Sec.~7.10 therein). For $\tan \beta = 30$ ($50$), the values basically remain at $(0.016,0.015)$~eV ($(0.016,0.015)$~eV), while still only NO is allowed.

\subsection{\label{sec:SR3} Sum Rule 3: $\mathbf{\tilde m_1=2 \tilde m_2+ \tilde m_3}$}

The parameters for this sum rule are $(d, c_1, c_2, \Delta \chi_{13}, \Delta \chi_{23})=(1, 1, 2, \pi, 0)$, and the corresponding plots look like:
\begin{center}
\begin{tabular}{lll}
\hspace{-1cm}
\includegraphics[width=5.4cm]{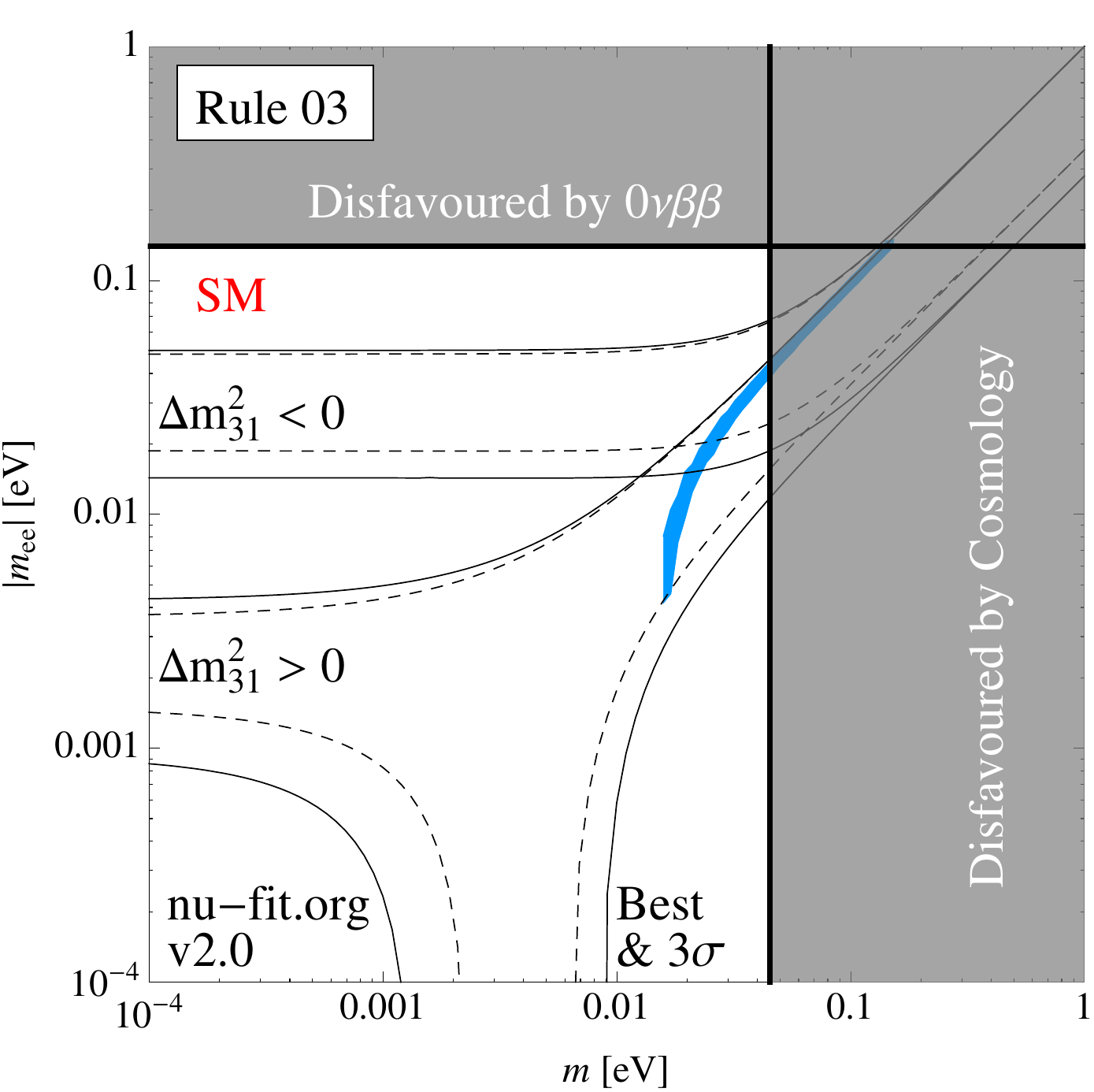} &
\includegraphics[width=5.4cm]{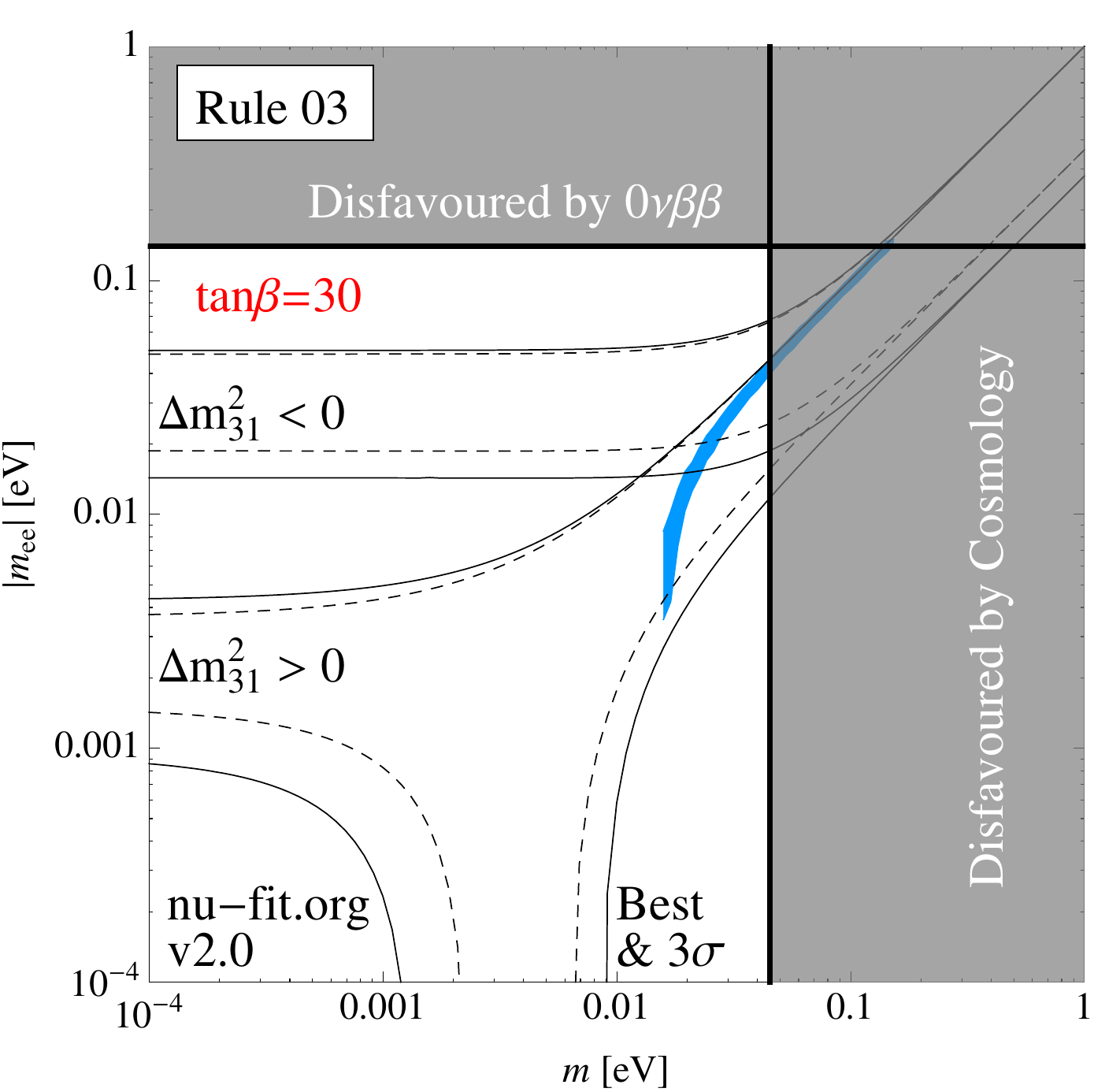} &
\includegraphics[width=5.4cm]{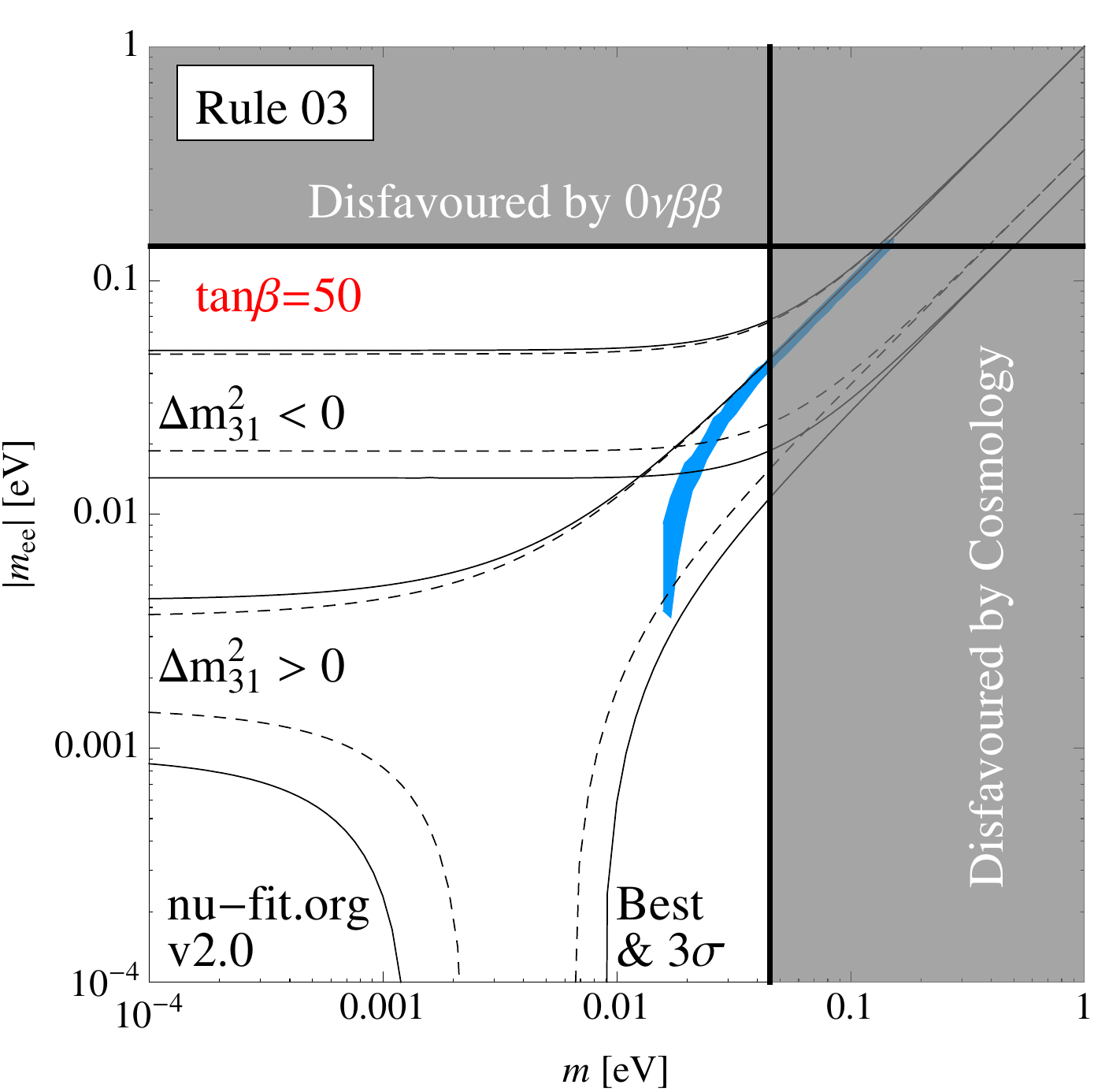}
\end{tabular}
\end{center}
This sum rule predicts normal ordering only, and with the SM particle content it yields $(m_{\rm min}, |m_{ee}|_{\rm min}) = (0.016,0.0042)$~eV, if the running is applied. These numbers are consistent with the values obtained in Ref.~\cite{King:2013psa} (see discussion in Sec.~7.9 therein).\footnote{However, note that our computation just misses the cancellation region, in contrast to the one presented in Ref.~\cite{King:2013psa}. Nevertheless there is no real discrepancy, since the question whether or not all parameters can conspire to yield $|m_{ee}|$ practically zero depends strongly on the actual oscillation parameters used~\cite{mee-references}, and our values are updated compared to the ones uses in publications two years ago.} For $\tan \beta = 30$ ($50$), the values change to $(0.016,0.0036)$~eV ($(0.016,0.0036)$~eV), while still only NO is allowed.

\subsection{\label{sec:SR4} Sum Rule 4: $\mathbf{\tilde m_1+ \tilde m_2=2 \tilde m_3}$}

The parameters for this sum rule are $(d, c_1, c_2, \Delta \chi_{13}, \Delta \chi_{23})=(1, 1/2, 1/2, \pi, \pi)$, and the corresponding plots look like:
\begin{center}
\begin{tabular}{lll}
\hspace{-1cm}
\includegraphics[width=5.4cm]{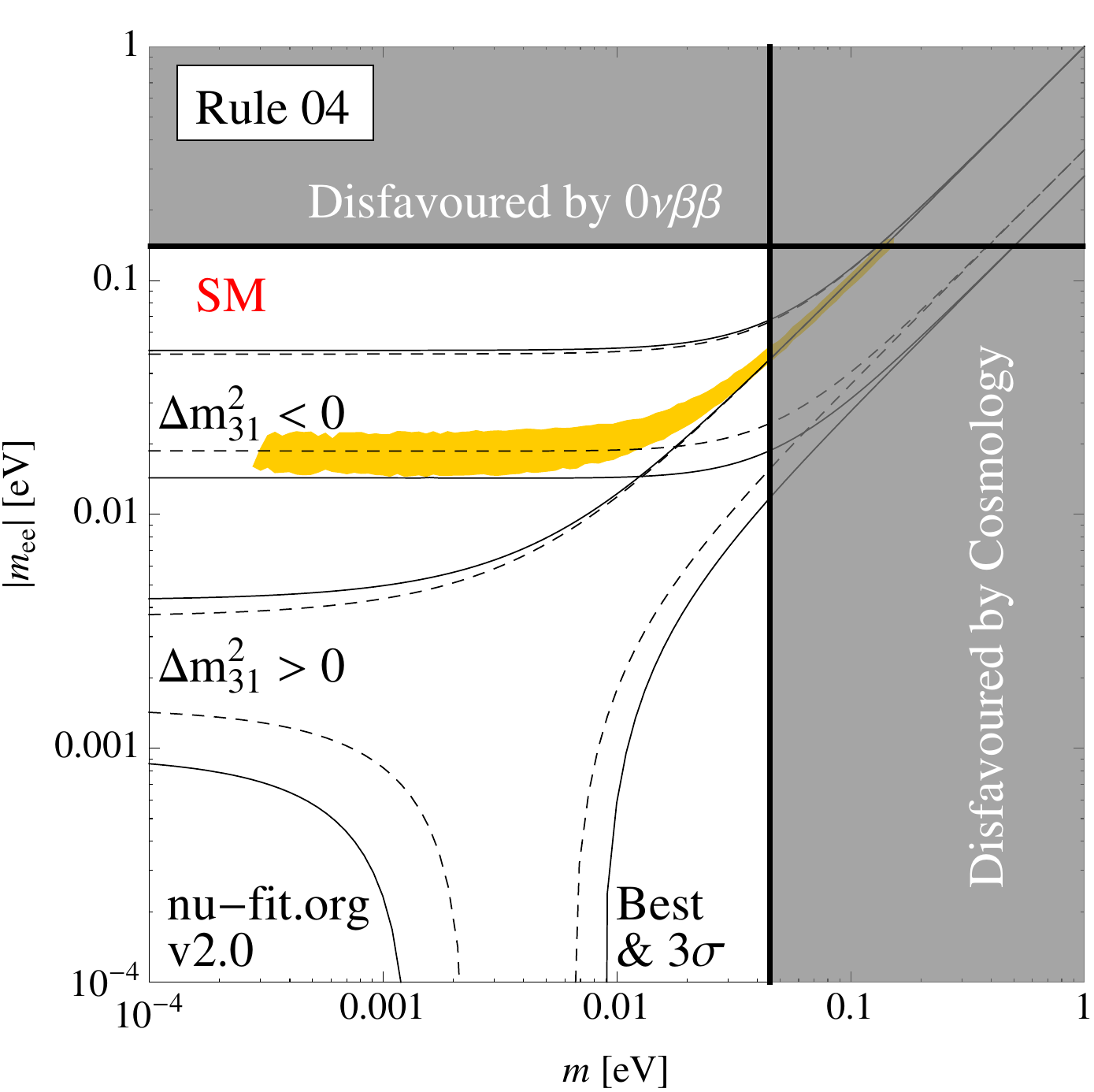} &
\includegraphics[width=5.4cm]{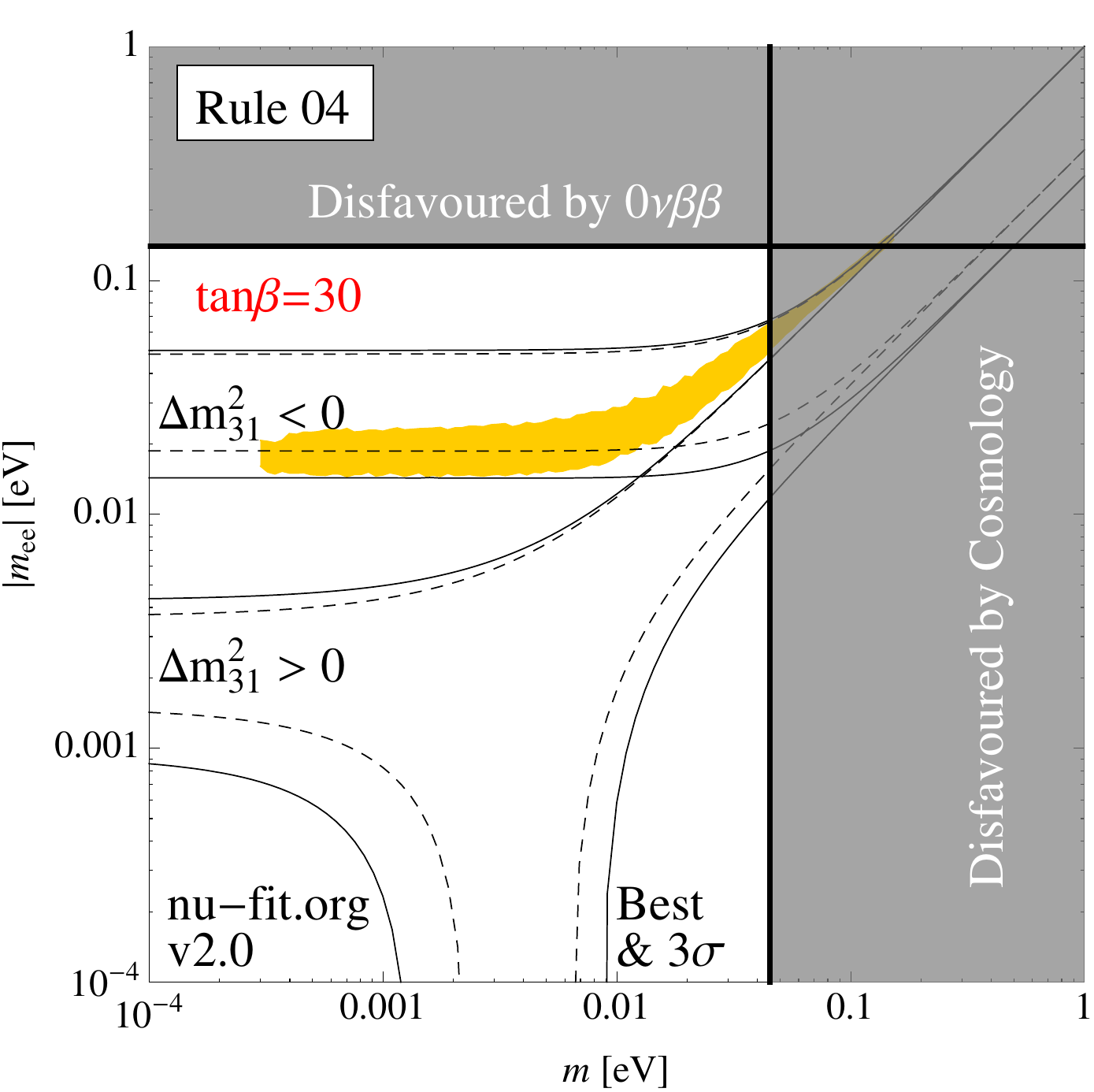} &
\includegraphics[width=5.4cm]{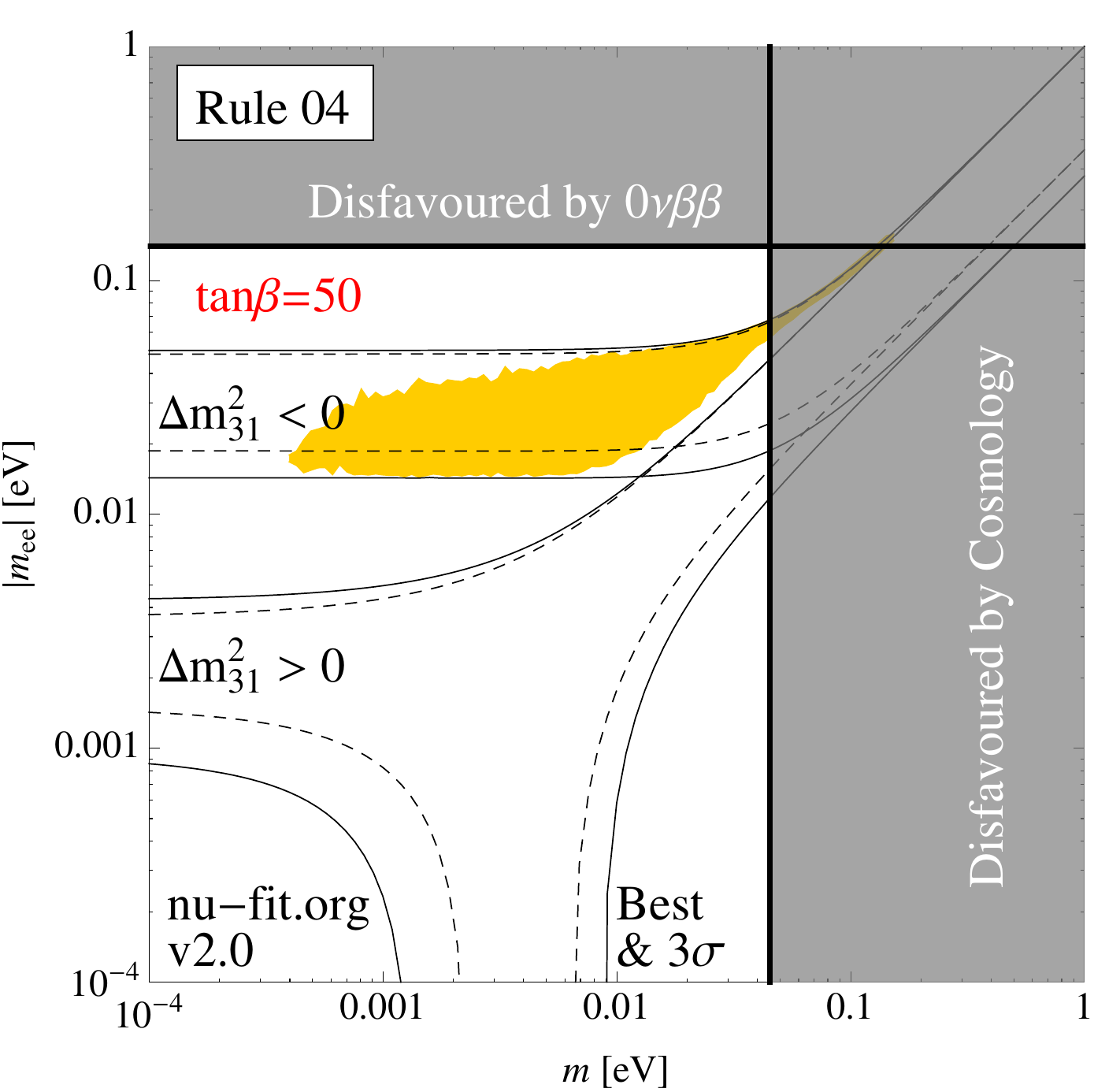}
\end{tabular}
\end{center}
This sum rule predicts inverted ordering only, and with the SM particle content it yields $(m_{\rm min}, |m_{ee}|_{\rm min}) = (0.00028,0.015)$~eV, if the running is applied. These numbers are consistent with the values obtained in Ref.~\cite{King:2013psa} (see discussion in Sec.~7.12 therein). For $\tan \beta = 30$ ($50$), the values change to $(0.00030,0.014)$~eV ($(0.00040,0.014)$~eV), while still only IO is allowed.

\subsection{\label{sec:SR5} Sum Rule 5: $\mathbf{\tilde m_1-\frac{\sqrt{3}-1}{2} \tilde m_2+\frac{\sqrt{3}+1}{2} \tilde m_3=0}$}

The parameters for this sum rule are $(d, c_1, c_2, \Delta \chi_{13}, \Delta \chi_{23})=(1, \frac{2}{\sqrt{3}+1}, \frac{\sqrt{3}-1}{\sqrt{3}+1}, 0, \pi)$, and the corresponding plots look like:
\begin{center}
\begin{tabular}{lll}
\hspace{-1cm}
\includegraphics[width=5.4cm]{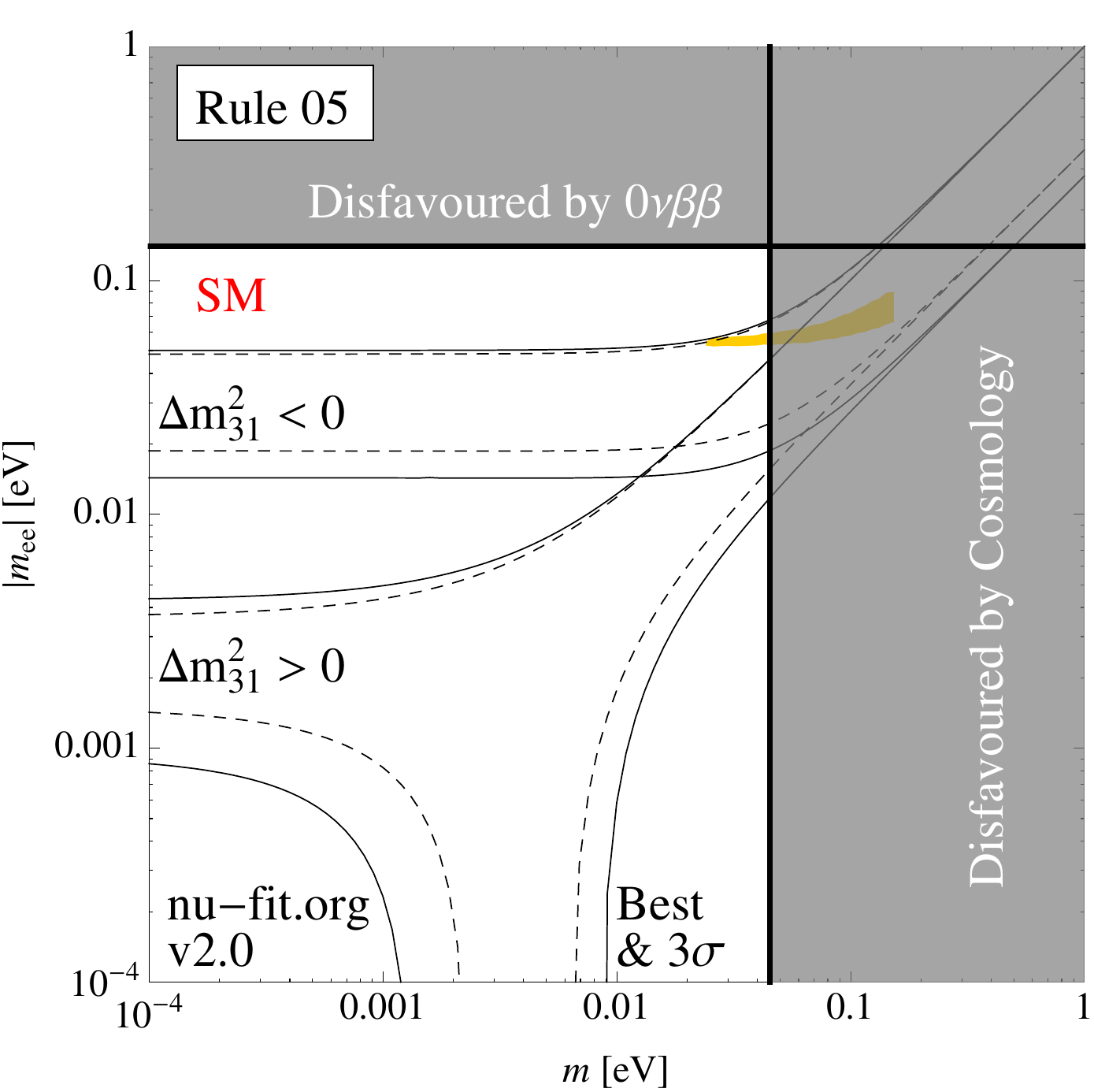} &
\includegraphics[width=5.4cm]{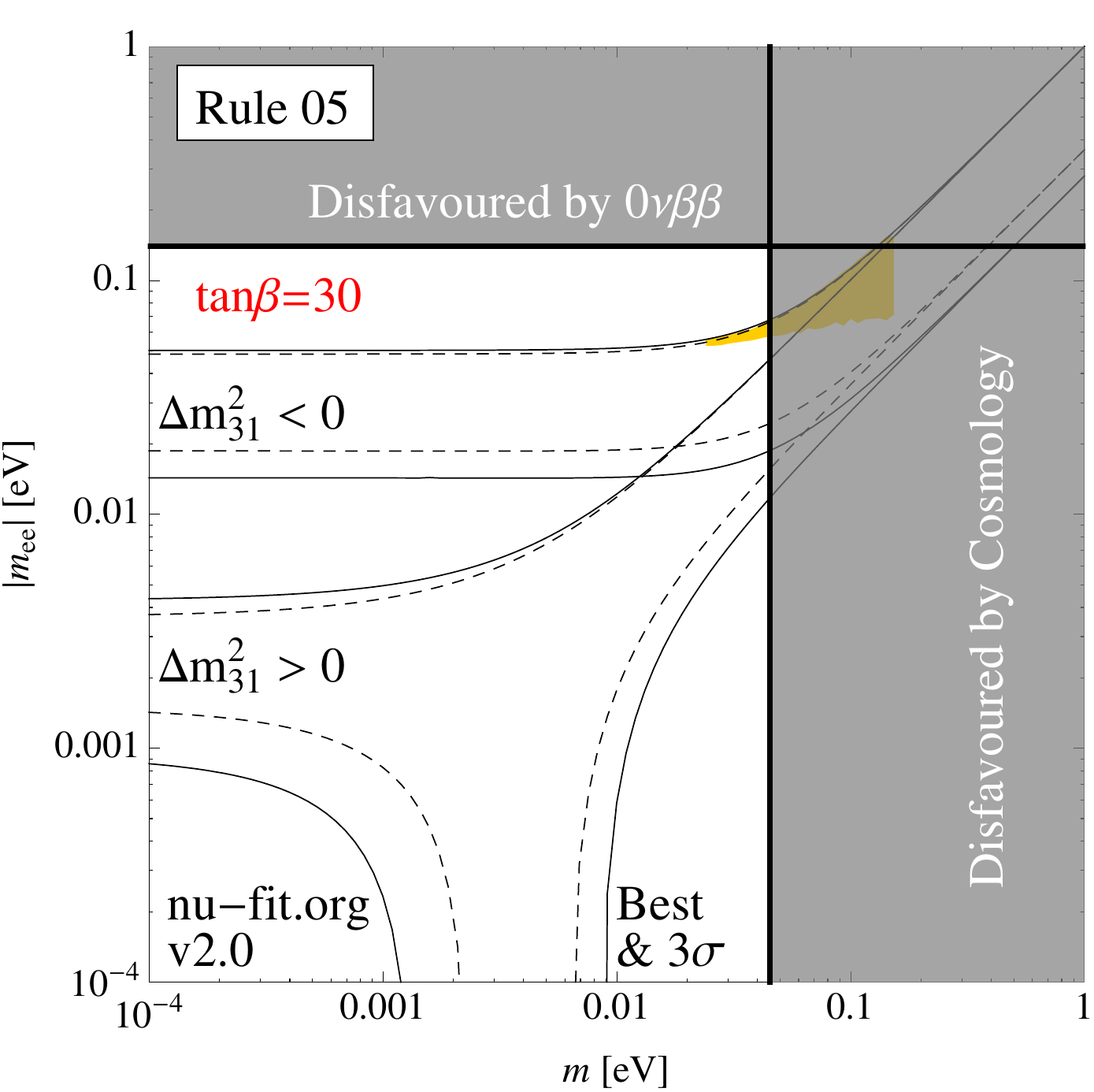} &
\includegraphics[width=5.4cm]{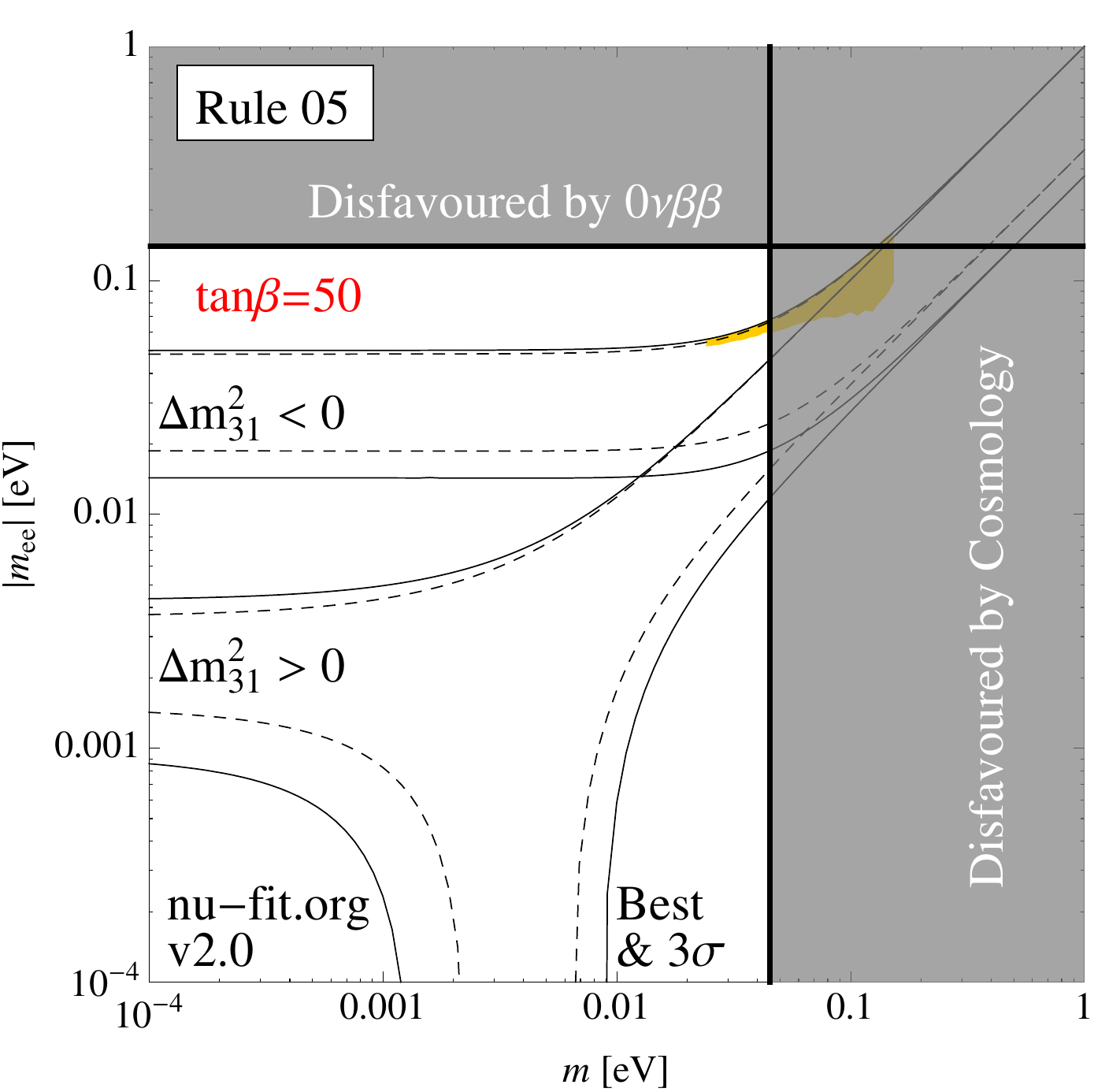}
\end{tabular}
\end{center}
This sum rule predicts inverted ordering only, and with the SM particle content it yields $(m_{\rm min}, |m_{ee}|_{\rm min}) = (0.024,0.053)$~eV, if the running is applied. These numbers are consistent with the values obtained in Ref.~\cite{King:2013psa} (see discussion in Sec.~7.14 therein).\footnote{Note the typo in the value for $m_{\rm min}$ in table~4 of that reference.} For $\tan \beta = 30$ ($50$), the values practically remain at $(0.024,0.053)$~eV ($(0.024,0.053)$~eV), while still only IO is allowed.

\subsection{\label{sec:SR6} Sum Rule 6: The sum rule $\mathbf{1/\tilde m_{1}+1/\tilde m_{2}=1/\tilde m_{3}}$}

The parameters for this sum rule are $(d, c_1, c_2, \Delta \chi_{13}, \Delta \chi_{23})=(-1, 1, 1, \pi, \pi)$, and the corresponding plots look like:
\begin{center}
\begin{tabular}{lll}
\hspace{-1cm}
\includegraphics[width=5.4cm]{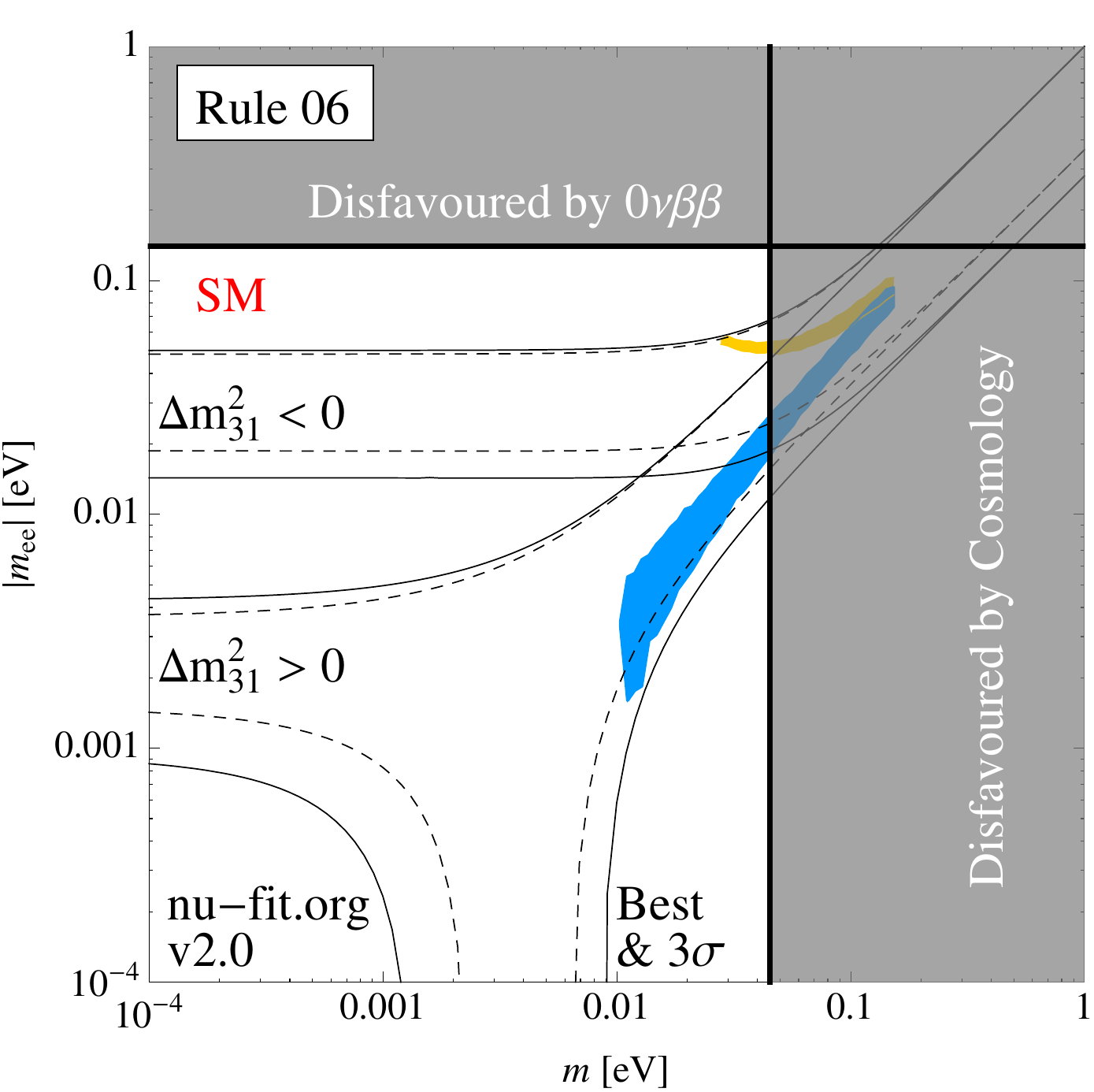} &
\includegraphics[width=5.4cm]{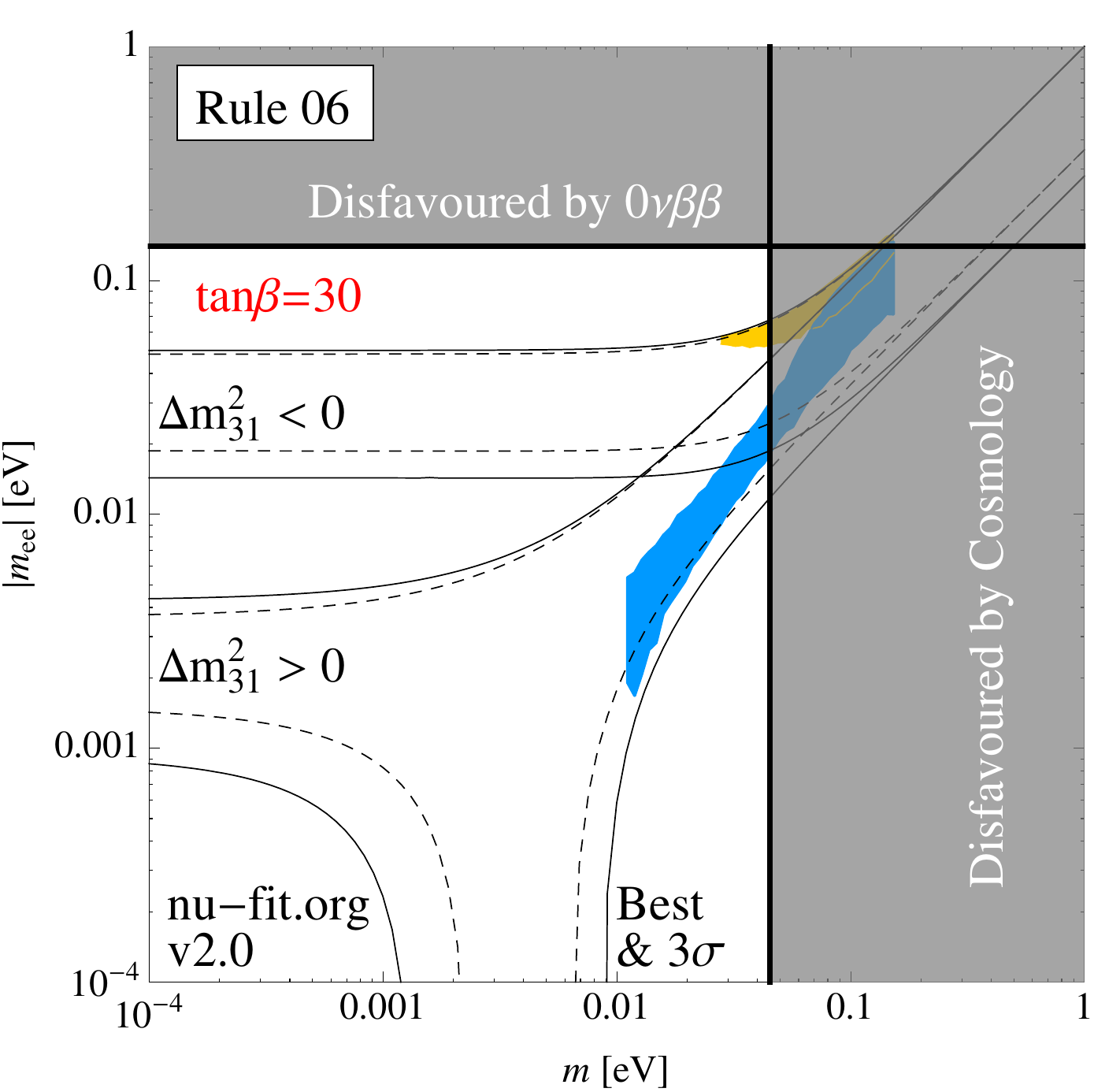} &
\includegraphics[width=5.4cm]{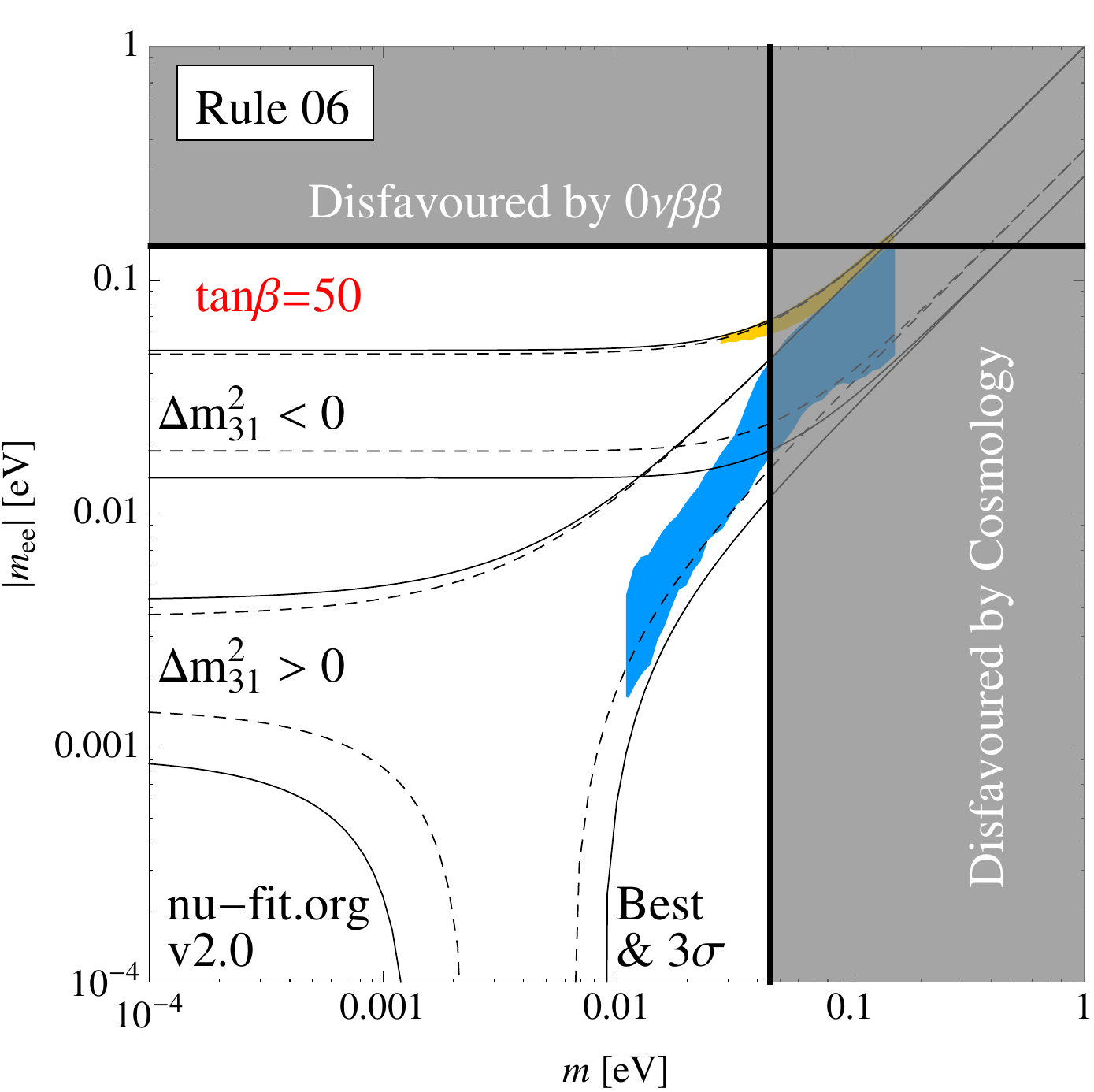}
\end{tabular}
\end{center}
This sum rule yields $(m_{\rm min}, |m_{ee}|_{\rm min}) = (0.010,0.0016)$~eV ($(0.028,0.048)$~eV) for normal (inverted) mass ordering, if running with the SM particle content is applied, which is consistent with the values obtained in Ref.~\cite{King:2013psa} (see discussion in Sec.~7.1 therein). For $\tan \beta = 30$ ($50$), the values change to $(0.011,0.0017)$~eV ($(0.011,0.0017)$~eV) for NO and to $(0.028,0.052)$~eV ($(0.028,0.054)$~eV) for IO, respectively.

\subsection{\label{sec:SR7} Sum Rule 7: $\mathbf{1/\tilde m_{1}-2 /\tilde m_{2}-1/\tilde m_{3}=0}$}

The parameters for this sum rule are $(d, c_1, c_2, \Delta \chi_{13}, \Delta \chi_{23})=(-1, 1, 2, \pi, 0)$, and the corresponding plots look like:
\begin{center}
\begin{tabular}{lll}
\hspace{-1cm}
\includegraphics[width=5.4cm]{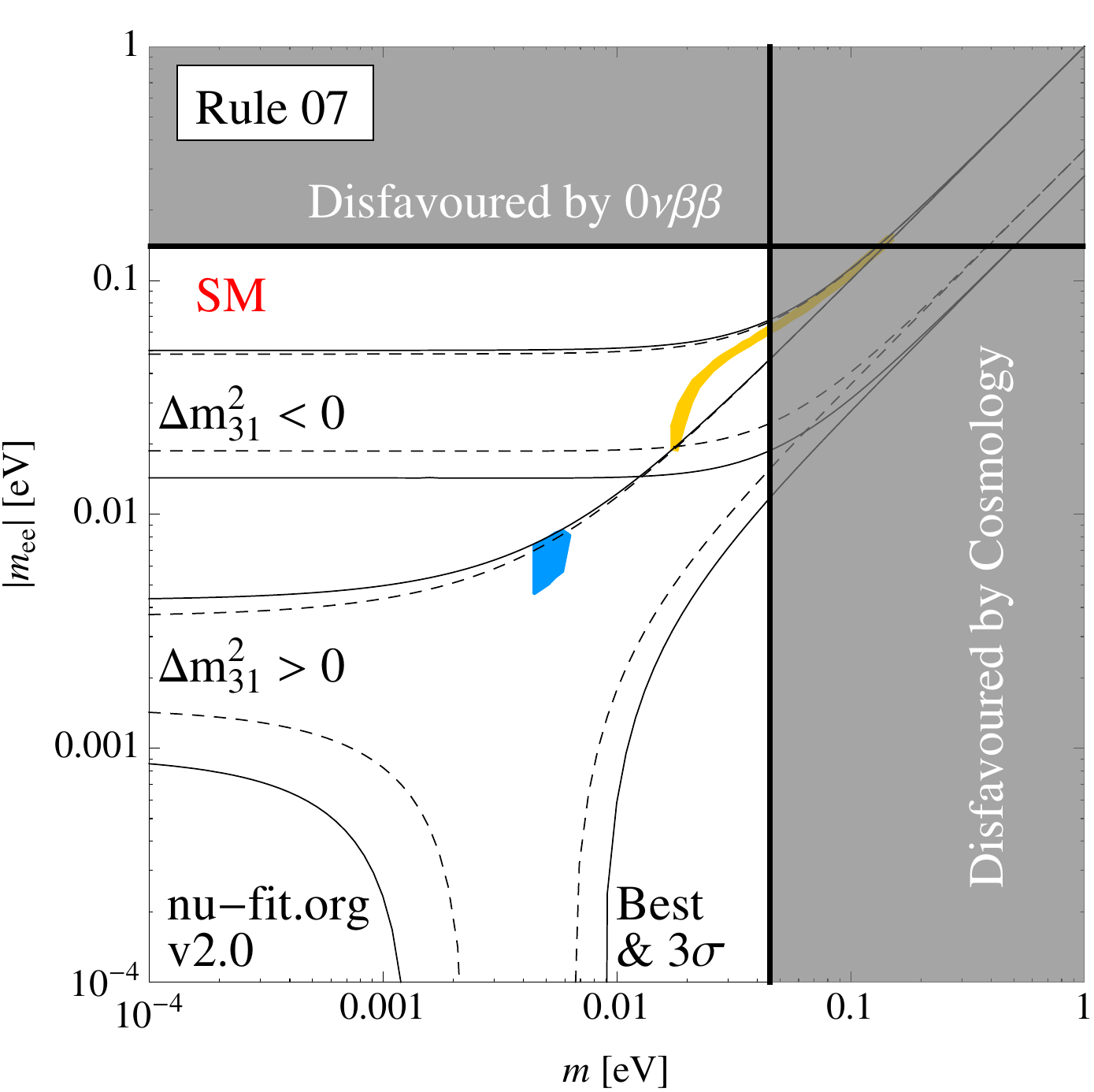} &
\includegraphics[width=5.4cm]{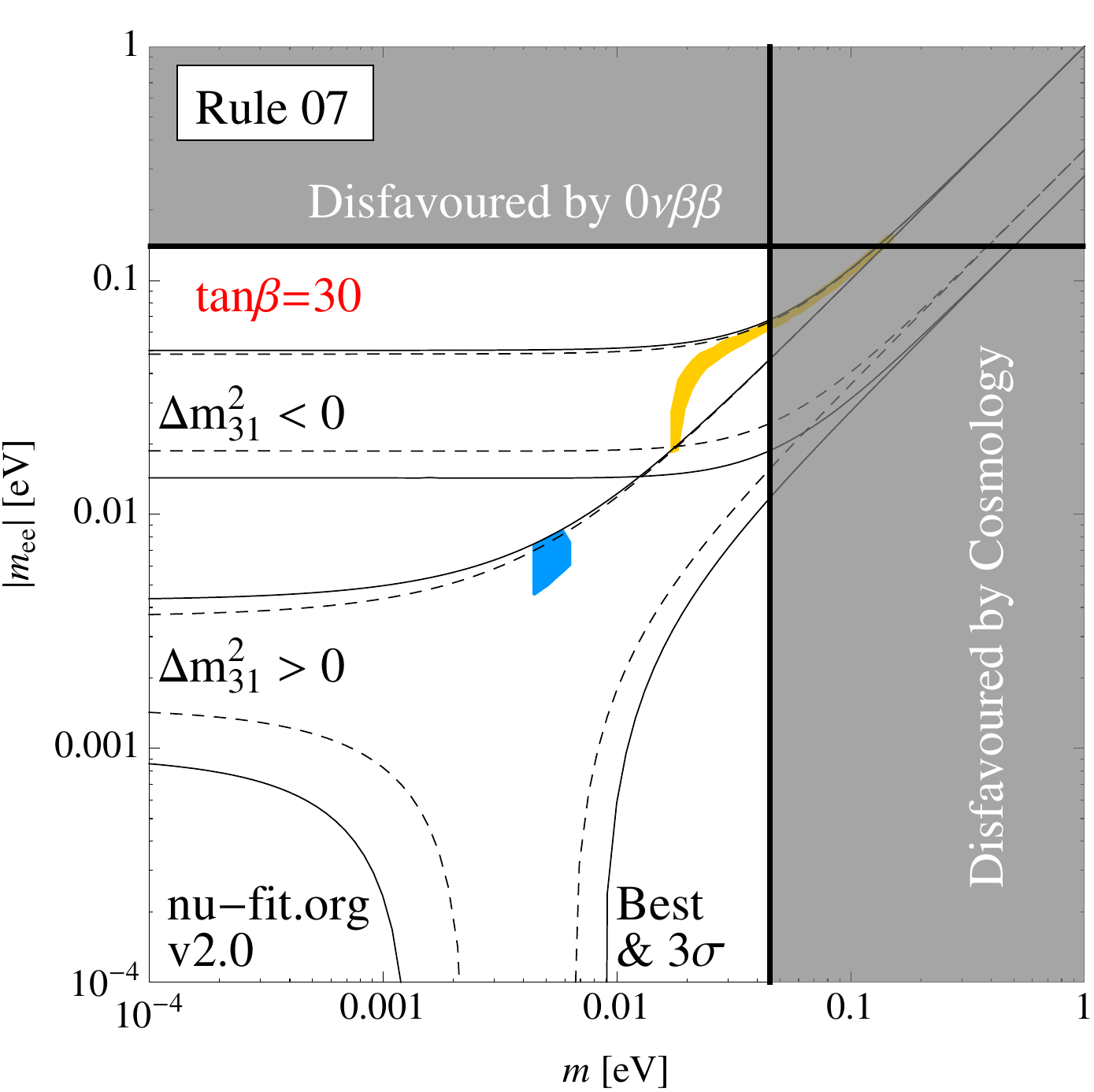} &
\includegraphics[width=5.4cm]{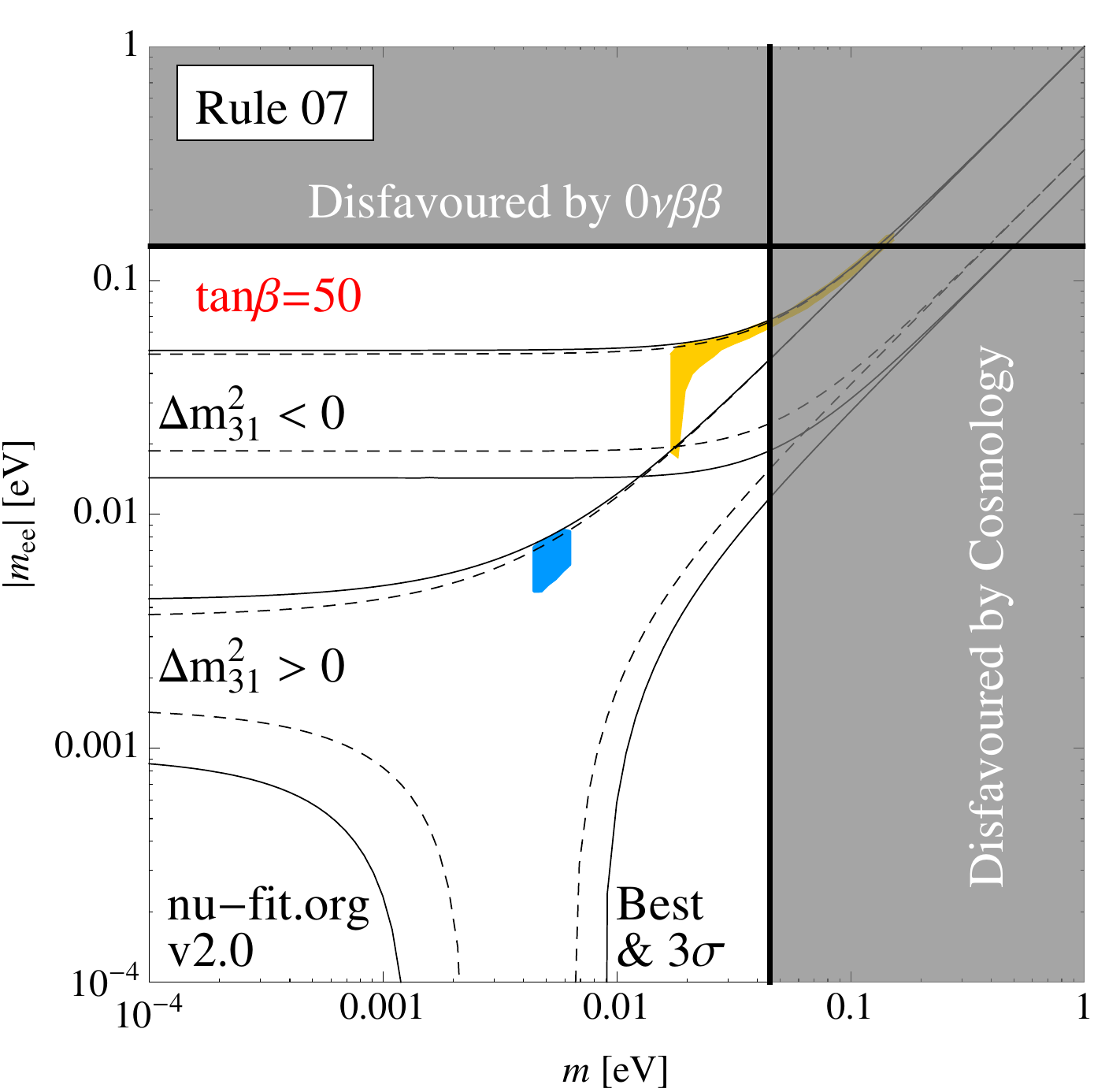}
\end{tabular}
\end{center}
This sum rule yields $(m_{\rm min}, |m_{ee}|_{\rm min}) = (0.0044,0.0046)$~eV ($(0.017,0.019)$~eV) for normal (inverted) mass ordering, if running with the SM particle content is applied, which is consistent with the values obtained in Ref.~\cite{King:2013psa} (see discussion in Sec.~7.8 therein). For $\tan \beta = 30$ ($50$), the values change to $(0.0044,0.0045)$~eV ($(0.0044,0.0047)$~eV) for NO and to $(0.017,0.018)$~eV ($(0.017,0.017)$~eV) for IO, respectively.

\subsection{\label{sec:SR8} Sum Rule 8: $\mathbf{2/\tilde m_2=1/\tilde m_1+1/\tilde m_3}$}

The parameters for this sum rule are $(d, c_1, c_2, \Delta \chi_{13}, \Delta \chi_{23})=(-1, 1, 2, 0, \pi)$, and the corresponding plots look like:
\begin{center}
\begin{tabular}{lll}
\hspace{-1cm}
\includegraphics[width=5.4cm]{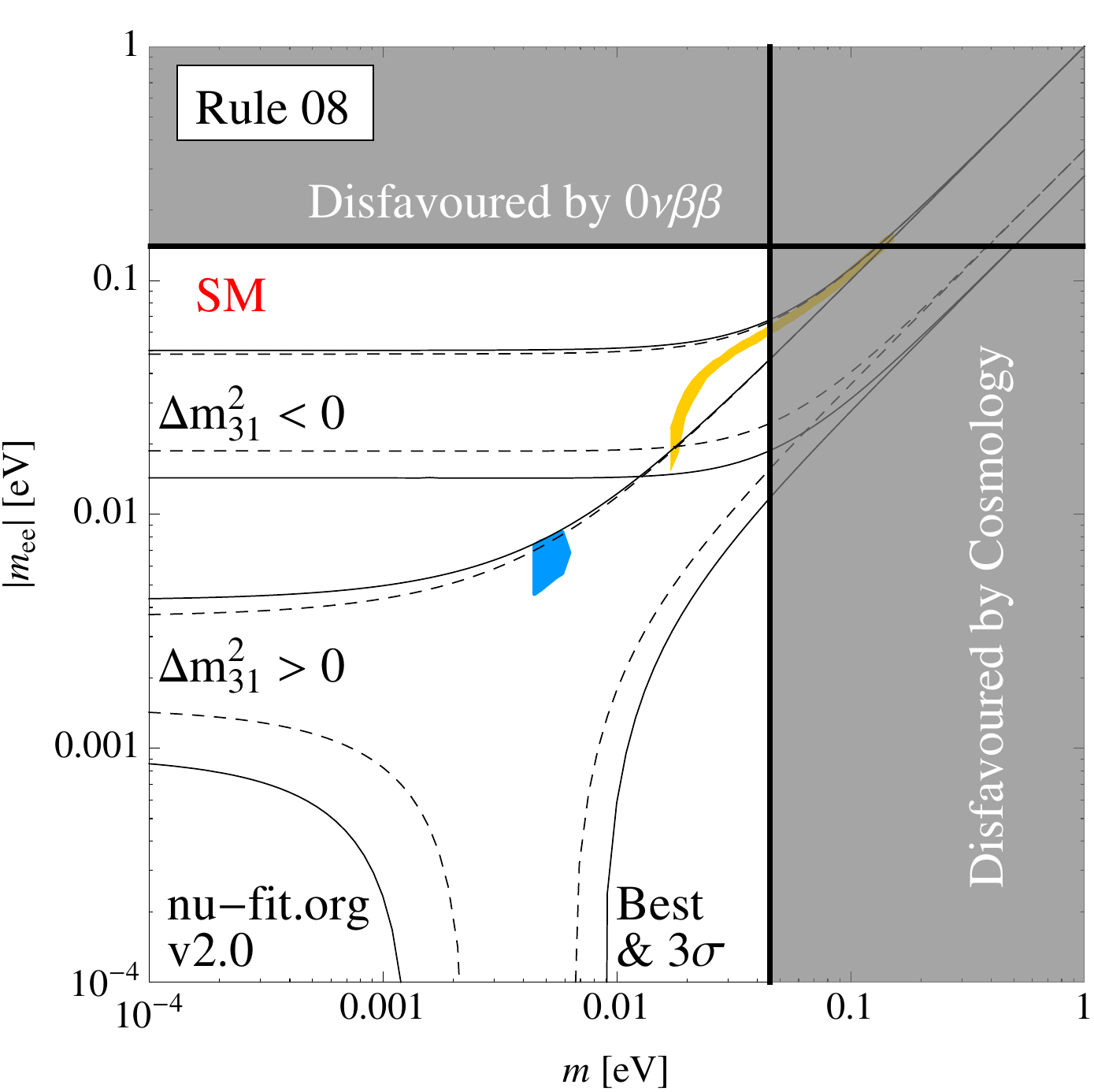} &
\includegraphics[width=5.4cm]{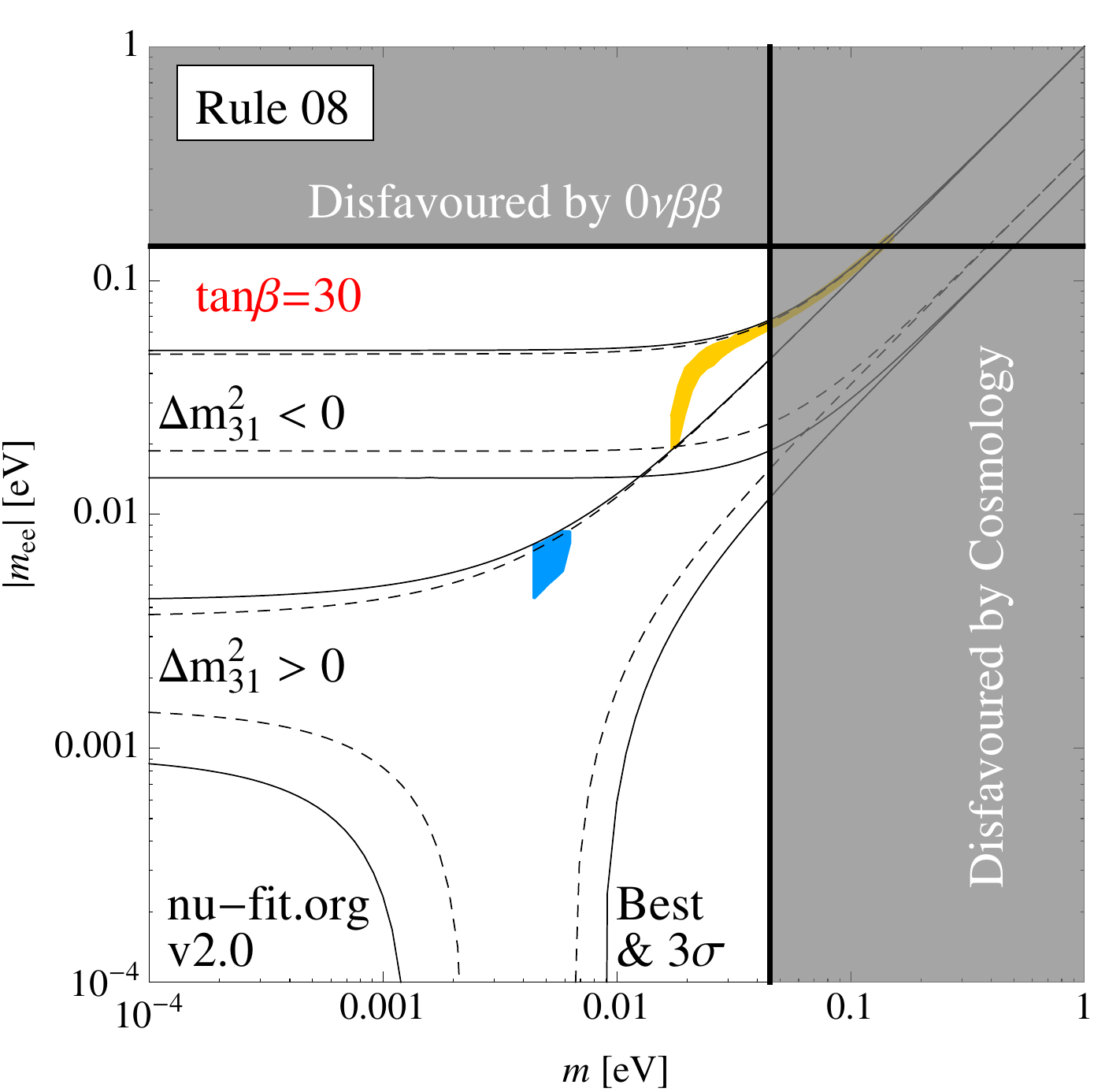} &
\includegraphics[width=5.4cm]{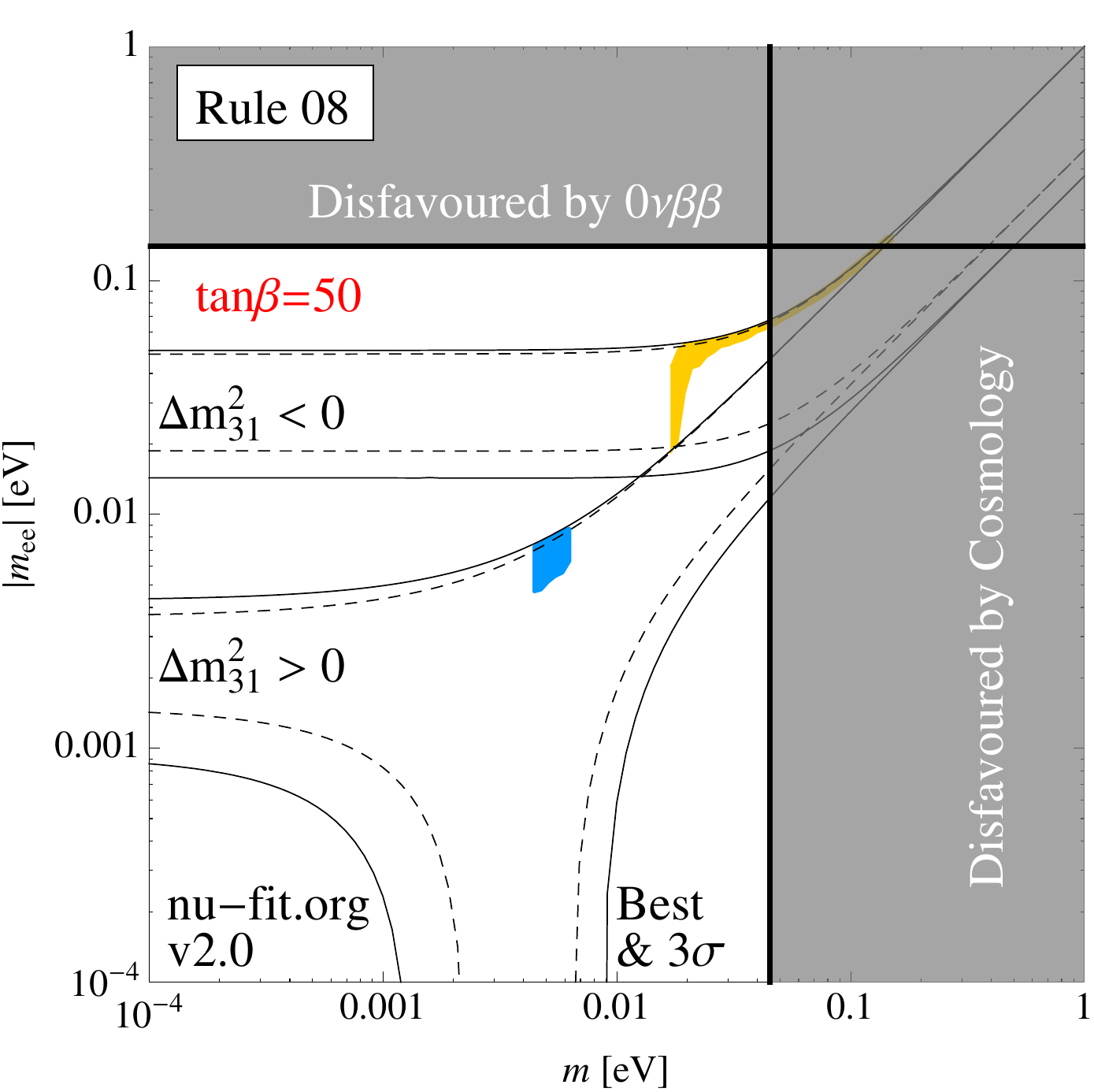}
\end{tabular}
\end{center}
This sum rule yields $(m_{\rm min}, |m_{ee}|_{\rm min}) = (0.0044,0.0045)$~eV ($(0.017,0.015)$~eV) for normal [inverted] mass ordering, if running with the SM particle content is applied, which is consistent with the values obtained in Ref.~\cite{King:2013psa} (see discussion in Sec.~7.6 therein). For $\tan \beta = 30$ ($50$), the values change to $(0.0044,0.0044)$~eV ($(0.0044,0.0047)$~eV) for NO and to $(0.017,0.019)$~eV ($(0.017,0.018)$~eV) for IO, respectively.

\subsection{\label{sec:SR9} Sum Rule 9: $\mathbf{1/\tilde m_{3}+2 \text{i}(-1)^\eta/\tilde m_{2}=1/\tilde m_{1}}$}

The parameters for this sum rule are $(d, c_1, c_2, \Delta \chi_{13}, \Delta \chi_{23})=(-1, 1, 2, \pi, \pi/2\ {\rm or}\ 3\pi/2)$, depending on whether $\eta = 0$ or $1$, and the corresponding plots look like:
\begin{center}
\begin{tabular}{lll}
\hspace{-1cm}
\includegraphics[width=5.4cm]{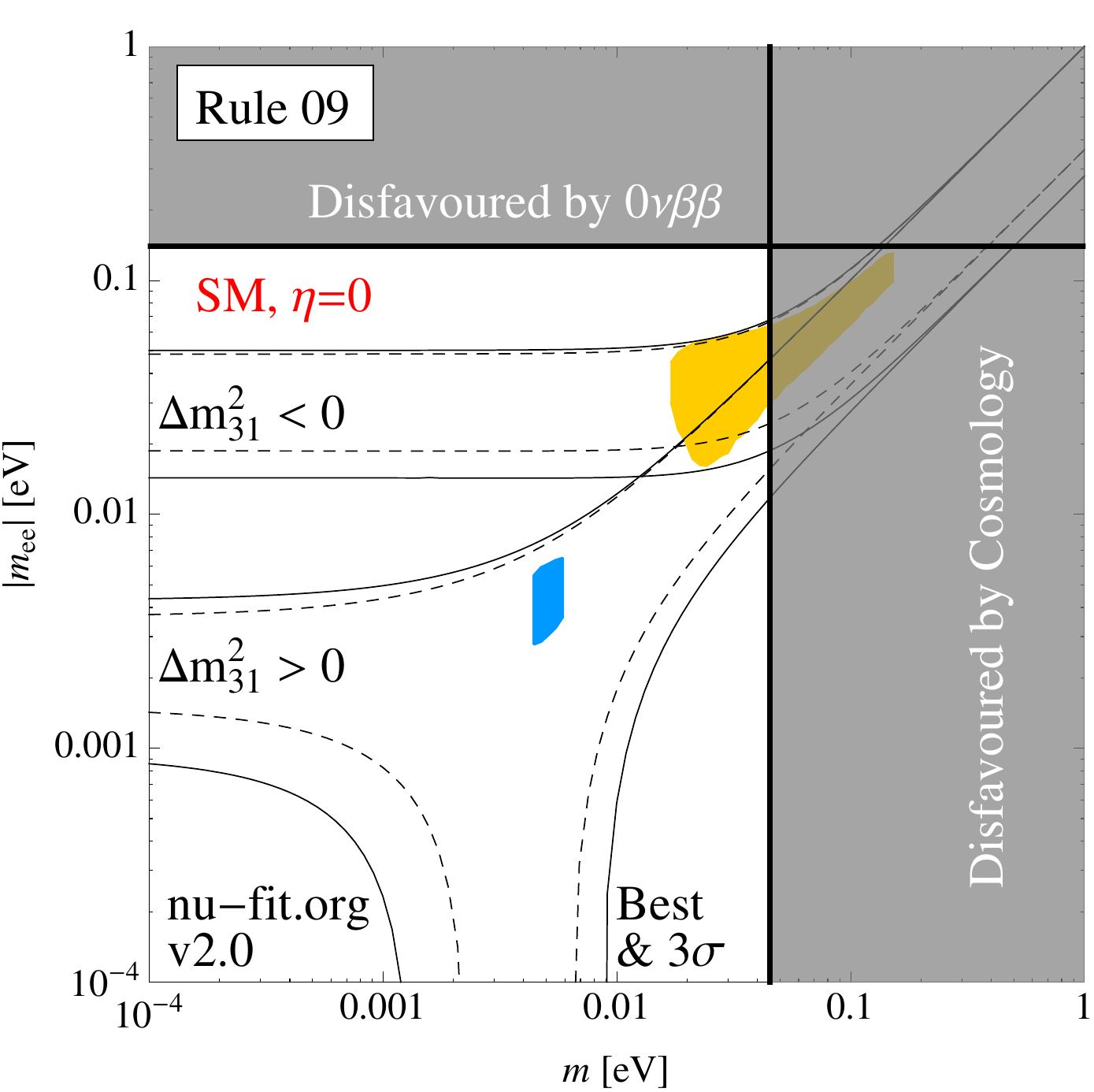} &
\includegraphics[width=5.4cm]{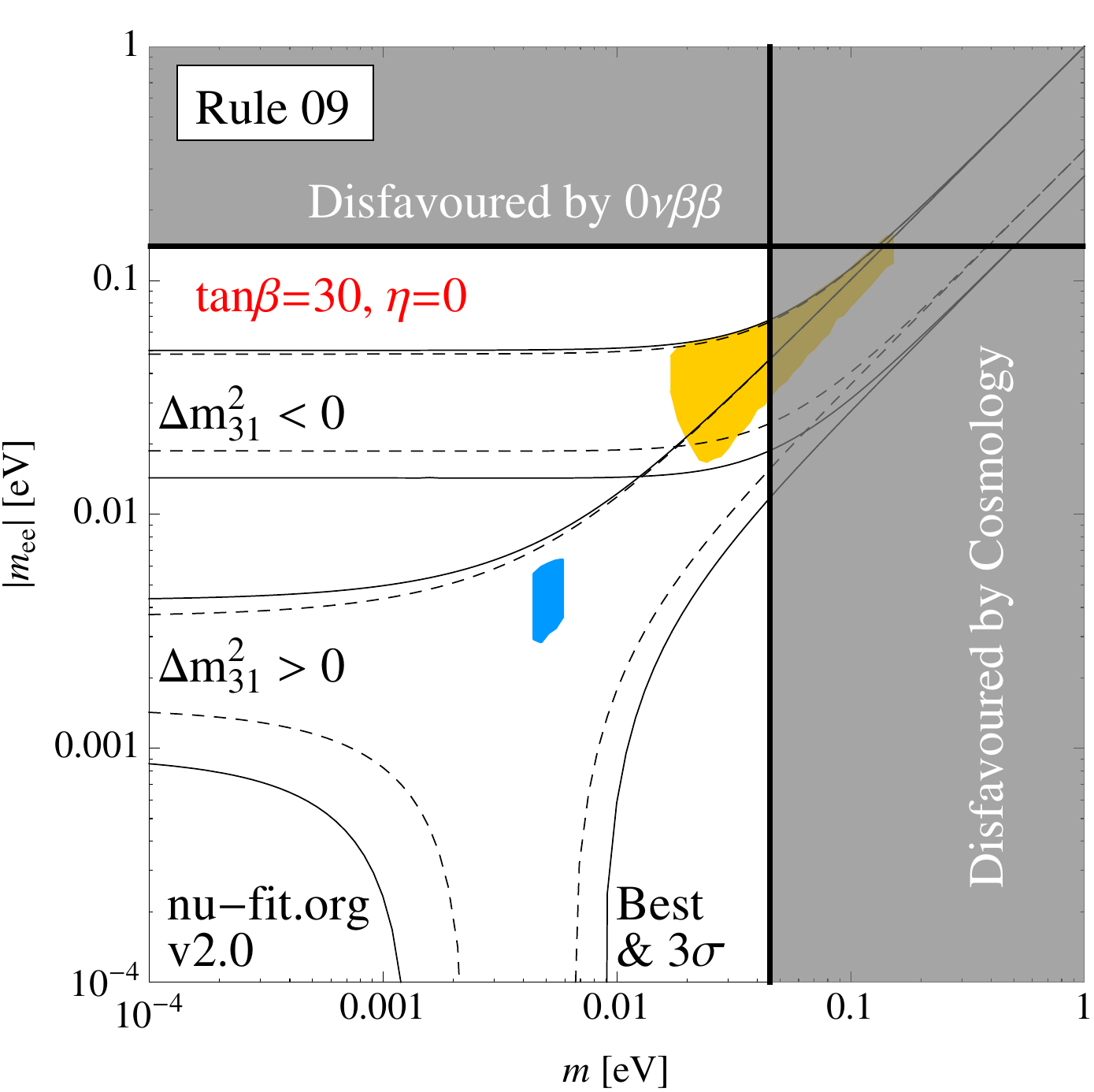} &
\includegraphics[width=5.4cm]{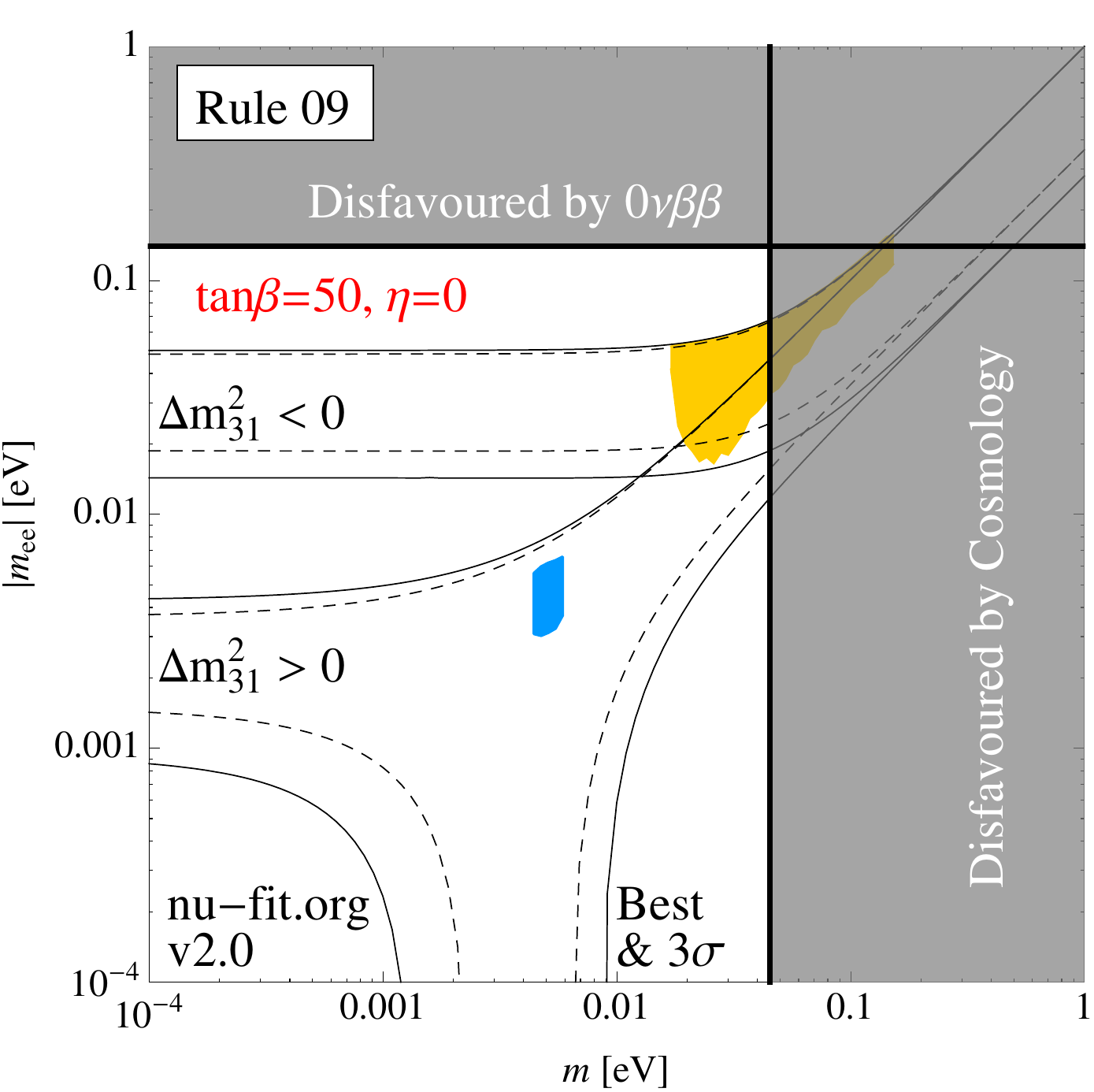}\\
\hspace{-1cm}
\includegraphics[width=5.4cm]{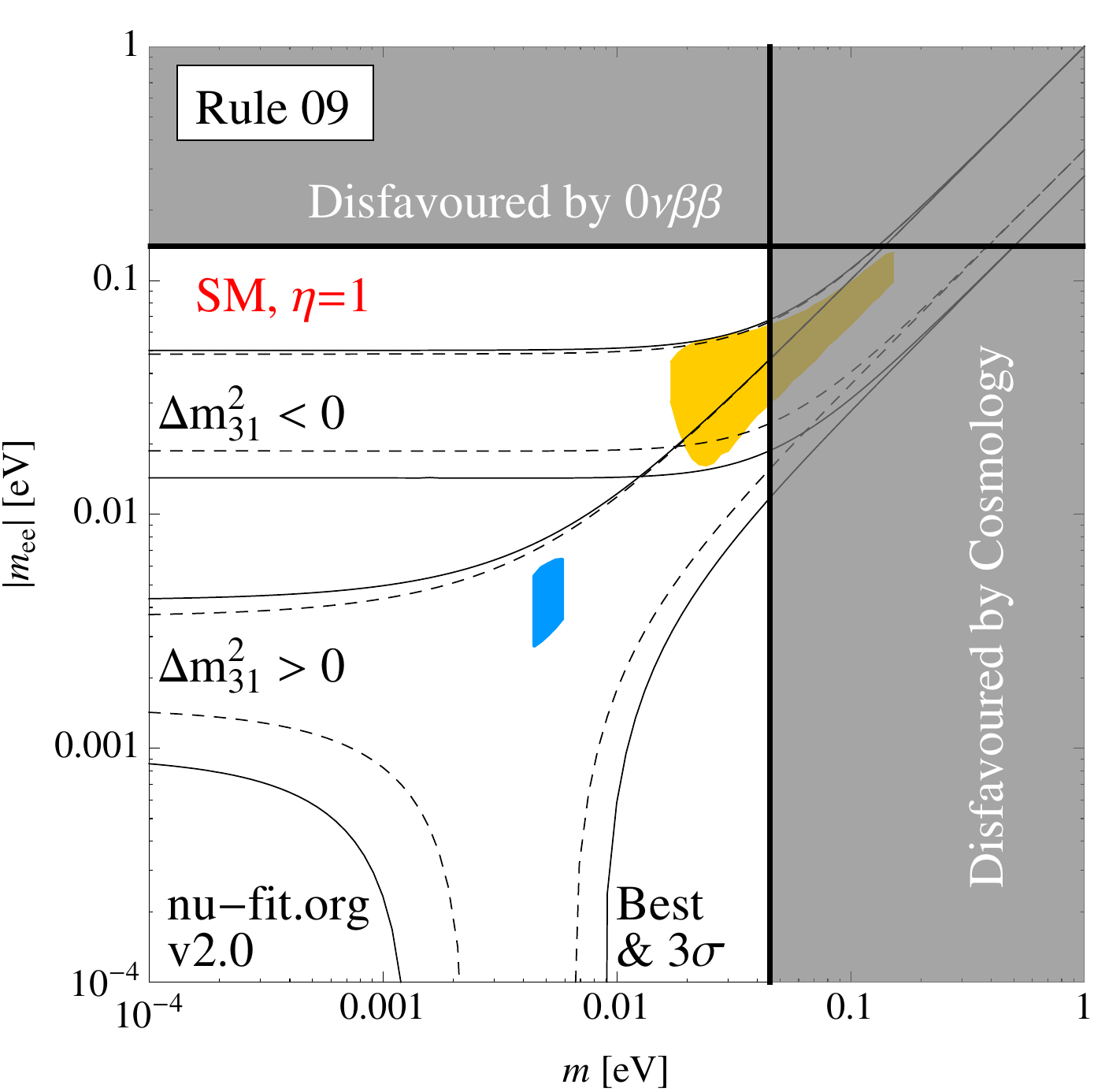} &
\includegraphics[width=5.4cm]{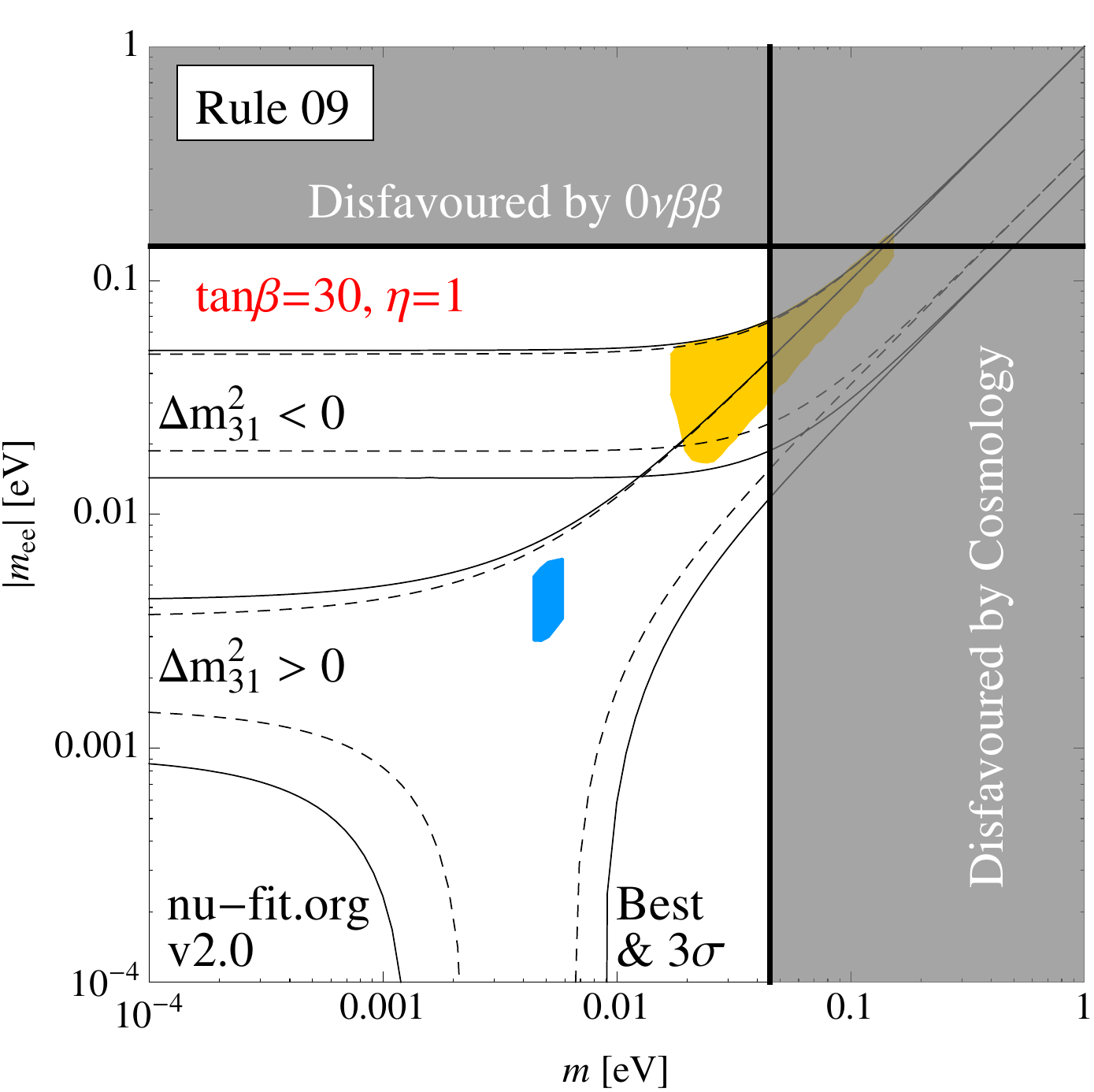} &
\includegraphics[width=5.4cm]{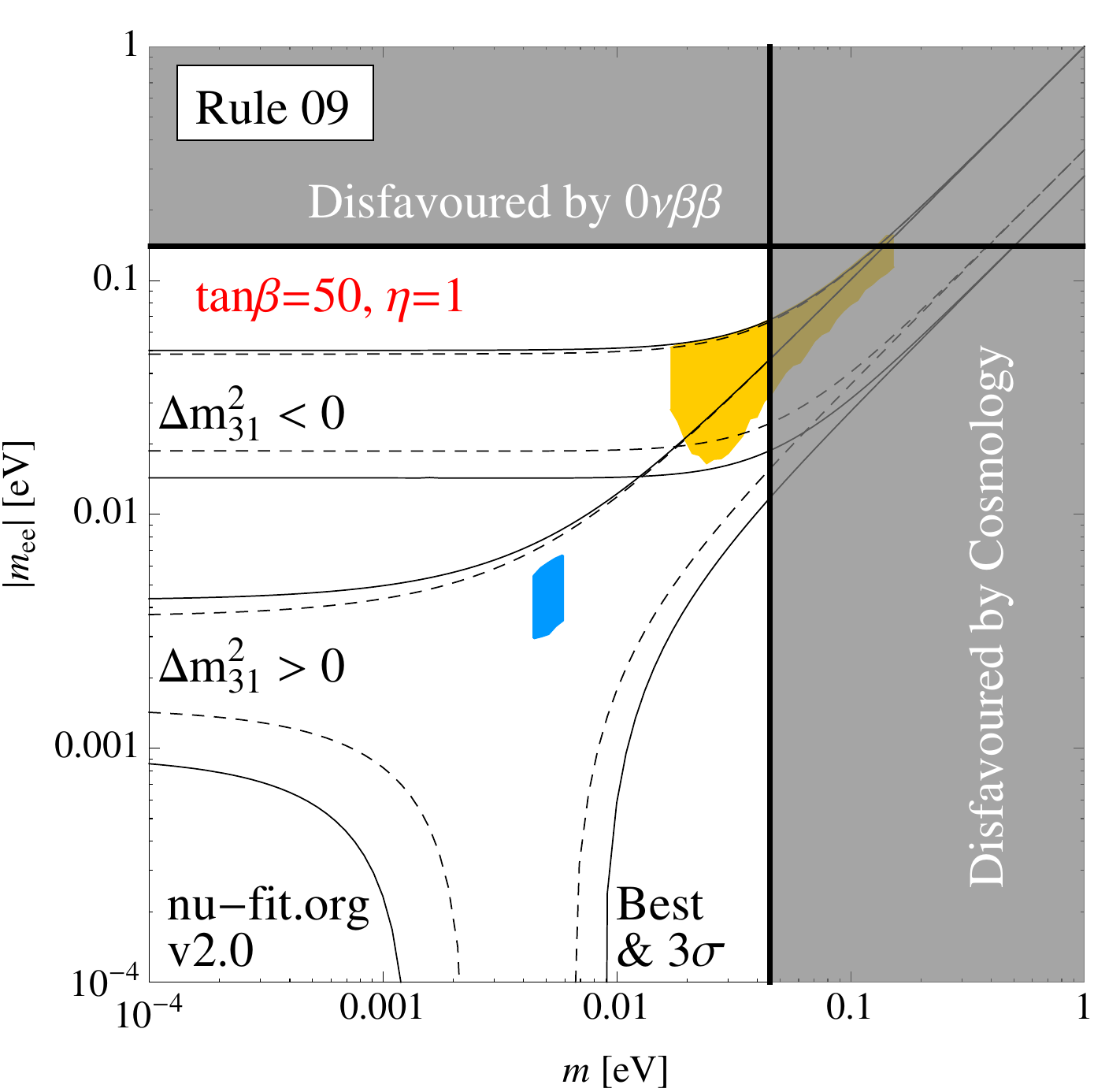}
\end{tabular}
\end{center}
For both $\eta=0,1$, this sum rule yields $(m_{\rm min}, |m_{ee}|_{\rm min}) = (0.0044,0.0028)$~eV ($(0.017,0.016)$~eV) for normal (inverted) mass ordering, if running with the SM particle content is applied, which is consistent with the values obtained in Ref.~\cite{King:2013psa} (see discussion in Sec.~7.2 therein). For $\tan \beta = 30$ ($50$), the values change to $(0.0044,0.0028)$~eV ($(0.0044,0.0030)$~eV) for NO and to $(0.017,0.017)$~eV ($(0.017,0.016)$~eV) for IO, respectively.

This may at first look surprising, however, even though the RGEs for the Majorana phases (and hence the corresponding predictions) are different in both cases, this information gets lost when varying over the Dirac CP phase $\delta$, as we have checked numerically. Turning the argument round, if $\delta$ was known at least to some extend, we would potentially be able to distinguish the two versions of this sum rule.

\subsection{\label{sec:SR10} Sum Rule 10: $\mathbf{\sqrt{\tilde m_1} + \text{i}^\eta \sqrt{\tilde m_3}=2\sqrt{\tilde m_2}}$}

The parameters for this sum rule are $(d, c_1, c_2, \Delta \chi_{13}, \Delta \chi_{23})=(1/2, 1, 2, \pi\ {\rm or}\ 0\ {\rm or}\ \pi/2, 0\ {\rm or}\ \pi\ {\rm or}\ \pi/2)$, depending on $\eta = 0,1,2$, and the corresponding plots look like:
\begin{center}
\begin{tabular}{lll}
\hspace{-1cm}
\includegraphics[width=5.4cm]{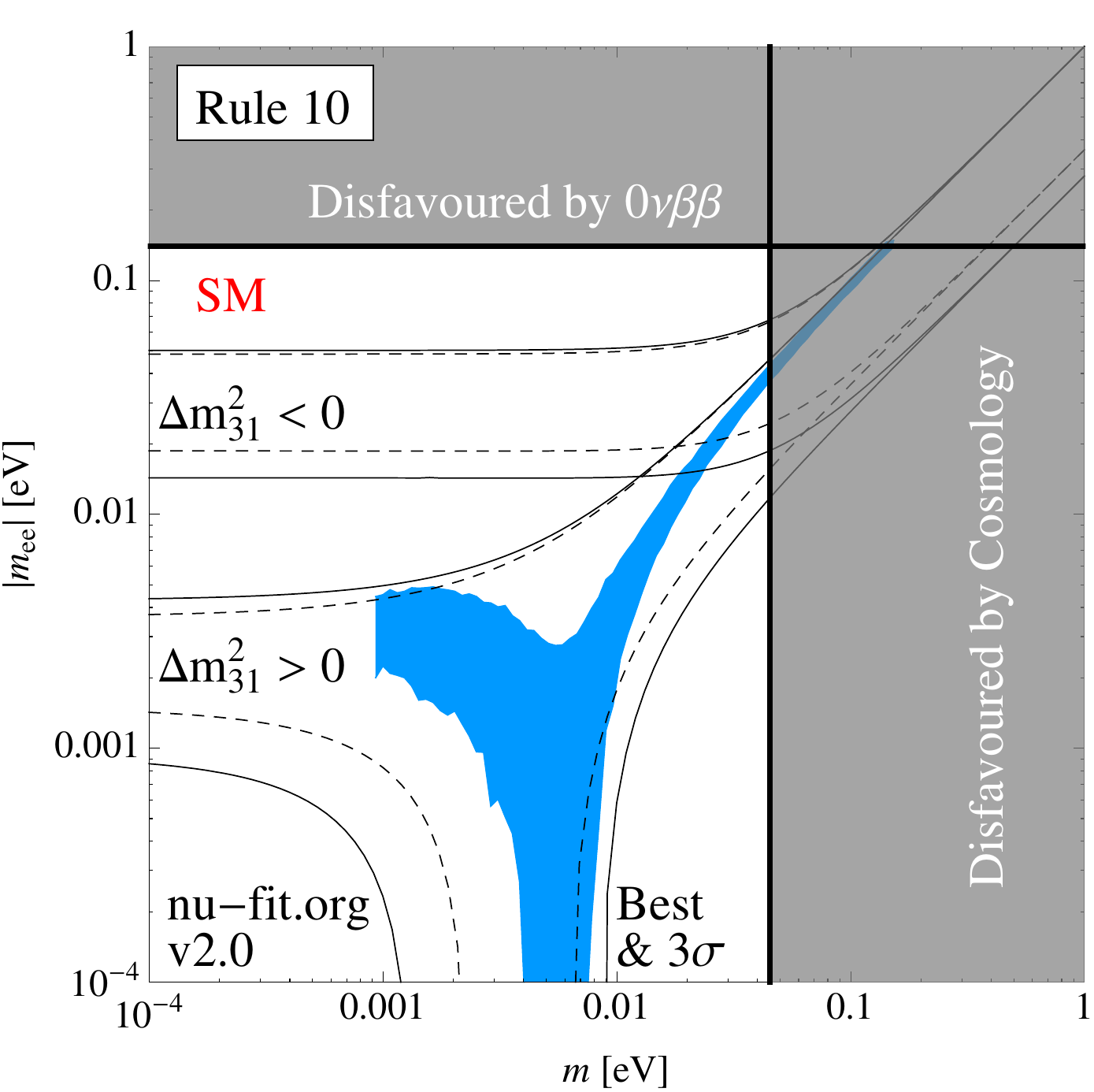} &
\includegraphics[width=5.4cm]{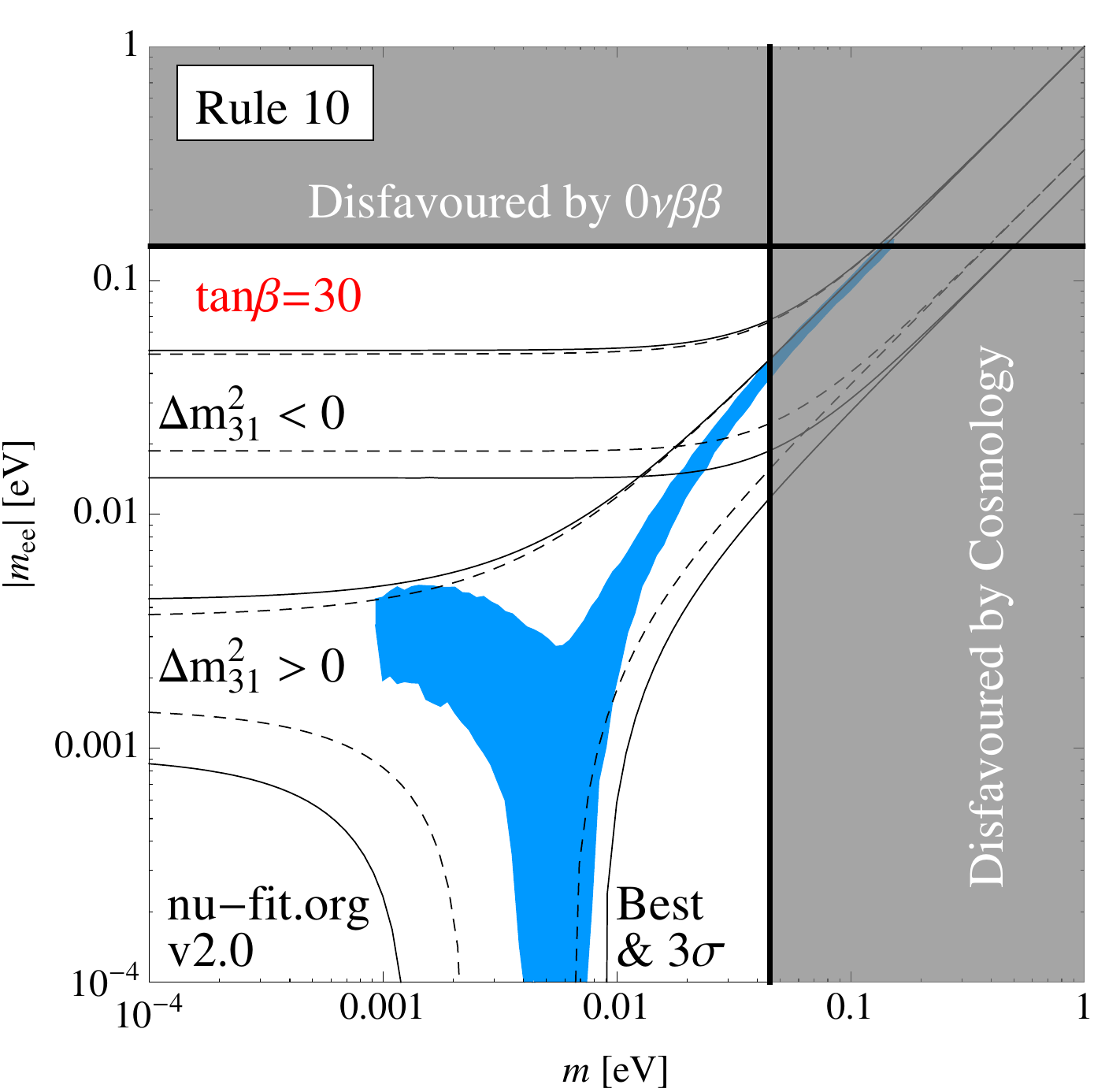} &
\includegraphics[width=5.4cm]{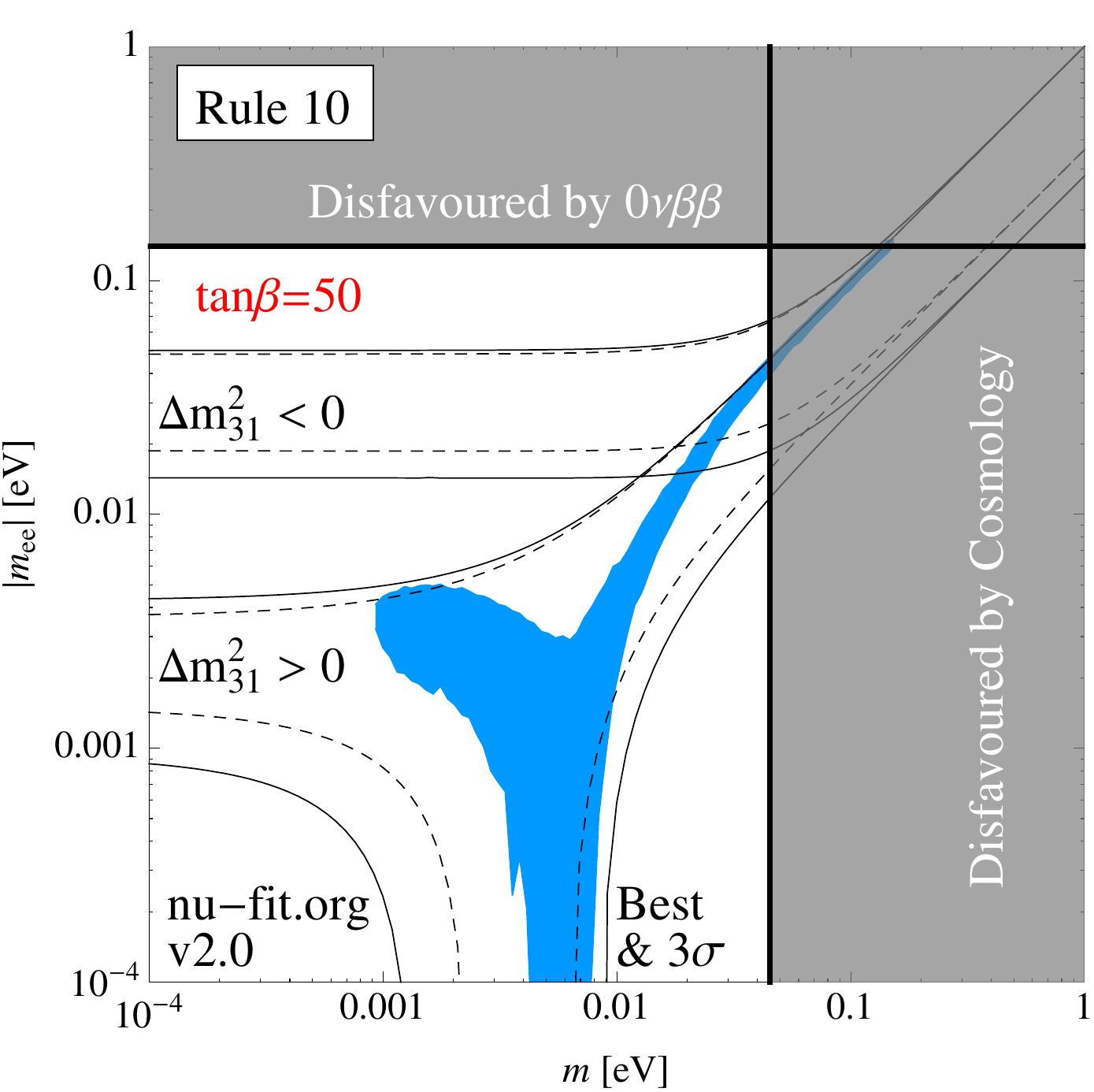}
\end{tabular}
\end{center}
For each value of $\eta$, this sum rule predicts normal ordering, and with the SM particle content it yields $(m_{\rm min}, |m_{ee}|_{\rm min}) = (0.00093,0.000014)$~eV, if the running is applied. These numbers are consistent with the values obtained in Ref.~\cite{King:2013psa} (see discussion in Secs.~7.4 and~7.13 therein). For $\tan \beta = 30$ ($50$), the values change to $(0.00093,0.000025)$~eV ($(0.00093,1.5\cdot 10^{-6})$~eV), while still only NO is allowed.

\subsection{\label{sec:SR11} Sum Rule 11: $\mathbf{3\sqrt{\tilde m_2}+3\sqrt{\tilde m_3}=\sqrt{\tilde m_1}}$}

The parameters for this sum rule are $(d, c_1, c_2, \Delta \chi_{13}, \Delta \chi_{23})=(1/2, 1/3, 1, \pi, 0)$, and the corresponding plots look like:
\begin{center}
\begin{tabular}{lll}
\hspace{-1cm}
\includegraphics[width=5.4cm]{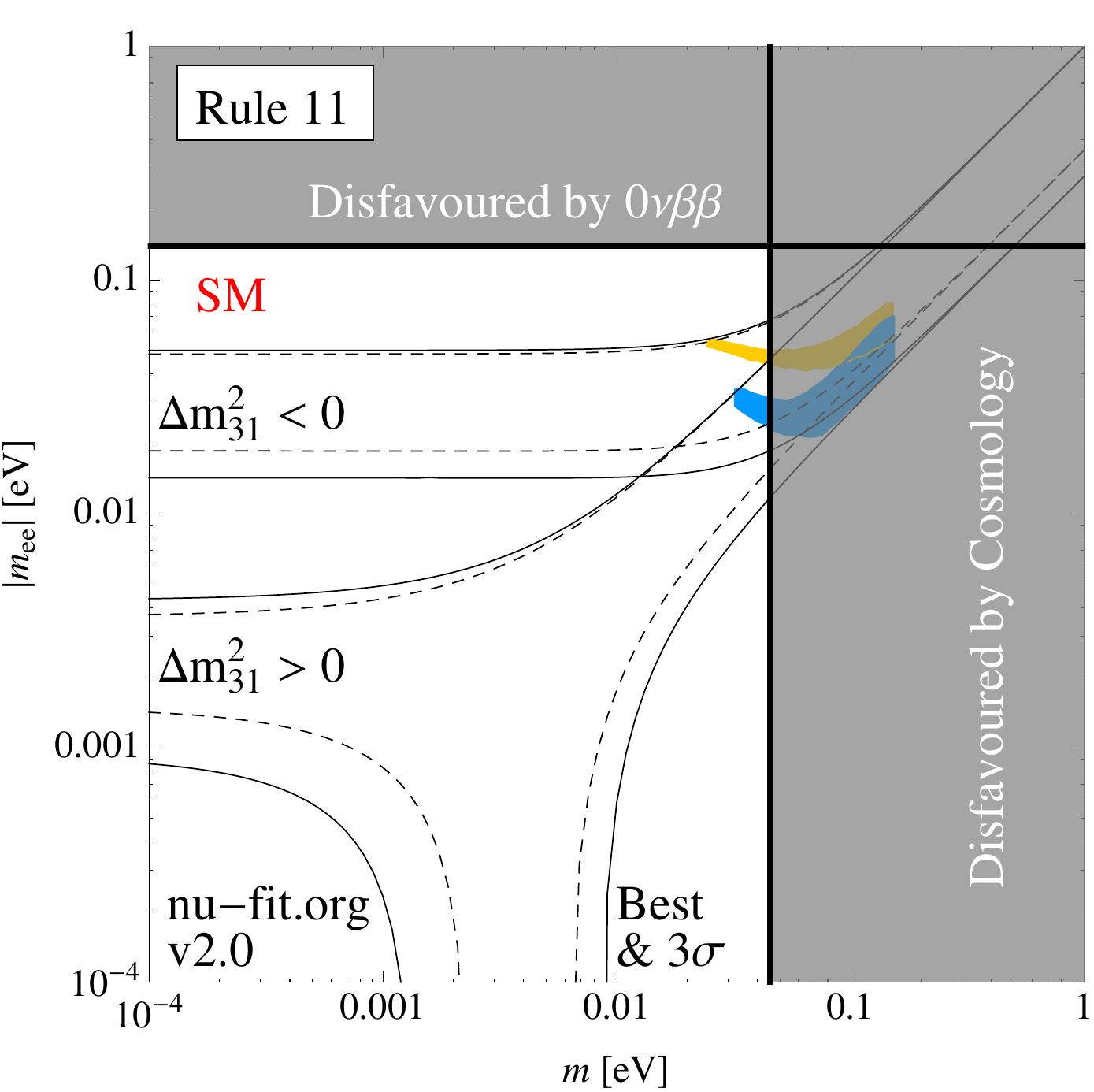} &
\includegraphics[width=5.4cm]{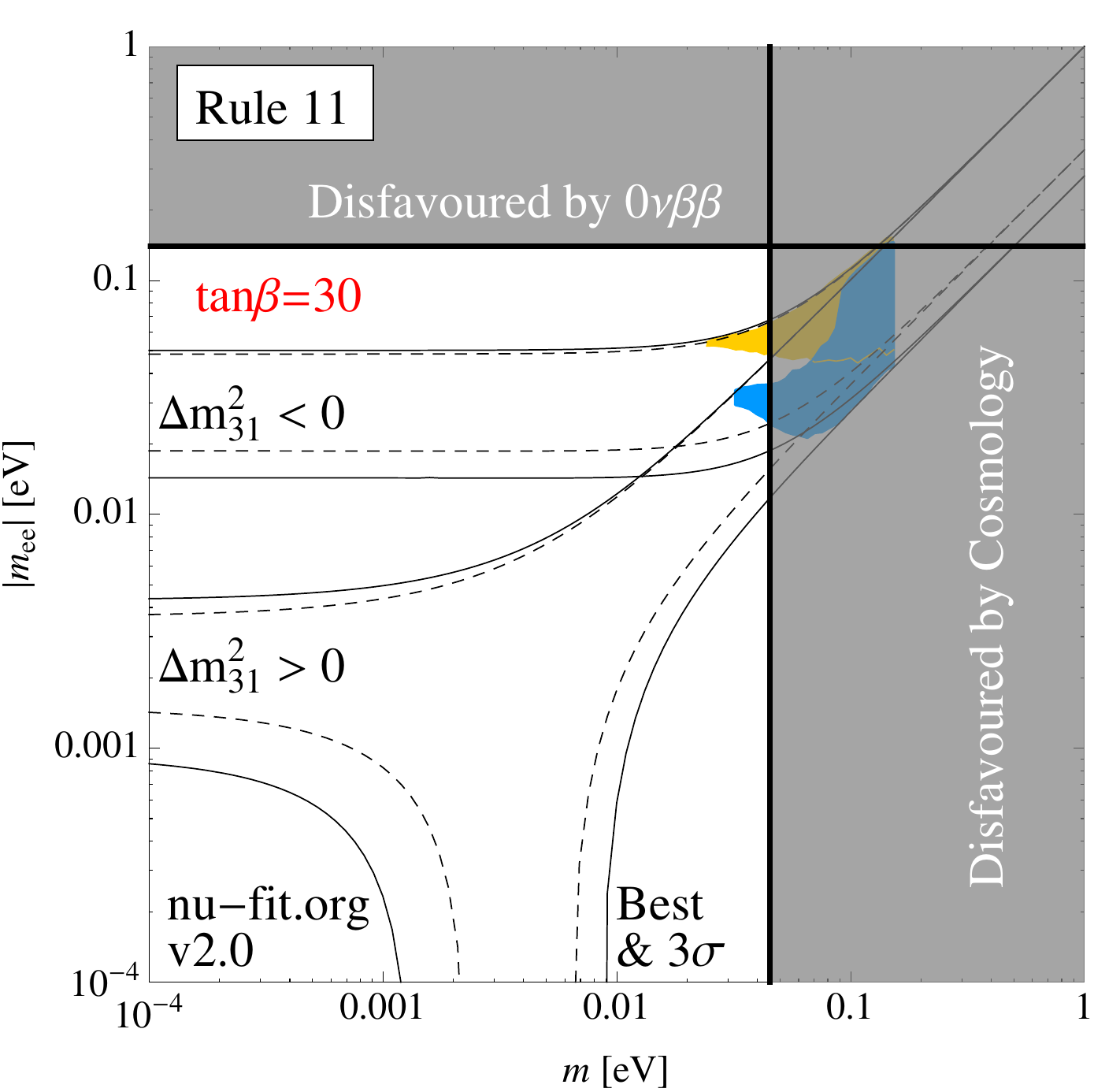} &
\includegraphics[width=5.4cm]{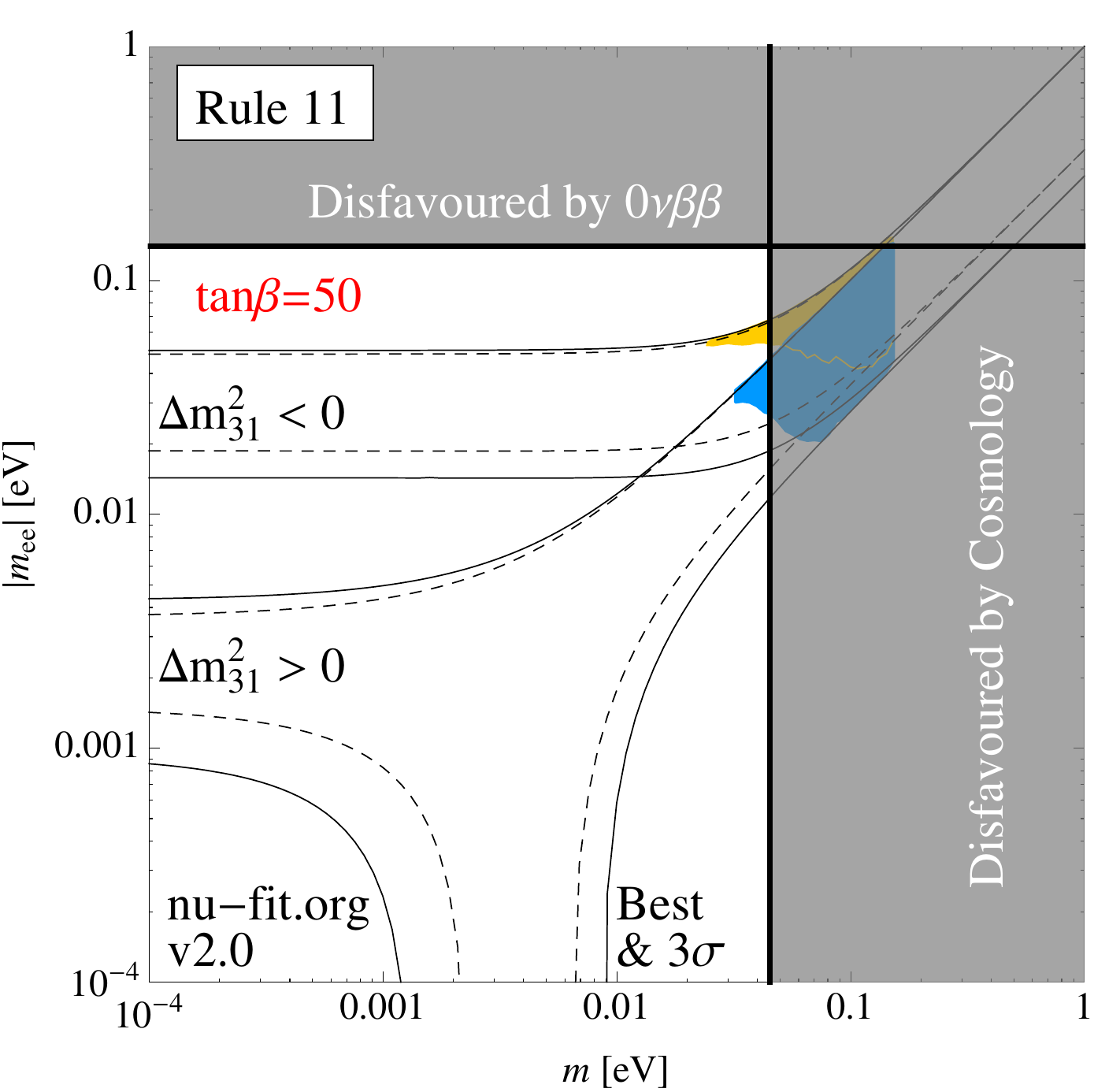}
\end{tabular}
\end{center}
This sum rule yields $(m_{\rm min}, |m_{ee}|_{\rm min}) = (0.032,0.022)$~eV ($(0.024,0.041)$~eV) for normal (inverted) mass ordering, if running with the SM particle content is applied. Since this sum rule is hypothetical, in the sense that no explicit underlying model is known yet, no numerical predictions have been listed in Ref.~\cite{King:2013psa} (see discussion in Sec.~7.5 therein). However, the left plot seems quasi identical to the one presented in that reference. For $\tan \beta = 30$ ($50$), the values change to $(0.032,0.021)$~eV ($(0.032,0.021)$~eV) for NO and to $(0.024,0.044)$~eV ($(0.024,0.042)$~eV) for IO, respectively.

\subsection{\label{sec:SR12} Sum Rule 12: $\mathbf{1/\sqrt{\tilde m_{1}}=2 /\sqrt{\tilde m_{3}}-1/\sqrt{\tilde m_{2}}}$}

The parameters for this sum rule are $(d, c_1, c_2, \Delta \chi_{13}, \Delta \chi_{23})=(-1/2, 1/2,  1/2, \pi, \pi)$, and the corresponding plots look like:
\begin{center}
\begin{tabular}{lll}
\hspace{-1cm}
\includegraphics[width=5.4cm]{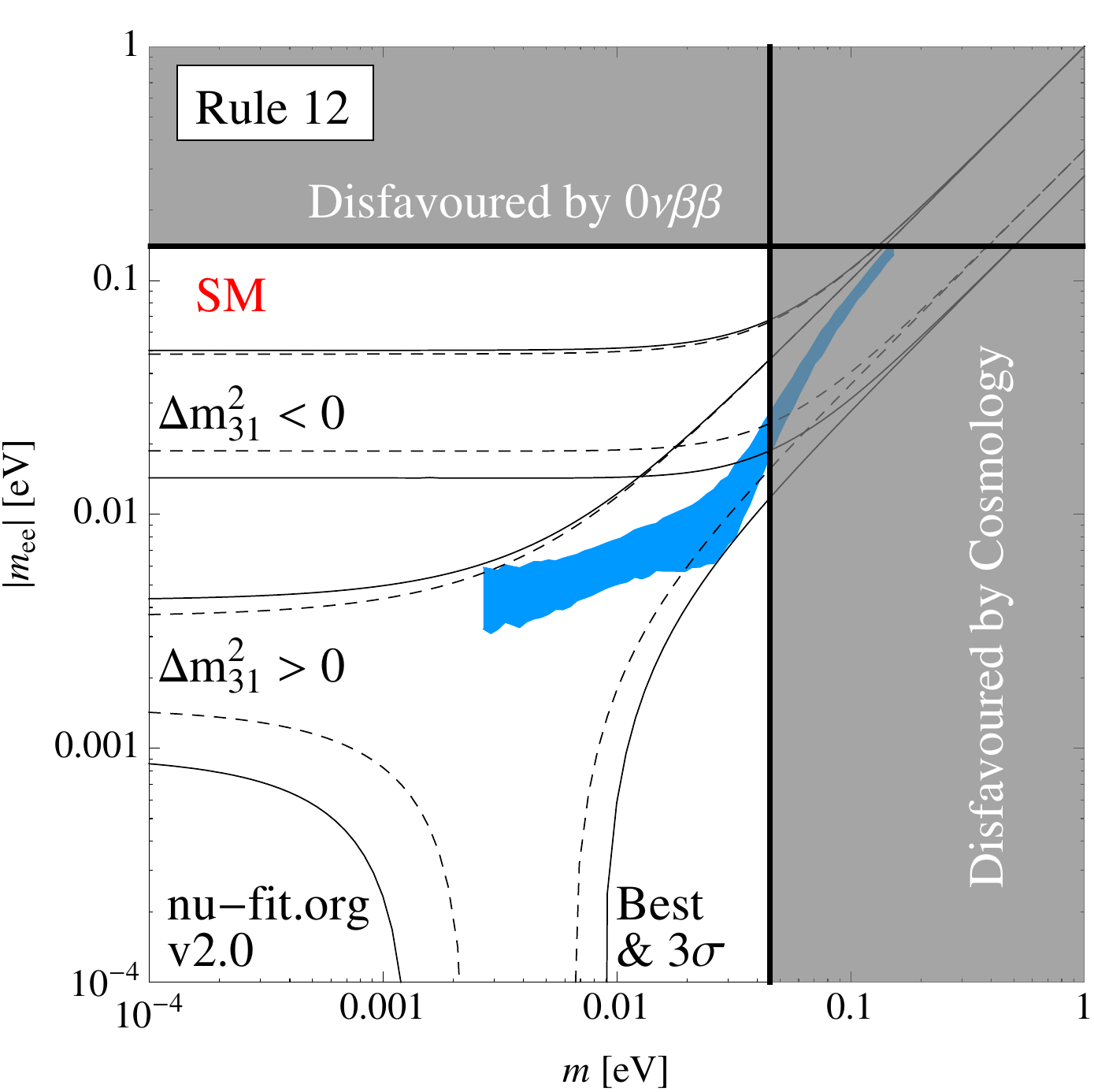} &
\includegraphics[width=5.4cm]{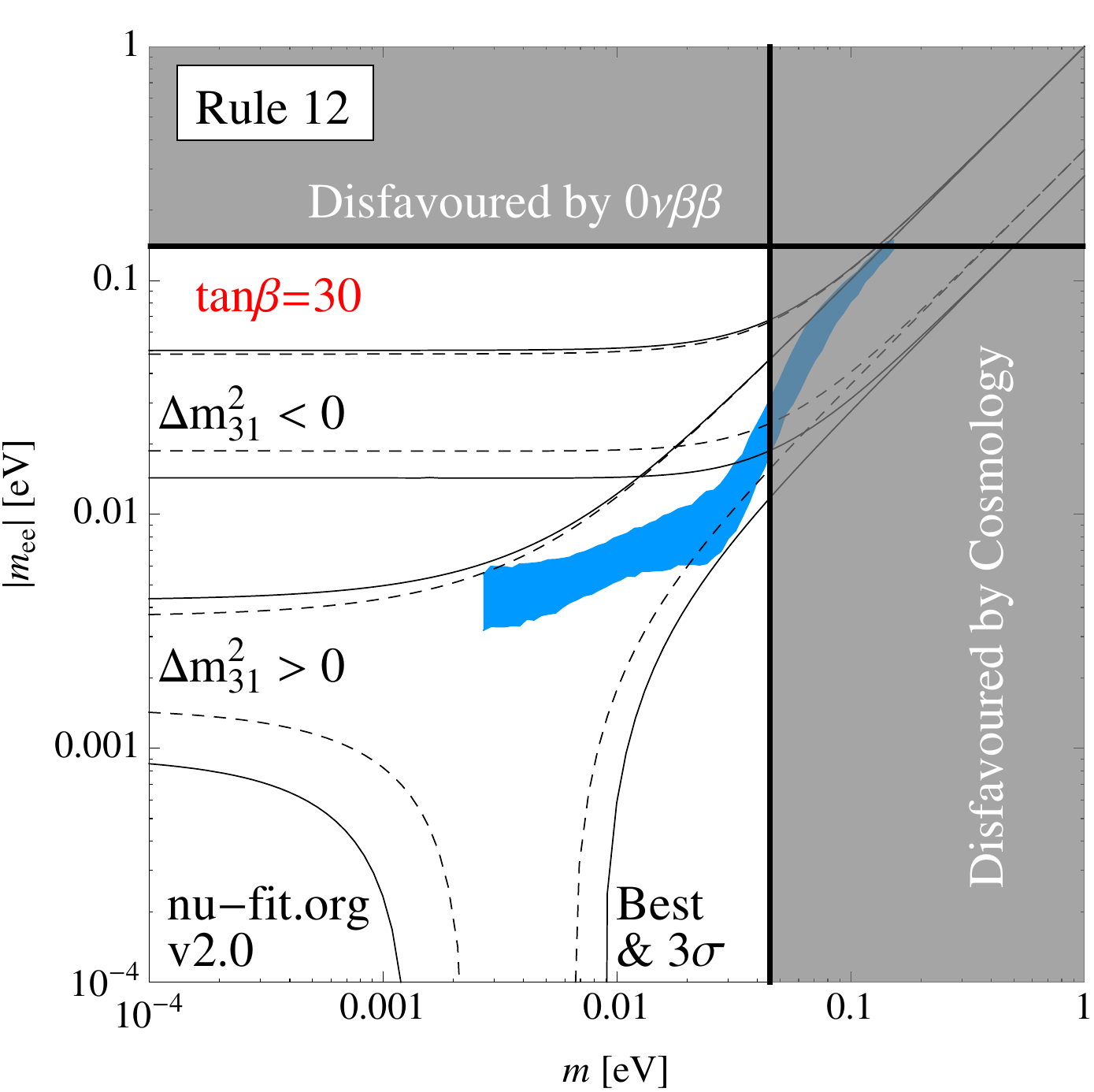} &
\includegraphics[width=5.4cm]{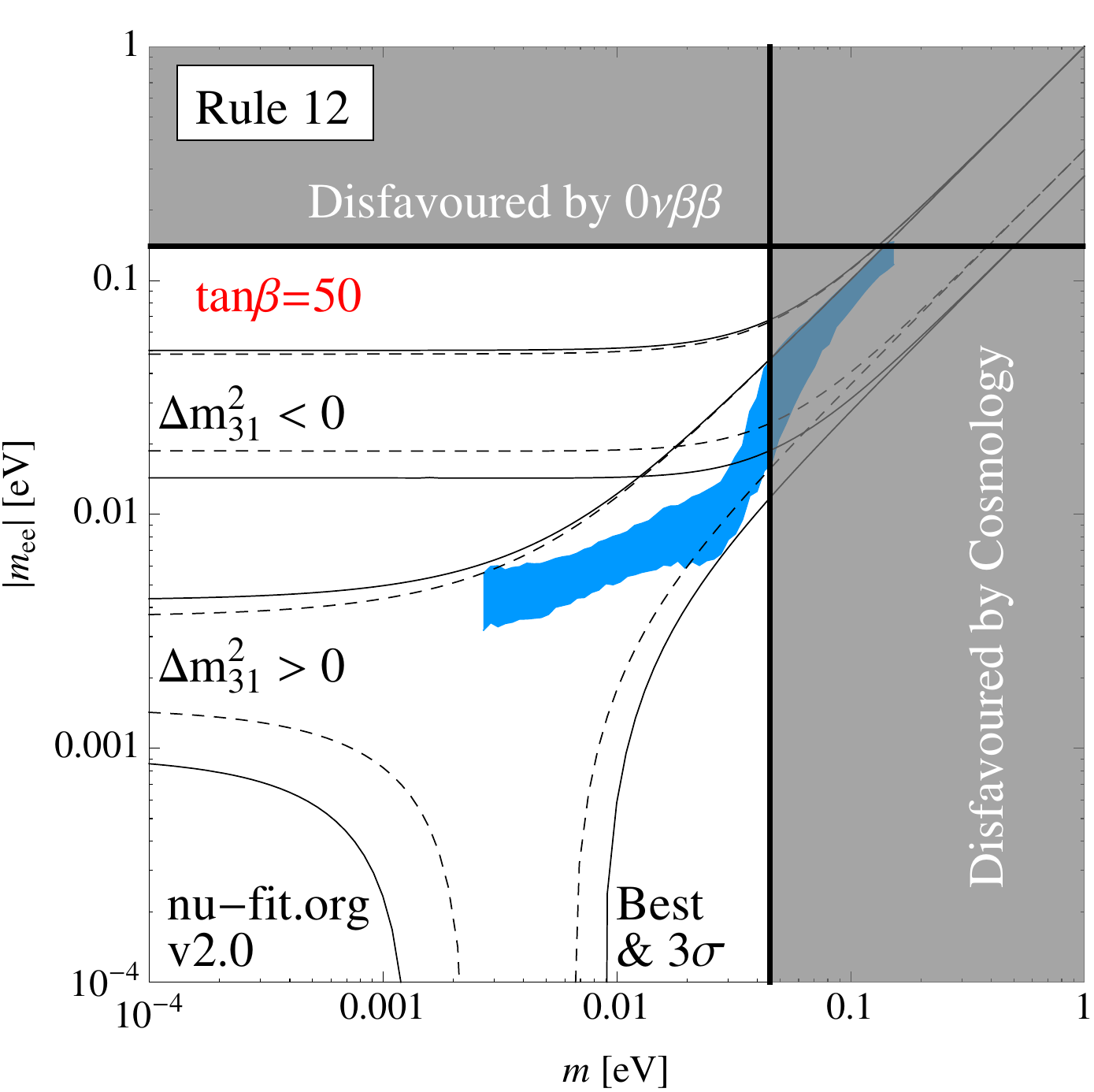}
\end{tabular}
\end{center}
This sum rule predicts normal ordering, and with the SM particle content it yields $(m_{\rm min}, |m_{ee}|_{\rm min}) = (0.0027,0.0031)$~eV, if the running is applied. These numbers are consistent with the values obtained in Ref.~\cite{King:2013psa} (see discussion in Sec.~7.3 therein). For $\tan \beta = 30$ ($50$), the values change to $(0.0027,0.0032)$~eV ($(0.0027,0.0032)$~eV), while still only NO is allowed.

\subsection{\label{sec:SR-discussion} Discussion}

As can be seen, RGEs can have a non-trivial effect on the regions allowed by certain sum rules. Although we ``only'' presented scatter plots (for a good reason though, see the discussion in Sec.~\ref{sec:numerical}), the tendency is clearly visible. In most cases (i.e., sum rules 1, 2, 4, 5, 6, 7, 8, 11, 12) the effect of the RGEs is to broaden the allowed regions, although in one case the broadening occured only within the parameter region that is already disfavoured by current neutrino mass bounds (sum rule~5) and in two cases it only appeared for inverted mass ordering, since only a small mass range for the normal ordering is allowed in these sum rules where we do not have an enhancement effect of the RGEs (sum rules 7 and 8). In a few cases (sum rules 3, 9, 10), there is no visible effect in the plots, even though -- numerically -- the parameters do run. These three sum rules at least naively seem to have nothing in common, so that a simple ``accidental'' cancellation of the running effects is unlikely. Rather, there is a more fundamental reason: in these sum rules, large RGE effects are suppressed by the values of the angles and the phases at the high scale.

Furthermore, as anticipated in Sec.~\ref{sec:numerical}, indeed we have in no case found points corresponding to a mass ordering that would be forbidden if the sum rule held exactly. More generally we have seen that the running has in many cases a visible but not a dramatic effect. The simple and intuitive reason for this is that the parameters in the neutrino sector are generally known to run relatively weakly (although exceptions do exist, see Ref.~\cite{Ma-running} for an example). Thus, even though the sum rules are in reality not anymore valid at the low scale, the running effects are sufficiently weak that the sum rules are nevertheless approximately fulfilled for all the points displayed, and thus their predictions are not spoiled. The small differences seen are negligible compared to the uncertainties coming from nuclear physics, which are however still not big enough as to destroy the testability of many groups of neutrino flavour models~\cite{King:2013psa,SR-GERDA}.

Hence, we have shown that the RGE effects do not change the qualitative predictions of the sum rules, but it should nevertheless not be neglected because they can even have an impact for a small mass scale. Especially in the regime with a large mass scale, we have shown that the running effects do broaden the allowed region whereas the absence of a visible broadening should be regarded as an exception where ``accidental'' cancellations take place.

Thus, while there may be further model-dependent corrections present in case a neutrino flavour model yields a sum rule, at least RGE corrections do not change the qualitative predictions of the sum rules. In most case, even the quantitative predictions are hardly changed, in particular when taking into account that the nuclear physics uncertainties will always dominate in a realistic measurement. In turn, the predictions by a certain sum rule are safe up to possible model-dependent effects, whose size can however typically be estimated or even computed exactly for a given realistic flavour model.

\section{\label{sec:conc}Summary and conclusions}

We have presented the first explicit and systematic study of the effect that radiative corrections have on the validity of neutrino mass sum rules. Since sum rules are able to yield very concrete predictions that are realistically testable with near-future experiments, it is important to take into account possible modifications if we are to truly put the models developed over more than a decade to the test. We have started this endeavour by numerically computing how the regions in the parameter space allowed by certain sum rules are affected if renormalisation group running is taken into account.

After briefly reviewing the most general form of a neutrino mass sum rule and a discussion of the general effect of renormalisation group running, we have explicitly computed the resulting allowed regions for all neutrino mass sum rules known if we assume the rules to hold exactly only at the seesaw scale, while correction terms appear when going to lower energies. The concrete settings we have used were a Standard Model-like scenario (where running effects are expected to be very small) and two scenarios corresponding to the minimal supersymmetric Standard Model (with $\tan \beta = 30$ and $50$, where we expect running effects to become stronger with larger $\tan \beta$). While we have explicitly verified these general tendencies, our results nevertheless show that the predictions derived from neutrino mass sum rules, although visibly changed by the corrections, are nevertheless quite stable due to the smallness of the effect (this holds unless the running was unusually strong). Three sum rules do not run because of cancellations in the $\beta$ functions, or at most by such a small amount that the resulting changes in the prediction regions in the $m$--$|m_{ee}|$ plane are basically invisible in the plots. In fact, not only experiments looking for neutrinoless double beta decay have an impact on neutrino mass sum rules. If accelerator experiments determine the neutrino mass ordering some cases are directly excluded. Furthermore the LHC can shed light on the question whether one has to apply the MSSM or the SM $\beta$ functions for the neutrino parameters. This is quite a difference because not only the size of the corrections differs but also the sign of the corrections changes.

Our findings considerably strengthen the position of neutrino flavour models featuring mass sum rules, since the predictions derived prove to be relatively insensitive to radiative corrections. This leads to a big advantage of such models compared to others not predicting any correlation between observables. The only caveat, apart from having a setting where the running is very strong, is that some concrete models may induce other big corrections that are completely unrelated to the running effects discussed here. While such effects may still be able to change the regions predicted by that specific sum rule (or maybe to even entirely destroy their validity) in that particular setting, typically both their origin and size would be clear in a concrete model, to the point that their strength may even be computed or estimated at least.

Thus, our results show that the most generic corrections one could think of are, in fact, not a problem for neutrino mass sum rules. These types of correlations hence exhibit a strong handle that can be used to realistically probe many neutrino flavour models already with upcoming experiments on neutrinoless double beta decay, without the need to wait for the precision era in neutrino flavour physics.

\section*{Acknowledgments}

We thank Michael A.~Schmidt for deeper insights into the \texttt{REAP/MPT} packages. JG acknowledges support by the DFG-funded research training group GRK 1694 ``Elementarteilchenphysik bei h\"ochster Energie und h\"ochster Pr\"azision''. AM acknowledges partial support from the European Union FP7 ITN-INVISIBLES (Marie Curie Actions, PITN-GA-2011-289442).

\appendix
\section{\label{sec:squareroot}Taking the square root of a complex number}

There are some subtleties in treating sum rules which include the square root of the masses.

For a positive real number $x$ one has to take both possibilities of the sign of the square root into account, i.e.~$\sqrt{x}=\pm|\sqrt{x}|$. For a complex number $z=\rho~ \text {e}^{\text{i}\chi}$, $\chi \in[0,2\pi]$, in turn, one encounters further subtleties. For example, one could either define $\sqrt{z}\equiv|\sqrt{\rho}|~ \text{e}^{\text{i}\chi/2}$, where $\chi \in[0,2 \pi]$, or one could alternatively define $\sqrt{z}\equiv|\sqrt{\rho}|~ \text{e}^{\text{i}\chi/2}$, where $\chi \in[-\pi,\pi]$. In the first case the result lies within the upper half of the complex plane, where Im($\sqrt{z})>0$ whereas in the second example the result is in the right half of the complex plane, with Re($\sqrt{z})>0$. Depending on the definition, the final results will differ. And special care has to be taken that in a numerical setup the two definitions are not messed up. For instance, REAP suggests the convention Im($\sqrt{z})>0$ while Mathematica uses Re($\sqrt{z})>0$.

 To cover the whole complex plane one should furthermore consider solutions with a negative sign of the square root. In our first example, this means that we also have to consider $\sqrt{z}=-|\sqrt{\rho}|~ \text{e}^{\text{i}\chi/2}$. In the following we will employ the definition of the square root according to our first example: $\sqrt{z}\equiv\pm|\sqrt{\rho}|~ \text{e}^{\text{i}\chi/2}$, where  $\chi \in[0,2\pi]$. Especially in the case of mass sum rules which include the square root of the complex neutrino masses a proper definition of the square roots is essential since the phases of the neutrino masses have a physical meaning: they are the physical Majorana phases.

 \begin{figure}
\centering
\includegraphics[scale=0.9]{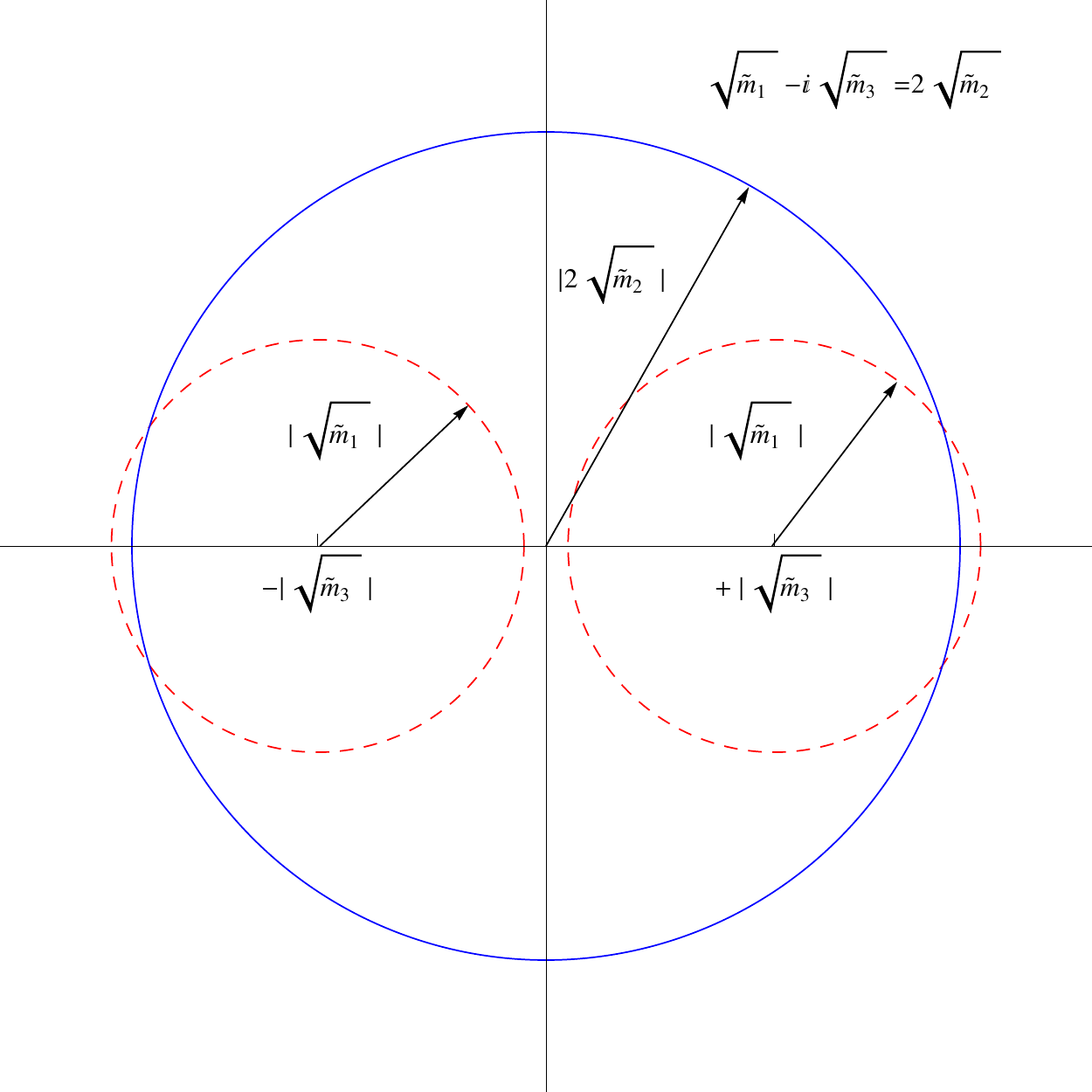}
\caption{Graphical representation of the sum rule $\sqrt{\tilde m_1}- \text{i} \, \sqrt{\tilde m_3}=2\sqrt{\tilde m_2}$. The small red dashed circles represent the left
hand side of this equation, the big blue circle represents the right hand side.}
\label{fig:sumrule10}
\end{figure}
 
An example for a mass sum rule including square roots of the masses is proposed in~\cite{Hirsch:2008rp}. For $\eta = 1$ the mass sum rule~10 (see Sec.~\ref{sec:SR10}) reads
\begin{align}
\sqrt{\tilde{m}_1} - \text{i}\,\sqrt{\tilde{m}_3}=2\sqrt{\tilde{m}_2},
\label{eq:sumrule10}
\end{align}
where the masses are all complex. They depend on the complex parameters  $a ,b$ defined
in the model 
\begin{align}
\tilde{m}_1&=(a+b)^2~,\\
\tilde{m}_2&=a^2~,\\
\tilde{m}_3&=-(a-b)^2~.
\end{align}
To get a graphical representation of the sum rule we can, e.g., choose the mass
$\tilde{m}_3$ to be real and positive, $\tilde{m}_3 = m_3$, since we can absorb one phase as a global phase factor. The phases of $\tilde{m}_1$ and $\tilde{m}_2$ are then the physical Majorana phases. The graphical representation is given in Fig.~\ref{fig:sumrule10}. The red dashed circles represent the left hand side of Eq.~\eqref{eq:sumrule10}, while the blue circle represents its right hand side.  We have taken into account both possible signs  for $\sqrt{\tilde m_3}$, which correspond to the centres of the small red circles with radius $|\sqrt{\tilde{m}_1}|$. The big blue circle with radius $|2 \sqrt{ \tilde{m}_2}|$ is centred around the origin. The sum rule is fulfilled if and only if the circles have an intersection. If we consider only the positive solution of $\sqrt{\tilde m_3}$, the intersections of the circles in the half-plane where Re($\sqrt{\tilde m_3})<0$ are absent. Since the angles in the triangles formed by the intersections of the circles are related to the Majorana phases whose interval is the whole complex plane, we would miss two physical solutions. As the circles have four intersection points, we therefore conclude that there are four solutions for the  Majorana phases which are in accordance with the sum rule, from which only two are physical (the other two solutions give the same results).

From this construction we can as well convince ourselves that the three values of
$\eta$ from sum rule 10 all give the same result. First of all, that the two cases
$\eta = 0$ and $\eta = 2$ are equivalent is obvious since by construction we have chosen
as the center of the red circles $\pm \sqrt{\tilde{m}_3}$ anyway. The third case with $\eta = 1$
can be rewritten to
\begin{equation}
 \sqrt{- \tilde{m}_1} + \sqrt{\tilde{m}_3} = 2 \sqrt{- \tilde{m}_2} \;,
\end{equation}
which just mirrors the blue and red circles along the horizontal axis (it adds $\pi$ to
the Majorana phases). So this simply interchanges the two physical and the two unphysical
(redundant) solutions with each other.

Similar considerations can be done for the other sum rules involving square roots,
such that there could be equivalent sum rules with additional signs and factors of $\text{i}$.
But in this study we have quoted only the sum rules we have found in the literature and the
underlying model fixes the concrete form of the sum rule and hence we do not claim that
we list all possible mass sum rules.

Regarding the sum rules which do not include square roots of masses we only obtain two solutions for Majorana phases, since we chose $\tilde{m}_3 = m_3$ to be positive via the redefintion of the phases. Hence we only find one circle around $\tilde m_3$ in the right complex half-plane with Re$(\tilde m_3)>0$.

In principle more sum rules can arise when taking different signs of the square roots of the masses 
into account. For all the reviewed sum rules which include square roots of the masses we have checked that there is only the quoted combination
of signs of the square roots which leads to a valid sum rule.
This means that there is only one possiblity to form a triangle when we interpret the sum rule geometrically.

\end{document}